\documentclass[a4paper,11pt]{article}
\pdfoutput=1 \usepackage{amssymb}
\usepackage{amsmath}
\usepackage{braket}
\usepackage{mathtools}
\usepackage[usenames,dvipsnames,svgnames,table]{xcolor}
\usepackage[utf8]{inputenc}
\usepackage{color}
\usepackage{cite}
\usepackage{subcaption}
\usepackage{lmodern}
\usepackage{footnote}
\usepackage[normalem]{ulem}
\usepackage{glossaries-extra}
\setabbreviationstyle[acronym]{long-short}
\glssetcategoryattribute{acronym}{nohyperfirst}{true}

\usepackage{slashed}
\usepackage[pdftex]{graphicx}
\usepackage{multirow}
\usepackage{jheppub}
\usepackage{here}
\usepackage{hyperref}
\usepackage{verbatim}
\usepackage[justification=centering,singlelinecheck=false]{caption}
\usepackage{cleveref}
\usepackage{listings}
\usepackage{fancyvrb}
\usepackage{lipsum}
\usepackage{mwe}
\usepackage{dsfont}
\usepackage{nicefrac,xfrac}
\usepackage{setspace}

\Crefname{equation}{eq.}{eqs.}
\Crefname{section}{section}{sections}
\Crefname{figure}{figure}{figures}
\Crefname{appendix}{appendix}{appendices}

\makeatletter
\AddToHook{cmd/appendix/before}{\def\cref@section@alias{appendix}\def\cref@subsection@alias{appendix}}
\makeatother

\newcommand{\cA}{\mathcal{A}}

\newcommand{\cD}[0]{\mathcal D}

\newcommand{\cK}[0]{\mathcal K}

\newcommand{\cM}[0]{\mathcal M}
\newcommand{\cO}[0]{\mathcal O}
\newcommand{\cP}[0]{\mathcal P}

\newcommand{\cY}[0]{\mathcal Y}

\newcommand{\mc}[0]{\mathcal}

\newcommand{\df}[0]{\mathrm{df}}

\newcommand{\Kdf}[0]{{\cK_{\df,3}}}

\newcommand{\bm}[0]{\boldsymbol}

\newcommand{\HSQCa}[0]{Hansen:2014eka}
\newcommand{\HSQCb}[0]{Hansen:2015zga}

\newcommand{\BHSnum}[0]{Briceno:2018mlh}

\newcommand{\dwave}[0]{Blanton:2019igq}

\newcommand{\largera}[0]{Romero-Lopez:2019qrt}

\newcommand{\threehadrons}[0]{Blanton:2021llb}
\newcommand{\implement}[0]{Blanton:2021eyf}

\newcommand{\BStwoplusone}[0]{Blanton:2021mih}
\newcommand{\threeN}[0]{Draper:2023xvu}

\newcommand{\DRS}[0]{Dawid:2024dgy}

\newcommand{\Npp}[0]{Hansen:2025oag}
\newcommand{\threepPRD}[0]{Dawid:2025doq}
\newcommand{\NumQCthree}[0]{%
Briceno:2018mlh,
Mai:2018djl,
Horz:2019rrn,
\dwave,
Blanton:2019vdk,
Mai:2019fba,
Culver:2019vvu,
\largera,
NPLQCD:2020ozd,
Fischer:2020jzp,
Hansen:2020otl,
Alexandru:2020xqf,
Brett:2021wyd,
Blanton:2021llb,
Mai:2021nul,
Garofalo:2022pux,
Draper:2023boj,
Yan:2024gwp,
Dawid:2024dgy,
Dawid:2025zxc,
Dawid:2025doq,
Yan:2025mdm,
Briceno:2025yuq}

\newcommand{\ThreeBody}[0]{%
Konig:2017krd,
Briceno:2017tce,
Hammer:2017uqm,
Hammer:2017kms,
Mai:2017bge,
Briceno:2018aml,
Blanton:2019igq,
Pang:2019dfe,
Briceno:2019muc,
Jackura:2019bmu,
Romero-Lopez:2019qrt,
Hansen:2020zhy,
Blanton:2020gha,
Blanton:2020jnm,
Pang:2020pkl,
Romero-Lopez:2020rdq,
Blanton:2020gmf,
Muller:2020vtt,
Muller:2020wjo,
Hansen:2021ofl,
Blanton:2021mih,
Mai:2021nul,
Muller:2021uur,
Blanton:2021eyf,
Jackura:2022gib,
Muller:2022oyw,
Draper:2023xvu,
Bubna:2023oxo,
Hansen:2024ffk,
Draper:2024qeh,
Yan:2024gwp,
Xiao:2024dyw,
Hansen:2025oag,
Yan:2025mdm}

\newcommand{\IntEqs}[0]{%
Jackura:2020bsk,
Hansen:2020otl,
Mai:2021nul,
Dawid:2023jrj,
Dawid:2023kxu,
Jackura:2023qtp,
Dawid:2024dgy,
Briceno:2024ehy,
Dawid:2025doq,
Dawid:2025wsn,
Yan:2025mdm,
Jackura:2025wbw,
Briceno:2025yuq}

\newacronym{CMF}{CMF}{center-of-momentum frame}

\DeclareFixedFont{\ttb}{T1}{txtt}{bx}{n}{9}
\DeclareFixedFont{\ttm}{T1}{txtt}{m}{n}{9}

\definecolor{deepblue}{rgb}{0,0,0.5}
\definecolor{deepred}{rgb}{0.6,0,0}
\definecolor{deepgreen}{rgb}{0,0.5,0}
\definecolor{jlab_red}{RGB}{192,39,45}
\definecolor{jlab_orange}{RGB}{249,102,0}
\definecolor{jlab_blue}{RGB}{47,122,121}
\definecolor{jlab_green}{RGB}{65,125,10}
\definecolor{jlab_blue}{RGB}{47,122,121}

\newcommand{\steve}[0]{\color{magenta}}

\title{Implementing the three-neutron quantization condition}

\author[a]{Wilder Schaaf}
\affiliation[a]{Physics Department, University of Washington, Seattle, WA 98195-1560, USA}

\author[a]{and~Stephen R. Sharpe}

\emailAdd{wschaaf@uw.edu}
\emailAdd{srsharpe@uw.edu}

\abstract{

We describe in detail the implementation of the relativistic three-neutron finite-volume quantization condition
derived in Ref.~\cite{\threeN}.
In particular, we show how the complications due to Wigner rotations acting on spins are included, 
and present concrete formulas for the case when the angular momenta within pairs is restricted to be less than $2$.
We describe the symmetries of the matrices appearing in the quantization condition, and decompose solutions
into irreducible representations of the appropriate doubled finite-volume symmetry groups.
We present an implementation of the three-particle K matrix, 
keeping the two lowest-order terms in the threshold expansion.
We provide numerical predictions for the finite-volume spectrum for a setup with nearly physical parameters,
including two-particle interactions that are based on experimental results.
This exploratory study shows the how lattice QCD calculations of the three-neutron spectrum with sufficient precision can provide detailed information on both two- and three-particle interactions.
}

\allowdisplaybreaks

\begin{document}
\today
\begin{spacing}{1.08}
\maketitle
\end{spacing}
\flushbottom
\clearpage

\section{Introduction}
\label{sec:intro}

The determination of multinucleon interactions from the underlying theory of the strong interactions, QCD, 
is a major theoretical challenge.
A first-principles approach using lattice simulations holds great promise, but faces significant numerical, algorithmic, and theoretical challenges. In particular, the extraction of two-nucleon scattering amplitudes using lattice QCD (LQCD) has had a long and controversial history, although recently a consensus picture appears to be emerging at heavier than physical quark masses~\cite{Detmold:2024iwz,Green:2025rel,Aoki:2025abn,BaSc:2025yhy}. 
This is based on both the 
L\"uscher approach---converting finite-volume spectra into scattering amplitudes~\cite{Luscher:1986n2,Luscher:1991n1,Luscher:1991n2}---and the HALQCD approach, which determines inter-nucleon potentials from Bethe-Salpeter wavefunctions~\cite{Ishii:2006ec,Ishii:2012ssm}.
We stress that for two-particle interactions, the theoretical formalism in both approaches is relatively mature;
progress has been held back largely by the challenge of obtaining reliable energy levels.

In this work we look into the future, and assume that LQCD methods will improve to the point that precise results for the finite-volume spectrum of systems of {\em three} neutrons will become available.
We then ask the following question: What precision is required in order that LQCD results for the spectrum (or other quantities) can provide detailed information on  two- and three-neutron interactions?
We are particularly interested in the three-neutron interaction, since there is no direct experimental input for this quantity, and yet it plays an important role in determining the properties of large nuclei and neutron stars~\cite{Hoppe:2019uyw,Drischler:2020hwi,Machleidt:2024bwl}.

In order to address this question we choose to follow the generalization of the L\"uscher method to three particles,
using the generic relativistic field theory (RFT) approach. This was first used to study systems of three identical spinless particles~\cite{\HSQCa,\HSQCb}, where it was shown how to convert finite-volume energy levels for two and three particles into infinite-volume scattering amplitudes by following two steps.
In the first, one inputs two- and three-particle K matrices into quantization conditions, and adjusts these matrices until the predicted spectrum matches that obtained from LQCD. In the second step, one inputs these K matrices into integral equations, the solutions of which yield the physical scattering amplitudes.
The three-particle formalism has subsequently been generalized to nearly all systems of phenomenological interest~\cite{\ThreeBody},
and several implementations of the quantization conditions~\cite{\NumQCthree}
and integral equations~\cite{\IntEqs} have been developed.
The relevant generalization for this work is that of Ref.~\cite{\threeN}, where the quantization condition and integral equations for three neutrons was derived. The main new feature of this derivation is the inclusion of the spin degrees of freedom, and, in particular, their transformation as one boosts between the rest frames of different neutron pairs.\footnote{%
We stress that the formalism holds only for isosymmetric QCD, i.e. for $m_u=m_d$, and does not include QED effects. 
}

The analysis of Ref.~\cite{\threeN} provided the requisite formalism, but significant further  work is needed to turn this into a practical tool. The required work is quite different for the two steps described above, and in this paper we focus entirely on the first step, namely the implementation of the three-neutron quantization condition.

This paper is organized as follows. In 
\Cref{sec:recap} we recapitulate the quantization condition and its building blocks, streamlining the notation of Ref.~\cite{\threeN} in a few places, and adding more details on the cutoff function that is an essential part of the formalism.
Next, in \Cref{sec:implement}, we describe the detailed steps needed to implement each of the building blocks.
The most complicated quantities to implement are the ``switch matrix'' $G$ (see \Cref{sec:implementG}),
and the three-particle K matrix $\Kdf$ (see \Cref{sec:implementKdf}).
The effect of Wigner rotations acting on spin degrees of freedom is particularly complicated for $\Kdf$,
and we relegate some of the technical details to \Cref{app:Ibar,app:KB,app:KA}.

An important part of the implementation, which was not discussed in Ref.~\cite{\threeN}, is the decomposition of
energy levels into irreducible representations (irreps) of the appropriate finite-volume symmetry groups.
In \Cref{sec:symm} we describe the symmetry groups, and determine the irrep decomposition of the building blocks of the quantization condition, of noninteracting three neutron finite-volume states, and of $\Kdf$.
Some technical details are discussed in \Cref{app:projdims}.

With the implementation in hand, we present, in \Cref{sec:num}, the results of a numerical exploration of the predictions of the quantization condition, using two-particle interactions motivated by experimental results.
We study the level splittings, and their dependence on $\Kdf$, with an eye to providing an answer to the question raised above.
Some additional numerical results are collected in \Cref{app:moreframes}, and examples of
a class of unphysical solutions for $\Kdf\ne 0$ are discussed in \Cref{app:unphys}.

We close in \Cref{sec:conc} with a summary and outlook.

Preliminary results from this project were presented in Ref.~\cite{Schaaf:2024qer}.

\section{Recap of three-neutron quantization condition}
\label{sec:recap}

We assume a cubic box of side $L$ and periodic boundary conditions.
The quantization condition then takes the standard form in the RFT approach,
\begin{equation}
\det_{\boldsymbol k, \ell, m, \bm m^*_s} \left[1+\cK_{\rm{df},3}(E^*) F_3(E,\bm P,L)\right]=0\,.
\label{eq:QC3}
\end{equation}
Here we have changed the notation slightly compared to Ref.~\cite{\threeN},
in which matrices were given as bold-faced quantities and included additional factors of $i$ and $L^3$ compared to those we use below.\footnote{%
The explicit relations are $\textbf G = i G/L^3$, $\textbf F = i F/L^3$, $\textbf K_2 = i L^3 \cK_{2,L}$,
$\textbf F_3 = i F_3$, and $\textbf K_{\rm df,3} = i \Kdf$.}
We make this (essentially trivial) change as it brings subsequent expressions in line with earlier RFT works, and matches our explicit implementation.
In \Cref{eq:QC3}, $\cK_{\rm{df},3}$ is the three-particle K matrix, 
which parametrizes short-range three-particle interactions.
It is an infinite-volume quantity that depends on the center-of-mass frame (CMF) energy $E^*$,
a quantity given by $E^* = \sqrt{E^2-\bm P^2}$, where $E$ is the total energy and $\bm P$ the total momentum.
$F_3$ is given by
\begin{equation}
    F_3=\frac{F}{3}-F\frac{1}{\cK_{2,L}^{-1}+F+G}F \,, 
\label{eq:F3def}
\end{equation}
where $F$ and $G$ are known kinematic quantities that depend on $E$, $\bm P$, and $L$, 
while $\cK_{2,L}$ contains the two-particle K matrix. Explicit expressions will be given below.

The quantization condition is valid, up to corrections that vanish exponentially with $L$, only for a range of 
energies~\cite{\threeN}
\begin{equation}
M_N + \sqrt{4 M_N^2 - M_\pi^2} < E^* < 3 M_N + M_\pi\,.
\label{eq:Erange}
\end{equation}
These constraints avoid intermediate states that involve more than three particles, for which no formalism presently exists.
The upper limit is set by the inelastic threshold.
The lower limit is set by the presence of the left-hand cut in the two-neutron amplitude, which arises from a (virtual) pion exchange. This cut is discussed further in \Cref{sec:cutoff}.
We stress that the exponentially-suppressed corrections grow as one approaches the limits, so one must stay some distance away.
We comment in \Cref{sec:num} on how this impacts the applicability of the formalism in practice.

The four matrices entering into the quantization condition---$F$, $G$, $\cK_{2,L}$, and $\Kdf$---have 
indices $k, \ell, m, \bm m^*_s$, which we now explain. 
The first three are common to all RFT three-particle quantization conditions: 
$k$ is shorthand for $\boldsymbol k$, which is the momentum of one of the three neutrons---the ``spectator''---and 
is drawn from the finite-volume set: $\boldsymbol k \in (2\pi/L) \mathbb Z$;
$\ell, m$ describe the angular momentum of the remaining pair (or ``dimer'') of neutrons in their c.m.~frame.
The final index, $\bm m^*_s$, is special to the three-neutron system and 
describes the spin degrees of freedom.
The details of this index are somewhat subtle, and are discussed in great detail in Ref.~\cite{\threeN}.
Here we summarize the result of that discussion.

The three spin components are defined in what is referred to in Ref.~\cite{\threeN} as the ``dimer-axis frame''.
Specifically, the vector of spin components in this frame is given by\footnote{%
Here we abbreviate the notation compared to that of Ref.~\cite{\threeN}, replacing 
$m_s(\bm k)$ with $m_{sk}$, $m_s(\bm a^*)$ with $m^*_{sa}$, and $m_s(\bm b^*)$ with $m^*_{sb}$.}
\begin{equation}
\bm m^*_s =
\big (m_{sk}, m^*_{sa}, m^*_{sb} \big ) \,,
\label{eq:dimeraxisms}
\end{equation}
where the second subscript of each spin component indicates which of the neutrons is being considered:
$k$ for the spectator, and $a$ and $b$ for the two members of the pair. 
The asterisks on $m^*_{sa}$ and $m^*_{sb}$ indicate that these spins are defined in the pair c.m.~frame,
while the absence of an asterisk on $m_{sk}$ indicates that the spectator spin is defined in the lab frame.
The choice of frame matters because of the Wigner rotations that appear when one combines two boosts,
as explained in Ref.~\cite{\threeN}.
The reason for this hybrid choice of spin indices is so that the spin and orbital angular-momentum of the pair can be combined in a simple manner, as in nonrelativistic QM.

While the composite spin index runs over $2^3=8$ values, those for $\bm k$, $\ell$ and $m$ have, {\em a priori}, an infinite range. However, the formalism incorporates a cutoff function, $H(\bm k)$, to be described in \Cref{sec:cutoff} below, that truncates the sum over $\bm k$ to a finite set of values. For $\ell,m$ there is no intrinsic cutoff in the formalism, and in  practical applications one has to truncate these indices by hand. The justification for this approximation is the fact that amplitudes in higher waves are kinematically suppressed close to threshold.
In this work we use $\ell_{\rm max}=1$.
After this truncation, the matrices in the quantization condition are finite, and the solutions can be found by straightforward matrix manipulations.

In the following subsections we collect the expressions for the matrices entering the quantization condition.

\subsection{Form of $F$}
\label{sec:QCF}

We begin with $F$, which corresponds to a ``finite-volume cut'' on the pair, with the spectator simply spectating.
It is given by [Eq.~(3.36) of Ref.~\cite{\threeN}]
\begin{align}
F_{p \ell'm' \bm m'^*_s ; k\ell m \bm m^*_s}  &=
 \delta_{m_{sp} m_{sk}} \delta_{m^*_{sa'} m^*_{sa}} \delta_{m^*_{sb'} m^*_{sb}}
 F^{\sf scal}_{p\ell' m';k \ell m}\,,
\label{eq:Fmat}
\\
F^{\sf scal}_{p\ell' m';k \ell m} &= \delta_{\bm p \bm k}
 \frac{ H(\boldsymbol k)}{2 \omega_k}
\frac12 \bigg [\frac{1}{L^3} \sum_{\boldsymbol r}^{\rm UV} - {\text{p.v.}} \int^{\rm UV}_{\boldsymbol r} \bigg ]
\times \frac{\mathcal Y_{\ell' m'}(\boldsymbol r^*_k)
\mathcal Y^*_{\ell m}(\boldsymbol r^*_k)}{2 \omega_a (b_{kr}^2 - M_N^2) } \frac{1}{(q^*_{k})^{\ell'+\ell}} \,.
\label{eq:Fscal}
\end{align}
The final-state momentum is $\bm p$, and the labels for the members of the final-state pair are $a'$ and $b'$.
The final state composite spin vector is
\begin{equation}
\bm m'^*_s =
\big (m_{sp}, m^*_{sa'}, m^*_{sb'} \big ) \,.
\label{eq:dimeraxismsfinal}
\end{equation}
We denote the energy of an on-shell four-momentum by $\omega$;
for example $\omega_k=\sqrt{\bm k^2+M_N^2}$ is the energy of the spectator, with $M$ the neutron mass.
The sum over $\bm r$ runs over the finite-volume set $(2\pi/L) \mathbb Z^3$, 
where we are assuming periodic boundary conditions, while $\int_{\bm r} = \int d^3 r/(2\pi)^3$.
The integral over the pole is regulated with the principle value (p.v.) prescription.
The ``UV'' superscripts indicate that both sum and integral are regulated in the ultraviolet (UV) in the same manner; the choice of regulator is irrelevant as the sum-integral difference is dominated by pole in the infrared (IR).
The four-momentum $b_{kr}$ is given by $b_{kr}^\mu= P^\mu - k^\mu-r^\mu$, where $k^\mu$ and $r^\mu$ are on-shell 
four momenta, e.g. $k^\mu=(\omega_k,\bm k)$, while $P^\mu=(E,\bm P)$.
The quantity $q_k^*$ is the momentum of each member of the pair in its c.m.~frame 
{\em if all three particles are on shell}, and is given by
\begin{equation}
q_k^{*2} = \frac{\sigma_k - 4 M_N^2}{4}\,,\qquad
\sigma_k = (E-\omega_k)^2 - (\bm P-\bm k)^2\,.
\label{eq:q2k}
\end{equation}
The momentum $\bm r_k^*$ is the spatial part of the four-momentum $r^\mu=(\omega_r,\bm r)$ after boosting from the lab frame to the pair c.m.~frame, with the subscript a reminder of the choice of spectator, which itself determines the boost.
The harmonic polynomials $\cY_{\ell m}$ are defined with a nonstandard normalization,
\begin{equation}
\cY_{\ell m}(\bm r) = \sqrt{4\pi} r^\ell Y_{\ell m}(\hat r)\,.
\label{eq:cYdef}
\end{equation}
Finally, we note that the factor of $1/2$ preceding the sum-integral difference arises because we are considering identical particles.

\subsection{Form of $G$}
\label{sec:QCG}

$G$ corresponds to a cut through a process in which the spectator is different on the two sides: a ``switch state''.
The complications due to spin enter here, since the frame used to define the pair's spins changes. 
The result is [Eqs.~(3.27) and (3.30) of Ref.~\cite{\threeN}]
\begin{equation}
G_{p\ell' m' \bm m'^*_s; k \ell m \bm m^*_s}
=
\mathcal D^{(p,k)\dagger}_{\bm m'^*_s \bm m''_s} \
G^{\sf lab}_{p\ell' m' \bm m''_s; k \ell m \bm m'''_s} \
\mathcal D^{(k,p)}_{\bm m'''_s \bm m^*_s} \,,
\label{eq:Gmat}
\end{equation}
where
\begin{align}
G^{\sf lab}_{p\ell'm' \bm m'_s ; k\ell m \bm m_s} &=
- \delta_{ m_{sp} m_{sa}}
\delta_{ m_{sa'} m_{sk}}
\delta_{ m_{sb'} m_{sb}} 
G^{\sf scal}_{p \ell' m'; k \ell m}\,,
\label{eq:Glab}
\\ 
G^{\sf scal}_{p \ell' m'; k \ell m} &=
\frac{1}{4 \omega_p\; \omega_k L^3} \frac{H(\boldsymbol{p}) H(\boldsymbol{k})}{b^2-M_N^2}
\frac{ \mc{Y}_{\ell'm'}(\boldsymbol{k}^*_p) \mc{Y}^*_{\ell m}(\boldsymbol{p}^{*}_k) }{q_p^{*\ell'}\,q_k^{*\ell} } \,.
\label{eq:Gscal}
\end{align}
As the name suggests, $G^{\sf lab}$ is the form of $G$ if all spins are defined in the lab frame,
in which case the compound spin index for final and initial states becomes
\begin{equation}
\bm m'_s =
\big (m_{sp}, m_{sa'}, m_{sb'} \big ) \ \ {\rm and}\ \ 
\bm m_s =
\big (m_{sk}, m_{sa}, m_{sb} \big ) \,.
\label{eq:dimeraxismsfinal}
\end{equation}
Here $m_{sa}$ ($m_{sb}$) are the spin components of the first (second) members of the initial state pairs in the lab frame, with the primed versions being the corresponding members of the final-state pair.\footnote{%
These are abbreviated forms of the notation used in Ref.~\cite{\threeN}, where $m_s(\bm a)$ was used for $m_{sa}$,
$m_s(\bm a')$ for $m_{sa'}$, etc.
 }
The four-momentum $b$ is given by $b^\mu = P^\mu - k^\mu - p^\mu$,
 where the spectator four-momenta $k$ and $p$ are on shell.
The other new quantities are $\bm k_p^*$, and $\bm p_k^*$.
To define $\bm k_p^*$, we note that the subscript indicates that $\bm p$ is the spectator momentum;
$\bm k_p^*$ is then the spatial part of the four-momentum $k^\mu=(\omega_k,\bm k)$ after boosting
to the c.m.~frame of the corresponding pair.
$\bm p_k^*$ is defined similarly, with the roles of $p$ and $k$ interchanged.
Finally, we note that the overall sign in \Cref{eq:Glab} arises from Fermi statistics.

Returning to \Cref{eq:Gmat}, the matrices $\cD^{(p,k)\dagger}$ and $\cD^{(k,p)}$ perform unitary transformations
on the spin indices. Their definitions are exemplified by the result
\begin{align}
\mathcal D^{(k,p)}_{\bm m'_s \bm m^*_s} &
= \delta_{m'_{sk} m_{sk}}
\cD(R_{k,p}^{-1})^{(1/2)}_{m_{sa'} m^*_{sa} }
\cD(R_{k,b}^{-1})^{(1/2)}_{m_{sb'} m^*_{sb} }\,,
\label{eq:Dkp}
\end{align}
where $\cD^{(1/2)}_{m_s' m_s}$ are spin-$1/2$ Wigner matrices.
In $\cD^{(k,p)}$, the first argument in the superscript indicates that $\bm k$ is the spectator momentum, 
while the second argument denotes the first member of the pair, with momentum $\bm p$.
The other member of the pair has momentum $\bm b=\bm P-\bm k -\bm p$.
When treated as on-shell four-momenta, and boosted to the pair rest frame,
 these become $\bm p_k^*$ (as discussed above) and $\bm b_k^*$, respectively.
The arguments of the Wigner matrices  are Wigner rotations.
These are given by the following rotation axes and angles,
\begin{align}
R_{k,p} & : \qquad \hat {\boldsymbol n} = -\frac{\boldsymbol \beta_{P-k} \times \boldsymbol \beta_p }{|\boldsymbol \beta_{P-k} \times \boldsymbol \beta_p |},\qquad
\cos \theta = \frac{(1+\gamma_p + \gamma_{P-k} +\gamma'_{P-k,p})^2}{(1+\gamma_p)(1+\gamma_{P-k})(1+\gamma'_{P-k,p})}-1\,, 
\label{eq:Rpdef}
\end{align}
where
\begin{equation}
\boldsymbol \beta_p = \frac{\bm p}{\omega_p}\,,\quad
\boldsymbol \beta_{P-k} = \frac{\bm P - \bm k}{E-\omega_k}\,,\quad
\gamma_i = \sqrt{\frac1{1-\beta_i^2}}\,, \quad
\gamma'_{P-k,p}  = \gamma_{P-k} \gamma_p (1 - \boldsymbol \beta_{P-k} \cdot \boldsymbol \beta_p ) \,.
\label{eq:betas}
\end{equation}
The rotation $R_{k,b}$ is defined as for $R_{k,p}$, but with $\boldsymbol \beta_p$ replaced by $\boldsymbol \beta_b=\bm b/\omega_b$, and $\gamma_p$ replaced by $\gamma_b$.
Note that the first index in the subscript of $R_{k;b}$ indicates the spectator momentum.

The matrix $\cD^{(p,k)}$ in \Cref{eq:Gmat} is obtained from $\cD^{(k,p)}$ simply by interchanging the roles
of $\bm k$ and $\bm p$.

\subsection{Form of $\cK_{2,L}$}
\label{sec:QCK2L}

The two-particle K matrix enters the quantization condition as [eqs.~(3.39) and (3.40) of Ref.~\cite{\threeN}]
\begin{equation}
[\mathcal K_{2,L}]_{p \ell'm' \bm m'^*_s ; k\ell m \bm m^*_s} =  \delta_{\bm p \bm k} 2 \omega_k 
\delta_{m_{sp} m_{sk}} \
\mathcal K_{2}^{[\ell' m' m^*_{sa'} m^*_{sb'}] , \, [\ell m m^*_{sa} m^*_{sb} ]}(\sigma_k) \,.
\label{eq:K2mat}
\end{equation}
where the $\cK_2$ on the right-hand side of \Cref{eq:K2mat} is the infinite-volume two-neutron K matrix
expressed in the $\ell m m^*_{sa} m^*_{sb}$ basis.

A more useful basis is that in which the spins of the members of dimer are combined
into total dimer spin $s$, which takes values $s=0$ or $1$, along with $z$ component $\mu_s$.
Fermi statistics then implies that $\cK_2$ vanishes unless $\ell$ is even for $s=0$ and odd for $s=1$.
Inverting the equations given in Ref.~\cite{\threeN}, the relation between the $\cK_2$ appearing on the
right-hand side of \Cref{eq:K2mat}, and that in the $s$ basis, is given in terms of Clebsch-Gordon (CG) coefficients by
\begin{equation}
\mathcal K_{2}^{[\ell' m' m^*_{sa'} m^*_{sb'}] , \, [\ell m m^*_{sa} m^*_{sb} ]} =
\sum_{s, \mu'_s, \mu_s}
\braket{\tfrac12 m^*_{sa'} \tfrac12 m^*_{sb'} | s \mu'_s}
\mathcal K_{2}^{(s;\ell' m' \mu'_s;\ell m \mu_s )} 
\braket{s \mu_s| \tfrac12 m^*_{sa} \tfrac12 m^*_{sb}}\,.
\label{eq:K2ltos}
\end{equation}
Both factors of $\cK_2$ implicitly depend on $\sigma_k$.
We note that the K matrix on the right-hand side involves a fixed total spin $s$, rather than both $s'$ and $s$.
This implements the well-known result that, due to parity conservation, the two-neutron interaction conserves $(-1)^\ell$, and thus cannot interchange even and odd values of $\ell$.
Given the above-mentioned constraints due to Fermi statistics, this implies that $s$ is conserved.

The next stage is to combine $s$ and $\ell$ into the dimer total angular momentum, which is a conserved quantity.
This is accomplished by
\begin{equation}
\mathcal K_{2}^{(s , \ell', m' , \mu'_s, \ell, m, \mu_s )} (\sigma_k)
= \sum_{j, \mu_j} 
\braket{\ell' m', s \mu'_s | j \mu_j}
\mathcal K_2^{(j, \ell', \ell, s)}(\sigma_k)
\braket{j \mu_j| \ell m, s \mu_s}\,,
\label{eq:K2Ltoells}
\end{equation}
where we now make the dependence on $\sigma_k$ explicit.

We now apply the restriction to $\ell \le 1$, as this will be the case we explicitly implement.
Conservation of $(-1)^\ell$ then implies that $\ell'=\ell$ in $\mathcal K_2^{(j, \ell', \ell, s)}$, 
i.e.
\begin{equation}
\mathcal K_2^{(j, \ell', \ell, s)}(\sigma_k) =
\delta_{\ell' \ell}\, \mathcal K_2^{(j,  \ell, s)}(\sigma_k)\,.
\end{equation}
The absence of channel mixing allows us to parametrize each channel with a phase shift.
Combining Eqs.~(3.43) and (3.44) of Ref.~\cite{\threeN}, we find
\begin{equation}
\frac1{\mathcal K_2^{(j, s, \ell)}(\sigma_k)} = 
\frac{ q_k^* \cot \delta^{(j,s,\ell)}(q_k^*)}{16 \pi \sqrt{\sigma_k}}
+
\frac{ \vert q_k^* \vert \big (1 - H(\bm k) \big ) }{16 \pi \sqrt{\sigma_k}}  \,,
\label{eq:K2cotd}
\end{equation}
where $q_k^*$ is defined in \Cref{eq:q2k}.
The first term on the right-hand side is the standard two-particle K matrix, while the second term is a subthreshold modification that interpolates between the standard form at threshold and $\cM_2^{-1}$ when $H(\bm k)$ 
vanishes.\footnote{%
There is some sloppiness in the notation here, since $\bm k$ is an argument on the right-hand side, but not on the left. 
However, $H(\bm k)$ is actually a function of $\sigma_k$ [see \Cref{eq:cutoff} below], so there is no inconsistency.
}

\subsection{Form of $\Kdf$}
\label{sec:QCKdf}

The final matrix entering the quantization condition is $\Kdf$. 
As explained in Sec.~3.2.1 of Ref.~\cite{\threeN},
it is determined starting from the lab-frame, infinite-volume expression for the three-particle K matrix,
written as a matrix in spin space,
\begin{equation}
\left[\cK_{\rm df,3}^{\sf lab}(\{\bm p'_i\},\{\bm p_i\})\right]_{\bm m'_s,\bm m_s} \equiv
\cK_{\rm df,3}^{\sf lab}(\bm  p, m_{sp}; \bm a', m_{sa'}; \bm b', m_{sb'}|
\bm  k, m_{sk}; \bm a, m_{sa}; \bm b, m_{sb})\,.
\label{eq:Kdfmata}
\end{equation}
Here $\{\bm p_i\}=\{\bm k, \bm a, \bm b\}$ are the initial-state on-shell momenta 
(with $\bm k+\bm a+\bm b = \bm P$),
while $\{\bm p'_i\}=\{\bm p, \bm a', \bm b'\}$ are the corresponding final-state momenta.\footnote{%
Strictly speaking, for fixed $\bm P$, there are only two independent momenta, e.g. $\bm k$ and $\bm a$
in the initial state, but we list all three momenta as arguments for the sake of clarity.
}
Note that all spin components are defined in the lab frame, and thus do not carry asterisks. 
Explicit choices for the function $\cK_{\rm df,3}^{\sf lab}$ on the right-hand side of \Cref{eq:Kdfmata} will be discussed below.

The next step is to convert to dimer-axis spin variables, which is achieved using the unitary matrices defined in
\Cref{eq:Dkp},
\begin{equation}
[\Kdf(\{\bm p'_i\},\{\bm p_i\})]_{\bm m'^*_s, \bm m^*_s} =  \cD^{(p,a')\dagger}_{\bm m'^*_s \bm m'''_s}
 [\cK_{\rm df,3}^{\sf lab}(\{\bm p'_i\},\{\bm p_i\})]_{\bm m'''_s,\bm m''_s} \cD^{(k,a)}_{\bm m''_s, \bm m^*_s}\,.
\label{eq:Kdfmatb}
\end{equation}
Finally, to obtain the form that enters the quantization condition, we project the angular dependence in the dimer frame onto spherical harmonics, using the projection operators defined in Eq.~(2.6) of Ref.~\cite{\threeN},
\begin{equation}
\left[K_{\rm df,3}\right]_{k' \ell' m' \bm m'^*_s; k \ell m \bm m^*_s} 
=  [\cP_{\ell' m'}^{\hat a'^*}] \circ [ \cP_{\ell m}^{\hat a^*}]^\dagger
\circ [\Kdf(\{\bm p'_i\},\{\bm p_i\})]_{\bm m'_s, \bm m_s}\,,
\label{eq:Kdfdimer}
\end{equation}
where\footnote{%
This corrects a typographical error in expression for the projector in Eq.~(3.25) of Ref.~\cite{\threeN},
which has $4\pi$ in the denominator, instead of $\sqrt{4\pi}$.
The correct factor is determined by the convention that a quantity that is independent of $\bm a^*$
is unchanged by the action of $\cP_{00}^{\hat a^*}$.
In addition, the complex conjugation convention is changed to the more natural choice of Ref.~\cite{\DRS}.
}
\begin{align}
\cP_{\ell m}^{\hat a^*} f(\bm k, \bm a)
&=
\frac1{\sqrt{4\pi}} \int_{\Omega_{\hat a^*}} Y^*_{\ell m}(\hat a^*) f^*(\bm k, \bm a^*)\,,
\label{eq:projectellm}
\\
\int_{\Omega_{\hat a^*}} &= \int d\Omega_{\hat a^*} \equiv \int_{-1}^1 d\cos\theta_{a^*} \int_0^{2\pi} d\phi_{a^*}\,.
\end{align}
We recall that $\bm a^*$ is the spatial part of the on-shell four-momentum $(\omega_{a},\bm a)$ after boosting to the pair CMF.
We also note that the magnitude of $\bm a^*$ is fully fixed once  $\bm P$ and $\bm k$ are specified, so the only dependence in the pair CMF is on the direction vector $\hat a^*$.

An important practical consideration is that the unitary matrices $\cD^{(p',a')}$ and $\cD^{(p,a)}$ in 
\Cref{eq:Kdfmatb} depend on the momenta, so the the projections in \Cref{eq:Kdfdimer} 
{\em do not commute} with the multiplication by the unitary matrices. This will lead to complications in the
implementation, as discussed in \Cref{sec:implementKdf} below.

\subsection{Form of the cutoff function }
\label{sec:cutoff}

Here we discuss the cutoff function $H(\bm k)$ that appears in $F$, $G$, and $\cK_{2,L}$
(see \Cref{eq:Fmat,eq:Gmat,eq:K2cotd} above).
For fixed $\bm P$, as the spectator momentum increases, the pair invariant mass, given by $\sqrt{\sigma_k}$
[with $\sigma_k$ defined in \Cref{eq:q2k}],
can drop below the two-neutron threshold at $\sigma_{\rm thr}=4 M_N^2$.
One must include a range of subthreshold momenta in order to avoid enhanced exponentially-suppressed finite-volume corrections~\cite{\HSQCa}.
One then turns off the contributions by introducing the cutoff function.
This must be smooth, in order to avoid power-law finite-volume effects.
For the same reason, it must also cut off the subthreshold contributions while the pair two-particle K matrix remains analytic. The natural choice, then, is to use a smooth cutoff function that vanishes at, or above, the leading nonanalyticity in $\cK_2$. As noted in Ref.~\cite{\threeN}, this is the pole due to single pion exchange between the two neutrons. After projection onto pair angular momenta, this becomes a cut (the ``left-hand cut''), 
with the first branch point occurring at
\begin{equation}
\sigma_{\rm lhc} = 4 M_N^2 - M_\pi^2\,.
\label{eq:sigmalhc}
\end{equation}

With this is mind, we use the cutoff function introduced in Ref.~\cite{\BStwoplusone}
\begin{equation}
H(\bm k) = J(z(\sigma_k))\,,\quad
z(\sigma_k) = (1+\epsilon_H) \frac{\sigma_k - \sigma_{\rm lhc}}{\sigma_{\rm th}- \sigma_{\rm lhc}}\,.
\label{eq:cutoff}
\end{equation}
where
\begin{equation}
J(z) = \begin{cases} 0 & z\le 0\\ \exp\left(-\frac1x \exp\left[- \frac1{1-x} \right] \right) & 0 < z < 1 \\
1 & z\ge 1 \end{cases}\,.
\end{equation}
Here $\epsilon_H$ is a small positive constant that moves the start of the cutoff function slightly below threshold.
In practice, we mainly use $\epsilon_H=0$, since $J(z)$ remains very close to unity for some distance below $z=1$.
However, in order to study the dependence of results on the form of $H(\bm k)$, we also do some calculations with $\epsilon_H=1$.

In recent work studying the $N\pi\pi$ system, it was realized that, for some three-particle systems, the standard RFT cutoff function, such as that just described, does not remove all nonanalyticities~\cite{\Npp}.
In particular, when the invariant mass of one pair lies below threshold,
it is possible for the invariant mass of the other pairs to exceed the pair inelastic threshold.
This problem is present, for example, in the $N\pi\pi$ system.
Using the methods of Ref.~\cite{\Npp}, we have found, however, that this issue does not arise for the three-neutron system.

A potential confusion concerning $H(\bm k)$ arises from the fact that it should not be thought of as a cutoff like those that appear in effective field theories. In particular, the scale of the cutoff cannot be raised  arbitrarily high.
It is better thought of as a ``transition function'' that smoothly turns off the sum or integral over $\bm k$. In particular, the derivation of the quantization condition holds, up to exponentially-suppressed corrections, as long as $H(\bm k)$ has the properties enumerated above.

A concern that one might have with our choice of $H(\bm k)$ is the rapidity of the transition from $1$ to $0$. The exponent in the dropped exponentially-suppressed corrections is proportional to the width of the transition region. In terms of $\sigma_k$, the transition region has width $M_\pi^2$, so the suppression factor is of the form $\exp(-c M_\pi L)$ (up to powers of $L$), where $c$ is a numerical constant.
This is of the same form as appearing in all sources of exponentially-suppressed corrections. 
Nevertheless, the numerical factor $c$ may be smaller in the three-neutron system than in mesonic systems such as  $3\pi$ and $3K$.
This is because the nearest left-hand singularity in such cases is due to two-pion exchange, so that the transition region widens to $4 M_\pi^2$.
Thus we might expect that the exponentially-suppressed corrections are larger in the present case, with the constant $c$ reduced by a factor of $2$,
which could lead to problems such as unphysical solutions to the quantization condition.
For that reason, in the numerical examples to be discussed in \Cref{sec:num}, we mainly use the small value $M_\pi L=3$ in order to enhance, and thus make easier to find, potential unphysical effects, while also considering larger values of $M_\pi L$ to see how such effects disappear.

\section{Implementing the quantization condition}
\label{sec:implement}

The RFT three-particle quantization condition has been implemented previously for spinless particles,
both degenerate~\cite{\BHSnum,\largera} and nondegenerate~\cite{\implement}, and with multiple two-particle waves~\cite{\dwave,\DRS}.
The extension to three spin-$1/2$ particles, described in the previous section, introduces several new features, and these are the focus of this section.\footnote{%
An implementation for the $N\pi\pi$ system has been developed in parallel with this work, and involves some
of the same issues~\cite{\Npp}.}

We begin with a general comment that applies to all matrices entering the quantization condition.
Although it is more natural for $\cK_{2,L}$ to use the dimer total angular-momentum basis, as discussed in \Cref{sec:QCK2L}, we find it simpler for the overall implementation to use the $\{\ell, s\}$ basis.
A truncation in $\ell$ is necessary for any implementation of the quantization condition, 
and is justified close to threshold by the standard $q^{2\ell}$ barrier factors in amplitudes.
We choose to consider $\ell \le 1$, for which there are only two $\{\ell, s\}$ channels: $\ell=s=0$ and $\ell=s=1$,
with $1$ and $3^2=9$ components, respectively.
When combined into dimer total angular momentum, as is necessary for $\cK_{2,L}$, there are four 
scattering channels: the single component $j=\ell=s=0$ channel, 
and the three channels for $\ell=s=1$, with $j=0, 1, 2$, having $1$, $3$, and $5$ components, respectively.

A second general comment is that we use the form of the quantization condition, described in Appendix A of Ref.~\cite{\dwave} and given explicitly in Eq.~(A.13) of that work,
in which the factors of $q^*$ are reshuffled. Specifically, we introduce the diagonal matrix
\begin{equation}
\left[ Q \right]_{k' \ell' m' \bm m'^*_s; k \ell m \bm m^*_s}
=
\delta_{\bm k' \bm k} \delta_{\ell' \ell} \delta_{m' m} \delta_{\bm m'^*_s \bm m^*_s}
q_k^{*\ell}\,,
\end{equation}
and note that the quantization condition \Cref{eq:QC3} takes the same form when we make the replacements
\begin{multline}
F \to F^Q = Q F Q\,, \ \
G \to G^Q = Q G Q\,, 
\\
\cK_{2,L} \to \cK_{2,L}^Q = Q^{-1} \cK_{2,L} Q^{-1}\,,\ \
\Kdf \to {\cK}_{\rm df,3}^Q = Q^{-1} \Kdf Q^{-1}\,.
\end{multline}
Doing so has three advantages.
First, it ensures that all matrices remain hermitian below two-particle thresholds, so that, in particular, the
eigenvalues of the matrix appearing in the quantization condition are real. This simplifies root finding.
Second, it removes unphysical solutions to the quantization condition that arise when $q_k^*=0$~\cite{\dwave}.
And, finally, it simplifies the numerical implementation by shortening some expressions.
To save notational overload, we leave the superscript $Q$ implicit in the following, except where it is important for a particular result.

Finally, we follow Ref.~\cite{\dwave}, and use real spherical harmonics in our implementation.
However, we keep explicit all complex conjugations in the results that follow, 
so that they hold also with complex spherical harmonics.

In the rest of this section, 
we discuss the component matrices in turn, pointing out new features associated with the neutron spins.
We then describe the symmetries of these matrices, the projection of solutions to the quantization condition
onto fermionic irreps of the appropriate finite-volume little groups,
and the subduction of infinite-volume $J^P$ states into these irreps.
Finally, we give examples of the lowest lying free levels, including their irrep decompositions,
and describe the irrep decomposition of the contributions to $\Kdf$.

\subsection{Implementing $G$}
\label{sec:implementG}

The lab frame version, $G^{\sf scal}$, given in \Cref{eq:Gscal},
has the same form as for degenerate spinless particles, 
and can be implemented as for the three-pion system~\cite{\BHSnum,\dwave}.
The use of the ``$Q$-form'' $G^Q$ simply removes the factors of $q_p^*$ and $q_k^*$ from
the denominator of \Cref{eq:Gscal}.

The new features here are the presence of spin indices,
and the need to include the unitary matrices that convert $G$ to the dimer-axis frame,
see \Cref{eq:Gmat}. 
The spectator spin index is kept explicit, while the spin indices of the pair are combined into $\{s,\mu_s\}$.
In other words, for each choice of $\{\bm k,\ell, m\}$, we decompose the eight-dimensional spin space as
$m_s(\bm k) \otimes \left[ (s=0) + (s=1,m_s) \right]$.
Thus the matrix we actually implement numerically has the indices
\begin{equation}
G_{p \ell' m' m_{sp} s' \mu'_s; k \ell m m_{sk} s \mu_s}\,.
\end{equation}
Furthermore, as explained above, the $s=0(1)$ components are combined only with $\ell=0(1)$, respectively.

The projection onto $s=0$ is accomplished in the initial state
by inserting the matrix $i\sigma_2/\sqrt2$ between the two dimer spin indices, 
while, for $s=1$, one uses
$i\sigma_{\mu_s} \sigma_2/\sqrt2$, with $\sigma_{\mu_s} = (\sigma_x,\sigma_z,\sigma_y)$
for $m_s=(1,0,-1)$ in the initial state.
For both choices of dimer spin, the complex conjugate insertion is used for the final state.
These choices simply implement the Clebsch-Gordon coefficients of \Cref{eq:K2ltos} in the real
spherical-harmonic basis.
The result can be written in the general form
\begin{equation}
G_{p \ell' m' m_{sp} s' \mu'_s; k \ell m m_{sk} s \mu_s}
=
- G^{\sf spin}_{m_{sp} s' \mu'_s;m_{sk} s \mu_s}
G^{\sf scal}_{p\ell' m'; k \ell m}
\,.
\end{equation}

For $s'=s=0$, the spin factor is
\begin{align}
G^{\sf spin}_{m_{sp} 0 0;m_{sk} 0 0} & =
\frac12 \left[\cD^{(1/2)}(R_{k,p}^{-1}) \sigma_2 D^{(1/2)}(R_{k,b}^{-1})^{\rm T}
\cD^{(1/2)}(R_{p,b})^{\rm T} \sigma_2 \cD^{(1/2)}(R_{p,k}) \right]_{m_{sp} m_{sk}}\,,
\\
&= \frac12 \left[\cD^{(1/2)}(R_{k,p}^{-1}) D^{(1/2)}(R_{k,b})
\cD^{(1/2)}(R_{p,b}^{-1}) \cD^{(1/2)}(R_{p,k}) \right]_{m_{sp} m_{sk}}\,,
\end{align}
where the superscript T indicates transpose, and to obtain the second form we have used
the following property of spin-$1/2$ Wigner matrices:
$\sigma_2 \cD^{(1/2)}(R^{-1})^{\rm T} \sigma_2 = \cD^{(1/2)}(R)$.
For the other spin choices, we find
\begin{align}
G^{\sf spin}_{m_{sp} 1 \mu'_s;m_{sk} 0 0} & =
\frac12 \left[\cD^{(1/2)}(R_{k,p}^{-1}) D^{(1/2)}(R_{k,b})
\cD^{(1/2)}(R_{p,b}^{-1}) \sigma_{\mu_s'}^\dagger \cD^{(1/2)}(R_{p,k}) \right]_{m_{sp} m_{sk}}\,,
\\
G^{\sf spin}_{m_{sp} 0 0; m_{sk} 1 \mu_s} & =
\frac12 \left[\cD^{(1/2)}(R_{k,p}^{-1}) \sigma_{\mu_s} D^{(1/2)}(R_{k,b})
\cD^{(1/2)}(R_{p,b}^{-1}) \cD^{(1/2)}(R_{p,k}) \right]_{m_{sp} m_{sk}}\,,
\\
G^{\sf spin}_{m_{sp} 1 \mu'_s; m_{sk} 1 \mu_s} & =
\frac12 \left[\cD^{(1/2)}(R_{k,p}^{-1}) \sigma_{\mu_s} D^{(1/2)}(R_{k,b})
\cD^{(1/2)}(R_{p,b}^{-1}) \sigma_{\mu'_s} \cD^{(1/2)}(R_{p,k}) \right]_{m_{sp} m_{sk}}\,.
\end{align}
The expressions for the rotations themselves, exemplified by \Cref{eq:Rpdef}, are tedious but straightforward to implement.

\subsection{Implementing $F$}
\label{sec:implementF}

The scalar part of $F$, \Cref{eq:Fscal}, is identical to that for degenerate spinless particles and is implemented as
in Refs.~\cite{\dwave,\implement}.
Again, the (implicit) use of the $Q$ form simply removes the factors of $q_k^*$ from the expression.
The spin factor remains trivial when expressed in the $\{\ell,s\}$ basis, so that one obtains
\begin{equation}
F_{p \ell' m' m_{sp} s' \mu'_s; k \ell m m_{sk} s \mu_s}
=
\delta_{m_{sp} m_{sk} } \delta_{s' s} \delta_{\mu_s' \mu_s}
F^{\sf scal}_{p\ell' m';k \ell m}\,.
\end{equation}

\subsection{Implementing $\cK_{2,L}$}
\label{sec:implementK2}

The conversion to the $\{\ell,s\}$ basis is given by \Cref{eq:K2ltos}.
Using this, and inverting \Cref{eq:K2Ltoells}, we find
\begin{equation}
[\cK_{2,L}^{-1}]_{p \ell' m' m_{sp} s' \mu'_s; k \ell m m_{sk} s \mu_s}
=
\frac{\delta_{k' k} \delta_{m_{sp} m_{sk}}}{2\omega_k}
\sum_{j, \mu_j}
\braket{\ell' m', s \mu'_s | j \mu_j}
\frac1{\cK_2^{(j, \ell, s)}}
\braket{j \mu_j| \ell m, s \mu_s}\,,
\label{eq:K2Linvfinal}
\end{equation}
with $1/\cK_2^{(j, \ell, s)}$ given by \Cref{eq:K2cotd}.
It is important that we include all values of $j$ allowed by combining $\ell$ and $s$,
in particular $j=0,1,2$ for $\ell=s=1$, such that $\cK_{2,L}$ is invertible.
Here the (implicit) use of the $Q$ form cancels the leading barrier factors from $\cK_2$,
as will be discussed in more detail when we turn to numerical results.

Our choice of phase shifts for the four scattering channels will be discussed below when we display numerical results.

\subsection{Implementing $\Kdf$}
\label{sec:implementKdf}

$\Kdf$ is by far the most complicated quantity to implement.
In \Cref{sec:QCKdf} we presented the steps required
to obtain the matrix that enters the quantization condition,
starting with the lab-frame form, $\cK_{\rm df,3}^{\sf lab}$ [\Cref{eq:Kdfmata}].
In this section, we describe our implementation of these steps,
relegating many  technical details to appendices.
We divide the discussion into two parts, the first providing the explicit relation between $\Kdf$ in the
two frames in the $\ell, m$ basis,
and the second determining the form of the lab-frame $\Kdf$ in a threshold expansion.

\subsubsection{Relating $\Kdf$ in the dimer-axis and lab frames}
\label{sec:labtodimer}

Given the form of $\cK_{\rm df,3}^{\sf lab}$, we wish to implement \Cref{eq:Kdfmatb,eq:Kdfdimer},
which provide the relation to the dimer-axis frame.
We do so by first projecting $\cK_{\rm df,3}^{\sf lab}$ onto the $\{\ell,m\}$ basis,
\begin{multline}
\left[\cK^{\sf lab}_{\rm df,3}\right]_{p \ell' m' m_{sp} m_{sa'} m_{sb'} ;k\ell m m_{sk} m_{sa} m_{sb}} 
= 
\\
[\cP_{\ell' m'}^{\hat a'^*}] \circ [ \cP_{\ell m}^{\hat a^*}]^\dagger \circ 
[\cK_{\df,3}^{\sf lab}(\{\bm p'_i\},\{\bm p_i\})]_{m_{sp} m_{sa'} m_{sb'}; m_{sk} m_{sa} m_{sb}}\,.
\label{eq:Kdflab}
\end{multline}
The relation between dimer-axis and lab frames can then be written
\begin{equation}
\left[ \cK_{\rm df,3}^Q \right] = [\overline I]^\dagger \cdot \left[\cK^{Q, \rm lab}_{\rm df,3}\right] \cdot [\overline I]\,,
\label{eq:Kdf3dimervslab}
\end{equation}
where matrix indices are left implicit.
We have made explicit that we are using the $Q$ form of $\Kdf$,
which is important here, for otherwise the matrix on the left of $\cK^{Q, \rm lab}_{\rm df,3}$ is
not, in general, the hermitian conjugate of that on the right.
The conversion matrix is given explicitly by
\begin{align}
[\overline I]_{p\ell' m' m_{sp} m'_{sa} m'_{sb}; k \ell m m_{sk} m^*_{sa} m^*_{sb} } 
&= \delta_{\bm p \bm k} \delta_{m_{sp} m_{sk}} 
I^Q(\bm k)_{\ell' m' m'_{sa} m'_{sb}; \ell m m^*_{sa} m^*_{sb}} \,,
\label{eq:Ibardef}
\\
I^Q(\bm k)_{\ell' m' m'_{sa} m'_{sb}; \ell m m^*_{sa} m^*_{sb}}
&= (q_k^*)^{\ell'}
I(\bm k)_{\ell' m' m'_{sa} m'_{sb}; \ell m m^*_{sa} m^*_{sb}} (q_k^{*})^{-\ell}
\label{eq:IQdef}
\end{align}
with
\begin{multline}
I(\bm k)_{\ell' m' m'_{sa} m'_{sb}; \ell m m^*_{sa} m^*_{sb}} = 
\\
\int_{\Omega_{\hat a^*}} Y^*_{\ell' m'}( \hat a^*)
\cD^{(1/2)}(R_{k,a}^{-1})_{m'_{sa} m^*_{sa}}
\cD^{(1/2)}(R_{k,b}^{-1})_{m'_{sb} m^*_{sb}} 
Y_{\ell m}(\hat a^*)  \,.
\label{eq:Idef}
\end{multline}

We describe the evaluation of $I(\bm k)$ and $I^Q(\bm k)$ in \Cref{app:Ibar}.
Combining the result \Cref{eq:Ires} with the definition of $I^Q$ given in \Cref{eq:IQdef}, we obtain\footnote{%
We stress that the primes in $m'_{sa}$, $m''_{sa}$, etc. simply indicate an alternate index with the same meaning as
$m_{sa}$, i.e. the spin component in the lab frame of the particle with momentum $\bm a$.
They should be distinguished from $m_{sa'}$, etc., which refer to the spin of the particle with momentum $\bm a'$.
}
\begin{multline}
I^Q(\bm k)_{\ell' m' m'_{sa} m'_{sb}; \ell m m^*_{sa} m^*_{sb}} =   
\overline{\cD}^{\ell'}(R_k^{-1})_{m' m'_{sa} m'_{sb}; m''' m''_{sa} m''_{sb}}
\\
\times I'^Q(\bm k)_{\ell' m''' m''_{sa} m''_{sb}; \ell m'' m'^*_{sa} m'^*_{sb}} 
\overline{\cD}^{\ell}(R_k)_{m'' m'^*_{sa} m'^*_{sb}; m m^*_{sa} m^*_{sb}}\,.
\label{eq:IQresa}
\end{multline}
Here, the compound Wigner rotation matrices are
\begin{equation}
\overline{\cD}^{\ell}(R_k)_{m' m'_{sa} m'_{sb}; m m_{sa} m_{sb}}
=
\cD^{(\ell)}_{m' m}(R_k) \cD^{(1/2)}_{m'_{sa} m_{sa}}(R_k) \cD^{(1/2)}_{m'_{sb} m_{sb}}(R_k)
\,,
\end{equation}
with $R_k$ the rotation that satisfies $R_k (\bm P-\bm k) = |\bm P-\bm k| \hat z$.
The matrix $I'^Q(\bm k)$ is evaluated explicitly in the appendix for  $\ell',\ell \le 1$,
with the result given in \Cref{eq:IpQres}. 

We stress that in \Cref{eq:Kdf3dimervslab},
even though the $\ell$ indices of $\cK^{Q}_{\rm df,3}$ are constrained to satisfy $\ell \le 1$.
those of $\cK^{Q, \rm lab}_{\rm df,3}$ run, in principle, over all values.
However, as noted in \Cref{app:Ibar} following \Cref{eq:threshI}, offdiagonal terms in $I'$ are
proportional to $(a^*)^{|\ell'-\ell|}$, such that the threshold suppression of higher values of $\ell$ is propagated
from $\cK^{Q}_{\rm df,3}$ to $\cK^{Q, \rm lab}_{\rm df,3}$, and it is self-consistent to apply the same
truncation to the latter. This argument is not impacted by the use of $Q$ form quantities, since that only reshuffles factors of $a^*$ between terms.

Combining the results given above, we arrive at the master formula determining the dimer-axis frame
$Q$-form of $\Kdf$, which enters the quantization condition, in terms of the $\ell,m$-projected lab-frame form,
\begin{multline}
\left[{\cK}^Q_{\rm df,3}\right]_{p \ell';k \ell} 
=  \overline \cD^{(\ell')}(R_p^{-1})  \left[I'^{Q}(\bm p)^\dagger\right]_{\ell' \ell''} \overline \cD^{(\ell'')}(R_p)
\left[\cK_{\rm df,3}^{Q,\rm lab}\right]_{p \ell'',k \ell'''} 
\\
\overline \cD^{(\ell''')}(R_k^{-1}) \left[I'^Q(\bm k)\right]_{\ell''',\ell} \overline \cD^{(\ell)}(R_k)\,.
\label{eq:master}
\end{multline}
Here, all matrix indices are implicit except $k$ and $\ell$,
and the rotation $R_p$ is defined analogously to $R_k$:
$R_p (\bm P-\bm p) = |\bm P-\bm p| \hat z$.
In this formula, the matrix $I'^Q$ carries the information about the Wigner rotations arising from boosting between frames. If these rotations are set to the identity, 
then $I'^Q$ becomes the identity matrix, and the factors of 
$\overline \cD^{(\ell)}$ cancel.

The final step in our implementation of $\Kdf$ is to convert to the $\{\ell, s\}$ basis, and truncate
to the $\ell=s=0$ and $\ell=s=1$ subspace.
In fact, the truncation to this subspace is automatic, since it follows from antisymmetry under particle exchange, which is built into our starting expression for $\cK_{\rm df,3}^{\sf lab}$.
That this is indeed the case provides a check on our implementation.

For each choice of spectator momenta $\{\bm p, \bm k\}$, $\cK^Q_{\rm df,3}$ in \Cref{eq:master}
is a $32\times 32$ matrix once we truncate to $\ell \le 1$.
This is because there are four choices of $\ell,m$ for $\ell=0,1$, and for each of these there are $2^3=8$ values of
the spin components.
The subspace that we want has only $\ell=s=0$ (two components from $m_{sk}$)
and $\ell=s=1$ ($3\times 3\times 2=18$ components from $m$, $\mu_s$, and $m_{sk}$),
and thus has dimension 20.
We have constructed the required conversion matrix using Clebsch-Gordon coefficients; it takes the schematic form
\begin{equation}
C^{20\leftarrow32} = \sum_{\ell=s} \ket{\text{20-d basis}}\bra{\text{32-d basis}} \,.
\end{equation}
We have checked that it satisfies the expected relations,
\begin{equation}
C^{20\leftarrow32} \cdot C^{20\leftarrow32 \dagger} = \bm 1_{20\times 20}\,,
\qquad
C^{20\leftarrow32 \dagger} \cdot C^{20\leftarrow32} = P(\ell=s)\,,
\end{equation}
where $P(\ell=s)$ is the projector onto the $\ell=s$ subspace,
and that it is invariant under rotations.
We conjugate the matrix ${\cK}^Q_{\rm df,3}$ that is obtained from \Cref{eq:master} with this conversion matrix 
in order to obtain the form that enters the quantization condition.

\subsubsection{Threshold expansion for $\cK_{\rm df,3}$ in the lab frame}
\label{sec:Kdflabexp}

Our final task is to determine the form of $\cK_{\rm df,3}^{\sf lab}$. 
Here, we use the results of the threshold expansion worked out in Ref.~\cite{\threeN}. 
We stress that one can consider a more general form of $\Kdf$---as long as it is consistent with Lorentz covariance---but we choose to use the threshold expansion so as to provide a concrete example of the required manipulations.
We also note that $\Kdf$ includes, in principle, contributions that involve (multiple) pion exchanges between nuclei (as long as they do not involve $s$-channel three-neutron cuts), but these are considered short-distance contributions within our energy range [\Cref{eq:Erange}], for the pion can be integrated out.

The threshold expansion is an expansion in the dimension of local six-neutron operators,
i.e. in derivatives. In addition, to limit the number of terms, Ref.~\cite{\threeN} used a nonrelativistic expansion, assuming $\bm k^2 \ll M_N^2$. In this way, it was found that there were two terms at leading nontrivial order,
\begin{equation}
M_N^2 \cK_{\rm df,3}^{\sf lab} = \frac{1}{\Lambda^2_{\rm EFT}} \left( c_A \cK_A + c_B \cK_B\right)\,,
\label{eq:KdfEFT}
\end{equation}
where 
\begin{align}
\cK_A &= \overline{\cA} \left[\chi_{p}^\dagger \boldsymbol \sigma\cdot \bm p \, \boldsymbol \sigma \cdot \bm k \chi_k\
\chi_{a'}^\dagger \chi_a \ \chi_{b'}^\dagger \chi_b\right] \,,
\label{eq:KA}
\\
\cK_B &= \overline{\cA} \left[\bm p\cdot \bm k\ \chi_{p}^\dagger \chi_k\
\chi_{a'}^\dagger \chi_a \ \chi_{b'}^\dagger \chi_b\right] \,,
\label{eq:KB}
\end{align}
Here $\overline{\cA}$ indicates complete antisymmetrization separately over $\{\bm p, \bm a', \bm b'\}$
and $\{\bm k, \bm a, \bm b\}$,
while $\chi_k$, $\chi_a$, \dots are dimensionless 
two-spinors associated with the particle having momenta $\bm k$, $\bm a$, \dots.
Thus, for example, if $m_s(\bm k) = 1$, then $\chi_k^T = (1, 0)$.
The natural cutoff scale on momenta is set by the pion mass, $\Lambda_{\rm EFT} \sim M_\pi$,
and we expect the dimensionless coefficients $c_A$ and $c_B$ to be of order unity.

Using the relation 
\begin{equation}
\boldsymbol \sigma\cdot \bm p \,\boldsymbol \sigma \cdot \bm k = 
 \bm p \cdot \bm k + i  \boldsymbol \sigma \cdot (\bm p \times \bm k)\,,
\end{equation}
we can rewrite $\cK_A$ as
\begin{align}
\cK_A &= \cK'_A + \cK_B\,,
\\
\cK'_A &= \overline{\cA} \left[\chi_{p}^\dagger  i  \boldsymbol \sigma \cdot (\bm p \times \bm k) \chi_k\
\chi_{a'}^\dagger \chi_a \ \chi_{b'}^\dagger \chi_b\right] \,.
\label{eq:KAp}
\end{align}
It turns out to be computationally simpler to calculate $\cK'_A$ rather than $\cK_A$.

In order to obtain the results \Cref{eq:KA,eq:KB,eq:KAp}, contributions suppressed by $\bm p^2/M_N^2$ have
been dropped. This brings up two concerns.
First, what is the relative size of the dropped terms?
Since, at the inelastic $nnn\pi$ threshold, $\bm p^2 \sim M_\pi M_N$, the dropped terms are of size
$M_\pi/M_N$ or smaller in the range of applicability of the quantization condition.
This is indeed small for physical hadron masses, and, eventually, lattice QCD calculations of multinucleon systems will approach this value.
The second issue concerns the relativistic invariance of the formalism.
If results from several different frames (i.e. different values of $\bm P$) are combined---as is now standard practice in multiparticle lattice QCD calcualtions---then it is important to have a relativistically-invariant formalism.
While this is the case for the underlying quantization condition, are we not losing this important feature by dropping higher-order terms in $\Kdf$? 
The answer is certainly yes, in principle.
However, in practice, the contribution to the shifts in energy levels from their noninteracting values due to $\Kdf$ is small (suppressed by $\sim 1/L^3$ compared to the contribution of $\cK_{2,L}$, which is being included in a relativistically-invariant manner), so a small error made in this contribution is likely acceptable.
We note, furthermore, that there is no theoretical barrier to keeping higher order terms; the issue is one of computational simplicity.

To use the forms above in the quantization condition, we need to convert them to the
$\{k\ell m m_{sk} m_{sa} m_{sb}\}$ basis, using \Cref{eq:Kdflab}.
This is a straightforward but tedious exercise, which is carried out for $\cK_B$ and $\cK'_A$ in \Cref{app:KB,app:KA}, respectively.

\section{Symmetries of the quantization condition}
\label{sec:symm}

When implementing the quantization condition, it is advantageous to project the solutions onto the
irreps of the appropriate subgroup of the symmetry group of the finite spatial volume.
We assume a cubic volume, so that the full group consists of rotations of the cube together with the parity transformation. We refer to this full group as the cubic group.
The advantages of projecting onto irreps are that it
(a) allows the energy levels to be associated with (a subset of) infinite-volume quantum numbers, 
(b) automatically accounts for symmetry-based degeneracies,
and (c) simplifies solution-finding by reducing the density of solutions.
In this section we describe the symmetry of the matrices entering the quantization condition, and thus of
the quantization condition itself.
This extends the results of Refs.~\cite{\dwave,\implement} from scalars to particles with spin $1/2$.
The irreps will be discussed in the following section.

As will be explained in the following, the matrices that enter the quantization condition
are all invariant under the unitary transformations
\begin{equation}
M = U(R)^\dagger M U(R)\,, \qquad M = F,\ G,\ \cK_{2,L},\ \Kdf\,,
\label{eq:mastertransf}
\end{equation}
where matrix indices are implicit.
$R$ is an element of the little group associated with total momentum $\bm P$,
i.e. the subgroup of the cubic group that leaves $\bm P$ invariant.
The transformations are given by
\begin{equation}
U(R)_{p \ell' m' \bm m_s'^*;k \ell m \bm m_s^*}
=
\delta_{\bm p,R\bm k} \delta_{\ell' \ell}\cD^{(\ell)}_{m'm}(R)\,
\cD^{(1/2)}_{m_{sp} m_{sk}}(R)\, 
\cD^{(1/2)}_{m^*_{sa'} m^*_{sa}}\!(R)\, 
\cD^{(1/2)}_{m^*_{sb'} m^*_{sb}}\!(R)\,,
\label{eq:Udef}
\end{equation}
where we are using the $\bm m^*_s$ basis rather than the $s$ or $j$ bases.
In fact, the same invariance holds also in the lab-frame basis, with the indices changed as 
$m^*_{sa} \to m_{sa}$, etc, as we shall see in the following.
We note that, in practice, we have found that testing the invariance of the matrices
provides a nontrivial check of our numerical implementation.

This invariance is most easily understood for the lab-frame version of $\Kdf$, 
using the momentum-spin basis appearing on the right-hand side of \Cref{eq:Kdfmata}.
This infinite-volume quantity is covariant under rotations, which leads to
\begin{multline}
\cK_{\rm df,3}^{\sf lab}(\bm  p, m_{sp}; \bm a', m_{sa'}; \bm b', m_{sb'}|
\bm  k, m_{sk}; \bm a, m_{sa}; \bm b, m_{sb})
=
\\
\widetilde \cD(R)^\dagger_{\bm m'_s \bm m'''_s}
\cK_{\rm df,3}^{\sf lab}(R\bm p, m'''_{sp}; R\bm a', m'''_{sa'}; R\bm b',m'''_{sb'} |
R \bm  k, m''_{sk}; R\bm a, m''_{sa}; R \bm b,m''_{sb})
\widetilde \cD(R)_{\bm m_s'' \bm m_s}\,,
\label{eq:rotateKdflab}
\end{multline}
where
\begin{equation}
\widetilde \cD(R)_{\bm m_s' \bm m_s}
= 
\cD^{(1/2)}_{m_{sk}' m_{sk}}\!(R)\, 
\cD^{(1/2)}_{m_{sa}' m_{sa}}\!(R)\, 
\cD^{(1/2)}_{m_{sb}' m_{sb}}\!(R)
\,.
\label{eq:tildeD}
\end{equation}
For the sake of brevity, we are not distinguishing between $m_{s[Rk]}$ and $m_{sk}$, etc.---indeed,
the second entry in the subscript simply indicates which of the three particles the spin index corresponds to.
Now we convert to the dimer-frame basis using \Cref{eq:Kdfmatb}, which leads to
\begin{multline}
\Kdf(\bm  p, m_{sp}; \bm a', m^*_{sa'}; \bm b',m^*_{sb'} |
\bm  k, m_{sk}; \bm a, m^*_{sa}; \bm b, m^*_{sb}) =
\left[\widetilde{ \cD}'(R)^{(p,a')\dagger}\right]_{\bm m'^*_s \bm m'''^*_s} 
\\
\times \Kdf(R\bm  p, m'''_{sp}; R\bm a', m'''^*_{sa'}; R\bm b',m'''^*_{s b'} |
R \bm  k, m''_{sk}; R\bm a, m''^*_{sa}; R \bm b,m''^*_{sb})
\times \widetilde{\cD}'(R)^{(k,a)}_{\bm m_s''^* \bm m_s^*}\,,
\label{eq:rotateKdf}
\end{multline}
where
\begin{equation}
\widetilde{ \cD}'(R)^{(k,a)} = \cD^{(Rk,Ra)\dagger} \widetilde{\cD}(R) \cD^{(k,a)}\,.
\label{eq:Dtildep}
\end{equation}
Writing this out in detail, using \Cref{eq:Dkp,eq:tildeD}, we have
\begin{multline}
\widetilde{ \cD}'(R)^{(k,a)}_{\bm m'^*_s, \bm m^*_s} =
\cD^{(1/2)}_{m'_{sk} m_{sk}}(R)
\left[\cD^{(1/2)}(R_{Rk,Ra}) \cD^{(1/2)}(R) \cD^{(1/2)}(R^{-1}_{k,a})\right]_{m'^*_{sa} m^*_{sa}}
\\
\times \left[\cD^{(1/2)}(R_{Rk,Rb}) \cD^{(1/2)}(R) \cD^{(1/2)}(R^{-1}_{k,b})\right]_{m'^*_{sb} m^*_{sb}}\,.
\label{eq:Dtildepa}
\end{multline}
To evaluate the terms in square brackets, we need the result, which follows from \Cref{eq:Rpdef},
and the fact that $R$ leaves $\bm P$ unchanged,
that the Wigner rotation $R_{Rk,Ra}$ involves the same angle as $R_{k,a}$, but the axis is rotated by $R$.
Now we can use the result
\begin{equation}
\cD^{(1/2)}(R[\theta,R' \hat n]) =
\cD^{(1/2)}(R') \cD^{(1/2)}(R[\theta,\hat n]) \cD^{(1/2)}(R'^{-1}) \,,
\end{equation}
which allows us to write
\begin{equation}
\cD^{(1/2)}(R_{Rk,Ra}) = \cD^{(1/2)}(R) \cD^{(1/2)}(R_{k,a}) \cD^{(1/2)}(R^{-1})\,.
\end{equation}
Substituting this, and the corresponding result with $a\to b$, into \Cref{eq:Dtildepa},
we find the simple result
\begin{equation}
\widetilde{ \cD}'(R)^{(k,a)} = \widetilde { \cD}(R)^{(k,a)}\,,
\end{equation}
so that the transformation of the dimer-frame $\Kdf$ takes the same form as that in the lab-frame basis.

The final step is to project onto $\{k \ell m\}$ indices in the standard way. The rotation of $\bm a$ implies
that $\bm a^* \to R \bm a^*$, which can be represented 
by the action of the Wigner matrix $\cD^{(\ell)}$ in \Cref{eq:Udef}, 
as explained in Ref.~\cite{\dwave}. This step does not affect the rotation of the spin indices, so that
we end up with the unitary operator $U$ given in \Cref{eq:Udef}.

Now we turn to the invariance of the other matrices in the quantization condition.
The invariance of $F$ under \Cref{eq:mastertransf} follows from that established for the spinless case
in Ref.~\cite{\dwave,\implement}, since $F$, given in \Cref{eq:Fmat},
 contains a delta function in all spin variables, so that the 
conjugation by Wigner D-matrices acting on spin indices has no effect.

Turning to $G$, given in \Cref{eq:Gmat},
since the lab-frame and dimer-axis frame versions are related in the same way as
for $\Kdf$,  we know from the discussion above that it is sufficient to demonstrate the invariance
of  $G^{\sf lab}$, \Cref{eq:Glab}. 
The analysis for the non-spin part is then as in Refs.~\cite{\dwave,\implement}, and leads to the $\cD^{(\ell)}$ term in $U(R)$.
For the spin part, the factors of $\cD^{(1/2)}$ in $U(R)$ and of $\cD^{(1/2)\dagger}$ in $U(R)^\dagger$ cancel.
The only subtlety is that the initial spectator spin
is connected to that of one of the final-state pairs, etc. Nevertheless, since $U(R)$ has exactly
the same Wigner matrix acting on all three spin indices, the product of delta functions is invariant under
conjugation, and maintains the form of the connections between spins. 
Thus invariance follows.

Finally, we consider $\cK_{2,L}$, whose form in the $\{\ell m \bm m^*_s\}$ basis is given in \Cref{eq:K2mat}. 
Its invariance under \Cref{eq:mastertransf}
follows from the same argument as for $\Kdf$, since the spectator delta function and
the two-particle interaction together transform in the same manner as $\Kdf$.
Alternatively, one can do an explicit check as the form of $\cK_{2,L}$ is relatively simple.

As described in Ref.~\cite{\threeN}, and discussed in detail in \Cref{sec:QCK2L,sec:implementG,sec:implementK2},
in practice we project the total dimer spin onto $s=0$ or $1$, and then enforce Fermi statistics by
using only even $\ell$ for $s=0$ and odd $\ell$ for $s=1$.
The projection involves conjugation by appropriate Clebsch-Gordon (CG) matrices.
It is straightforward to see, using the CG series,
\begin{equation}
\cD^{(j_1)}_{m_1' m_1}(R) \cD^{(j_2)}_{m_2' m_2}(R) = \sum_{j,m_j', m_j}
\langle j_1 j_2 m_1' m_2' | j m_j'\rangle
\cD^{(j)}_{m_j' m_j}(R)
\langle j m_j| j_1 j_2 m_1 m_2\rangle\,,
\label{eq:CGseries}
\end{equation}
that, in the $\{\ell m_{sk} s \}$ basis, invariance still holds, but with the conjugation matrix becoming\footnote{%
We are abusing notation by using the same name for the matrix as in \Cref{eq:Udef}, but context will make clear which version is being referred to.}
\begin{align}
U(R)_{p \ell' m' m_{sp} s' \mu'_s ;k \ell m m_k s \mu_s}
&=
\delta_{\bm p,R\bm k} \delta_{\ell' \ell} \delta_{s' s} U(R)^{(\ell,s)}_{m' m_{sp} \mu'_s ; m m_k \mu_s}\,,
\label{eq:Udefells}
\\
U(R)^{(\ell,s)}_{m' m_{sp} \mu'_s ; m m_k \mu_s} &=
\cD^{(\ell)}_{m'm}(R)\,
\cD^{(1/2)}_{m'_{sk} m_{s k}}(R)\, 
\cD^{(s)}_{\mu'_{s} \mu_{s}}\!(R)\,.
\label{eq:Uells}
\end{align}

\subsection{Irrep decomposition}
\label{sec:irrep}

In this section we describe the decomposition of the solutions of the quantization condition into irreps
of the appropriate little group.
More precisely, since the three-neutron system has half-integer total angular momentum,
we must consider the doubled versions of the little groups,
which include the element corresponding to a $2\pi$ rotation about any axis. 
We use the presentation of the group, classes and character tables from Refs.~\cite{Morningstar:2013bda,\threehadrons}.
We need only the fermionic irreps, and we list these, together with their dimensions,
in Table~\ref{tab:fermionicirreps}.\footnote{%
The irreps for $\bm P=0$ (without considering parity)
have also be discussed in Ref.~\cite{Mandula:1983wb}, and in the notation of that work,
$G_{1g}=\tfrac12$, $G_{2g}=\overline{\tfrac12}$, and $H_g(4) = \tfrac32$.
}

\begin{table}[htp]
\begin{center}
\begin{tabular}{|c|c|c|}
\hline
$\bm P$ & Little Group(order) & Fermionic irreps \\
\hline
$(0,0,0)$ & $O_h^D(96)$ & $G_{1g}(2),\ G_{2g}(2),\ H_g(4),\ G_{1u}(2),\ G_{2u}(2),\ H_u(4)$ \\
$(0,0,a)$ & $C_{4v}^D(16)$ & $G_1(2),\ G_2(2)$ \\
$(0,a,a)$ & $C_{2v}^D(8)$ & $G(2)$ \\
$(a,a,a)$ & $C_{3v}^D(12)$ & $F_1(1),\ F_2(1),\ G(2)$ \\
$(0,a,b)$ & $C_s^D(4)$ & $F_1(1),\ F_2(1)$\\
$(a,a,b)$ & $C_s^D(4)$ & $F_1(1),\ F_2(1)$\\
$(a,b,c)$ & $Z_2(2)$ & $F(1)$\\
\hline
\end{tabular}
\caption{Fermionic irreps of double groups corresponding to different classes of total
lab momenta, $\bm P $, for which $a$, $b$, and $c$ are different, nonzero, components.
The dimensions of the irreps are given in parentheses. The subcsripts $g$ and $u$ indicated
positive and negative parity, respectively. }
\label{tab:fermionicirreps}
\end{center}
\end{table}

The matrices $U(R)$ appearing in \Cref{eq:mastertransf}
form a reducible representation of the little group.
As explained in Refs.~\cite{\dwave,\implement},
one can project the quantization condition onto the irreps that this representation contains.
We will use the $\{\ell m_{sk} s\}$-basis result of \Cref{eq:Udefells},
from which it is apparent that $\ell$ and $s$ are unchanged by the rotations.
A further simplification is that the $U(R)$ is diagonal in orbits, where an orbit
is a set of finite-volume momenta that are connected by little-group transformations:
$o_{\bm k} \equiv \{R\bm k| R \in {\rm LG}(\bm P)\}$. 
Together these results imply that the projectors are block-diagonal in $\ell$, $s$ and orbit space. 
Specifically, when we project onto irrep $I$, of dimension $d_I$, and having characters $\chi_I(R)$, using
\begin{align}
P_I &= \frac{d_I}{[{\rm LG}(\bm P)]} \sum_{R\in {\rm LG}(\bm P)}\chi_I(R)^* U(R)\,,
\end{align}
the projectors block diagonalize as
\begin{align}
P_I &= {\rm diag}\left(P_I^{(\ell=0,s=0)},P_I^{(\ell=2,s=0)}, \dots, P_I^{(\ell=1,s=1)}, P_I^{(\ell=3,s=1)},\dots \right)\,,
\\
P_I^{(\ell,s)} &= {\rm diag} \left(P_{I,o_1}^{(\ell,s)}, P_{I,o_2}^{(\ell,s)}, \dots \right)\,,
\end{align}
where the most fine-grained projectors are
\begin{align}
\left[P_{I,o}^{(\ell,s)}\right]_{p m_{sp} m' \mu_s'; k m_{sk} m \mu_s}
&= \frac{d_I}{[{\rm LG}(\bm P)]} \sum_{R\in {\rm LG}(\bm P)}\chi_I(R)^* \delta_{\bm p,R\bm k} 
U(R)^{(\ell,s)}_{m_{sp} m' \mu_s'; m_{sk} m \mu_s} \,.
\end{align}
Here $U(R)^{(\ell,s)}$ is given in \Cref{eq:Uells}, and the indices $p,k$ run over the elements of the orbit $o$.
The dimensions of the projectors for low-lying values of $\{\ell,s\}$ are given in \Cref{app:projdims}.

\subsection{Subductions of infinite-volume $J^P$ into irreps}

In this section we describe the subduction of irreps of the infinite-volume doubled rotation-parity group,
$SU(2)\times Z_2^P$,
which are described by the quantum numbers $J^P$, into irreps of the various finite-volume little groups.

First we list the total $J^P$ values that occur for three neutrons in infinite volume.
These are obtained by combining a pair into a definite value of $\{\ell,s\}$, 
and then combining that pair with the spectator (whose intrinsic $J^P=\frac12^+$) 
with relative angular momentum $L$. For the three cases with $\ell \le 2$, the resulting $J^P$ values are
\begin{align}
s=\ell=0: & \quad (L=0) \tfrac12^+;\ (L=1) \tfrac12^-, \tfrac32^-;\ (L=2) \tfrac32^+, \tfrac52^+;\ \dots
\label{eq:s0ell0}
\\
s=\ell=1: & \quad (L=0) \tfrac12^-, \tfrac32^-,\tfrac52^-;\ 
(L=1) \tfrac12^+, \tfrac32^+, \tfrac52^+, \tfrac72^+;\ \dots
\label{eq:s1ell1}
\\
s=0,\ell=2: & \quad (L=0) \tfrac32^+,\tfrac52^+;\ (L=1) \tfrac12^-, \tfrac32^-,\tfrac52^-,\tfrac72^-;\ 
(L=2) \tfrac12^+,\dots
\label{eq:s0ell2}
\end{align}
Thus all half-integer values of $J$ with either parity are allowed.

The subduction of the infinite-volume fermionic irreps into those of the full doubled cubic group is given in Table XVII of
Ref.~\cite{Morningstar:2013bda} and reproduced in \Cref{tab:subduct1}.
The subductions of the fermionic irreps of $O_h^D$ into the little groups for the other frames
is given in part in Table XVIII of Ref.~\cite{Morningstar:2013bda}, and can be completed
using the character tables given in Ref.~\cite{\threehadrons}. The result is given in \Cref{tab:subduct2}.
Combining these two tables one can subduce any infinite-volume value of $J^P$ into the irreps of the appropriate little group.
For example, we see that only in the $\{0,0,0\}$, $\{0,0,a\}$ and $\{a,a,a\}$ frames are there irreps than subduce from $J=3/2$ that do not include a $J=1/2$ component, although in the latter two frames there are no irreps that include
$J=1/2$ but not $J=3/2$.

\begin{table}[htb]
\centering
\begin{tabular}{c|ccc}
$J$ & $G_{1g/u}(2)$ & $G_{2g/u}(2)$ & $H_{g/u}(4)$ \\
\hline\hline
$\tfrac12^{\pm}$ & 1 & 0 & 0
\\[3pt]\hline
$\frac32^{\pm}$ & 0 & 0 & 1
\\[3pt]\hline
$\frac52^{\pm}$ & 0 & 1 & 1
\\[3pt]\hline
$\frac72^{\pm}$ & 1 & 1 & 1
\\\hline
\end{tabular}
\caption{Subductions from $SU(2)\times Z_2^P$ half-integer irreps into those of the doubled cubic group, $O_h^D$. 
The number of appearances of each irrep is listed. Positive (negative) parity irreps map into those with subscripts $g$ ($u$).
\label{tab:subduct1}}
\end{table}

\begin{table}[htb]
\centering
\begin{tabular}{c|ccccc}
\text{irrep} & $C_{4V}^D$ & $C_{3V}^D$ & $C_{2V}^D$ & $C_s^D$ & $Z_2$
\\ \hline
\text{frame} & $(0,0,a)$ & $(a,a,a)$ & $(0,a,a)$ & $(0,a,b)/(a,a,b)$ & $(a,b,c)$
\\[3pt]\hline\hline
$G_{1g/u}(2)$ & $G_1(2)$ & $G(2)$ & $G(2)$ & $F_1(1)\oplus F_2(1)$ & $2 F(1)$
\\[3pt]\hline
$G_{2g/u}(2)$ & $G_2(2)$ & $G(2)$ & $G(2)$ & $F_1(1)\oplus F_2(1)$ & $2 F(1)$
\\[3pt]\hline
$H_{g/u}(4)$ & $G_1(2)\oplus G_2(2)$ & $F_1(1)\oplus F_2(1) \oplus G(2)$ & $2G(2)$ 
& $2F_1(1)\oplus 2F_2(1)$ & $4 F(1)$
\\\hline
\end{tabular}
\caption{Subductions from fermionic irreps $O_h^D$ into the little groups of moving frames.
The subduction for both choices of parity is the same.
\label{tab:subduct2}}
\end{table}

\subsection{Irreps in lowest-lying free levels}
\label{sec:irrepsfree}

To get an idea of the density of the spectrum, and the distribution of levels into irreps,
we present results for the lowest-lying free energy levels in a cubic box in various frames.
We take $M_N L=20$, which, for a neutron with physical mass, corresponds to a box of length
$L=4.3\;$fm. This is a roughly the box size used in present LQCD studies of the dinucleon system~\cite{BaSc:2025yhy}.
The results are shown in \Cref{tab:free1} for the rest frame and the five moving frames with lowest momenta.
We also display the lab-frame energy $E$ and the CMF energy $E^*=\sqrt{E^2-\bm P^2}$.

\begin{table}[htb]
\centering
\begin{tabular}{c|cccccc}
$n_P^2$ & $\{n_1^2,n_2^2,n_3^3\}$ & degen & $E^*/M_N$  & $E/M_N$ & irreps \\
\hline
0 & 
$\{0,1,1\}$ & 24 & 3.0964 & 3.0964 & $G_{1g} \oplus  H_g \oplus 2G_{1u} \oplus G_{2u} \oplus 3H_u$
\\
 & $\{0,2,2\}$ & 48 & 3.1885 &  3.1885 &
 $G_{1g}\oplus  G_{2g} \oplus 2 H_g \oplus  3G_{1u} \oplus  3G_{2u} \oplus 6 H_u$
\\
 & $\{1,1,2\}$ & 96 & 3.1906 &  3.1906 &
 $4 G_{1g}\oplus  4G_{2g} \oplus 8 H_g \oplus 4 G_{1u} \oplus 4 G_{2u} \oplus 8 H_u$
 \\ \hline
1 & $\{0,0,1\}$ & 2 & 3.0320 & 3.0482 & $ G_{1}  $
 \\
& $\{0,1,2\}$ & 32 & 3.1267 &  3.1424 &$ 8G_{1}\oplus 8 G_2  $
 \\
& $\{1,1,1\}$ & 2 & 3.1288 &  3.1446 &$ G_{1}  $
\\
& $\{1,1,1\}$ & 16 & 3.1288 & 3.1446 &$ 4G_{1}\oplus  4G_2  $
\\ \hline
2 & $\{0,0,2\}$ & 2 & 3.0622 & 3.0943 & $ G  $
 \\
& $\{0,1,1\}$ & 8 & 3.0643 & 3.0964 & $ 4G  $
 \\
& $\{0,1,3\}$ & 16 & 3.1555 & 3.1867 & $ 8G  $
\\
& $\{0,2,2\}$ & 16 & 3.1574 & 3.1885 & $ 8G  $
\\
& $\{1,1,2\}$ & 4 & 3.1595 & 3.1906 & $ 2G  $
\\
& $\{1,1,2\}$ & 56 & 3.1595 & 3.1906 & $ 28G  $
\\ \hline
 3 & $\{0,0,3\}$ & 2 & 3.0909 & 3.1385 & $ G  $
 \\
 & $\{0,1,2\}$ & 24 & 3.0950 & 3.1424 & $ 4 F_1\oplus 4 F_2 \oplus 8 G  $
 \\
 & $\{1,1,1\}$ & 8 & 3.0971 & 3.1446 & $ 2 F_1\oplus 2 F_2 \oplus 2 G  $
 \\
 & $\{1,1,3\}$ & 48 & 3.1887 & 3.2348 & $ 8 F_1\oplus 8 F_2 \oplus 16 G  $
  \\
 & $\{1,2,2\}$ & 72 & 3.1906 & 3.2367 & $ 12 F_1\oplus 12 F_2 \oplus 24 G  $
 \\ \hline
 4 & $\{0,1,1\}$ & 2 & 3.0320 & 3.0964 & $ G_1  $
\\
 & $\{0,0,4\}$ & 2 & 3.1183 & 3.1810 & $ G_1  $
\\
 & $\{0,2,2\}$ & 16 & 3.1183 & 3.1885 & $ 4G_1 \oplus 4 G_2 $
\\
 & $\{1,1,2\}$ & 32 & 3.1282 & 3.1906 & $ 8G_1 \oplus 8 G_2 $
\\ \hline
5 & $\{0,1,2\}$ & 8 & 3.0629 & 3.1424 & $ 4 F_1 \otimes 4 F_2 $
 \\ 
 & $\{1,1,1\}$ & 2 & 3.0651 & 3.1446 & $  F_1 \otimes  F_2 $
 \\ 
 & $\{0,0,5\}$ & 2 & 3.1446 & 3.2221 & $  F_1 \otimes  F_2 $
 \\ 
 & $\{0,1,4\}$ & 8 & 3.1519 & 3.2292 & $  4F_1 \otimes  4F_2 $
  \\ 
 & $\{0,2,3\}$ & 16 & 3.1555 & 3.2327 & $  8F_1 \otimes  8F_2 $
  \\ 
 & $\{1,1,3\}$ & 16 & 3.1576 & 3.2348 & $  8F_1 \otimes  8F_2 $
   \\ 
 & $\{1,2,2\}$ & 2 & 3.1595 & 3.2367 & $  F_1 \otimes  F_2 $
\\
 & $\{1,2,2\}$ & 48 & 3.1595 & 3.2367 & $  24F_1 \otimes  24F_2 $
\\ \hline
\end{tabular}
\caption{Spectrum of noninteracting three neutron states for $M_N L=20$,
together with their decomposition into irreps of the corresponding
doubled little group.  $\bm n_P$ and $\bm n_i$ are defined in the text.
Both the CMF energy $E^*$ and the lab frame energy $E$ are shown, together with the degeneracy of the levels.
We show  levels satisfying $E^* \lesssim 3.2 M_N$.
\label{tab:free1}}
\end{table}

To obtain these results, we consider ordered momentum triplets,
$\{\bm n_1, \bm n_2, \bm n_3=\bm n_P-\bm n_1-\bm n_2\}$, where the integer vectors are 
$\bm n_i = \bm p_i (L/2\pi)$ and $\bm n_P = \bm P (L/2\pi)$.
We then determine the orbits of such triplets under the appropriate little group, and decompose into irreps
using methods similar to those described above.
Orbits with $\bm n_1=\bm n_2 = \bm n_3$ are forbidden by Fermi statistics, since complete
antisymmetry in the spin wavefunction is not possible. 
For orbits with two equal momenta, the spin of this pair must be zero, and so there is a single spin degree of freedom.
For orbits with three different momenta, there are $2^3=8$ spin degrees of freedom.
One subtlety here is that, because rotations not only mix the triplets but also permute the order within the triplet,
 this permutation must be taken into account in the action of the rotation on spin indices. 
In addition, there is an overall fermion sign given by the signature of the permutation.

The table shows levels up to approximately $E^* = 3.2 M_N$.
The rationale for this upper limit is based on the inelastic threshold at which the quantization condition breaks down, $E_{\rm inel}^* = 3 M_N+ M_\pi$.
If we set $M_\pi$ to its physical value, such that $M_\pi/M_N\approx 0.15$, then $E_{\rm inel}^*=3.15 M_N$,
while taking a somewhat heavier-than-physical value, $M_\pi/M_N=0.2$, leads to $E_{\rm inel}^*=3.2 M_N$.

There are several features of note in the results of \Cref{tab:free1}.
First, there are several instances of degeneracy between the listed levels.
This is due to our use of the label $\{n_1^2, n_2^2, n_3^2\}$, which is unique in most but not all cases.
For example, for $\bm n_P=(0,0,1)$, there are two $\{n_1^2, n_2^2, n_3^2\}=(1,1,1)$ orbits,
the two dimensional one with $\bm n_1=\bm n_2=(0,0,1)$ and $\bm n_3=(0,0,-1)$,
and the sixteen-dimensional orbit containing the state with $\bm n_1=(1,0,0)$, $\bm n_2=(-1,0,0)$,
and $\bm n_3 = (0,0,1)$.
Second, we observe that levels come in bands with nearly degenerate energies, 
with two bands apparent for each frame.
This arises because we are close to the nonrelativistic limit, in which
$E/M_N \approx 3  + 2\pi^2\sum_i n_i^2/ (M_N L)^2$,
and levels in each band have the same value of $\sum_i n_i^2$.
Finally, we note that to obtain states in all irreps requires only going to at most the second energy level.
This suggests that a future LQCD calculation should use all irreps when studying this system.

We have used these results to test our implementation of the matrices $F$ and $G$.
In particular, the residues of the poles in $F+G$ at the free energies should have the rank given by the
degeneracies listed in \Cref{tab:free1}, and should decompose into the listed irreps.
We have checked that this is the case.

\subsection{Irreps in the $\Kdf$ matrix}

As will be seen below, the two-particle interactions contained in $\cK_{2,L}$ are sufficient to
shift the energies of states in all irreps listed in \Cref{tab:free1}. 
It is interesting to determine, however, which irreps are shifted further when $\Kdf$ is included.
We find that $\cK_A$ and $\cK_B$ have, respectively, 2 and 6 nonzero eigenvalues in all frames, 
and these live in the irreps shown in \Cref{tab:Kdfirreps}.
Comparing to \Cref{tab:fermionicirreps}, we see that, in moving frames, all irreps are shifted by $\cK_B$,
while, in the rest frame, only two of the six irreps are shifted: those that subduce from $J^P=(1/2)^-$ and $(3/2)^-$.
As can be seen from \Cref{eq:s0ell0,eq:s1ell1}, these are total $J^P$ values that require 
at least one unit of relative angular momentum, which is consistent with $\cK_B$ being quadratic 
in momenta---see \Cref{eq:KB}.
As for $\cK_A$, it shifts only a subset of the irreps effected by $\cK_B$.

\begin{table}[htb!]
\centering
\begin{tabular}{c|cc}
frame & $\cK_A$ irreps & $\cK_B$ irreps \\
\hline\hline
$(0,0,0)$ & $G_{1u}$ & $G_{1u} \oplus H_u$ 
\\\hline
$(0,0,a)$ & $G_1$ & $2 G_1 \oplus G_2$
\\\hline
$(0,a,a)$ &  $G$ & $3 G$
\\\hline
$(a,a,a)$ &  $G$ & $F_1\oplus F_2 \oplus 2 G$
\\\hline
$(0,a,b)$ & $F_1 \oplus F_2$ & $3 F_1 \oplus 3 F_2$
\\\hline
$(a,a,b)$ & $F_1 \oplus F_2$ & $3 F_1 \oplus 3 F_2$
\\\hline
$(a,b,c)$ & $2F$ & $6F$
\\\hline
\end{tabular}
\caption{Irreps contained in the $\cK_A$ and $\cK_B$ terms in $\Kdf$ in the different classes of frame.
\label{tab:Kdfirreps}}
\end{table}

\section{Numerical examples}
\label{sec:num}

In this section we use the three-neutron quantization condition to predict the finite-volume spectrum for given choices of the two- and three-particle K matrices.
As for the noninteracting energies in \Cref{tab:free1}, we set $M_N L=20$ for most of the results.
The pion mass must also be specified, for it enters the cutoff function \Cref{eq:cutoff} through the value of 
$\sigma_{\rm lhc}$---see \Cref{eq:sigmalhc}.
We use $M_\pi/M_N=0.15$, which is close to the physical ratio.
This leads to an uncomfortably small value of the box size in pion units, $M_\pi L=3$, 
for which finite-volume effects proportional to $\exp(-M_\pi L)$ may not be sufficiently suppressed. In this regard, the quantization condition provides a built-in cross check on its applicability,
because unphysical solutions start to appear when $M_\pi L$ becomes too small~\cite{\BHSnum,\dwave}.
We will see several examples of this in the following, and, in those cases, study the effect of moving to a larger lattices with $M_\pi L$ ranging up to $6$.
We view the use  $M_\pi L=3$ as providing a stress test of the formalism, 
while, at the same time, showing results for a choice of parameters that is not implausible for an initial future attempt at studying the three-neutron system using LQCD.

All numerical results have been obtained and cross-checked using two independent {\sf Mathematica} codes. We have used laptop computers, which somewhat limits the resolution of curves in the following, but is adequate for this preliminary investigation.

\subsection{Form of $\cK_2$}
\label{sec:k2form}
As described above, truncating to $\ell \le 1$ implies that we need the two-particle K matrix in four channels:
 $[j=\ell=s=0]$ (${}^1\!S_0$), $[j=0,\ell=s=1]$ (${}^3\!P_0$), $[j=\ell=s=1]$ (${}^3\!P_1$), 
 and $[j=2,\ell=s=1]$ (${}^3\!P_2$).
 There is no mixing between these channels; in particular, the two $j=0$ channels cannot mix as they have
 opposite parities.
 The $j=2$ channel could mix with that having $[j=2,\ell=3,s=1]$ (${}^3\!F_2$), 
 but we assume that there are no interactions in the latter channel.

\begin{figure}
    \centering
    \includegraphics[width=\linewidth]{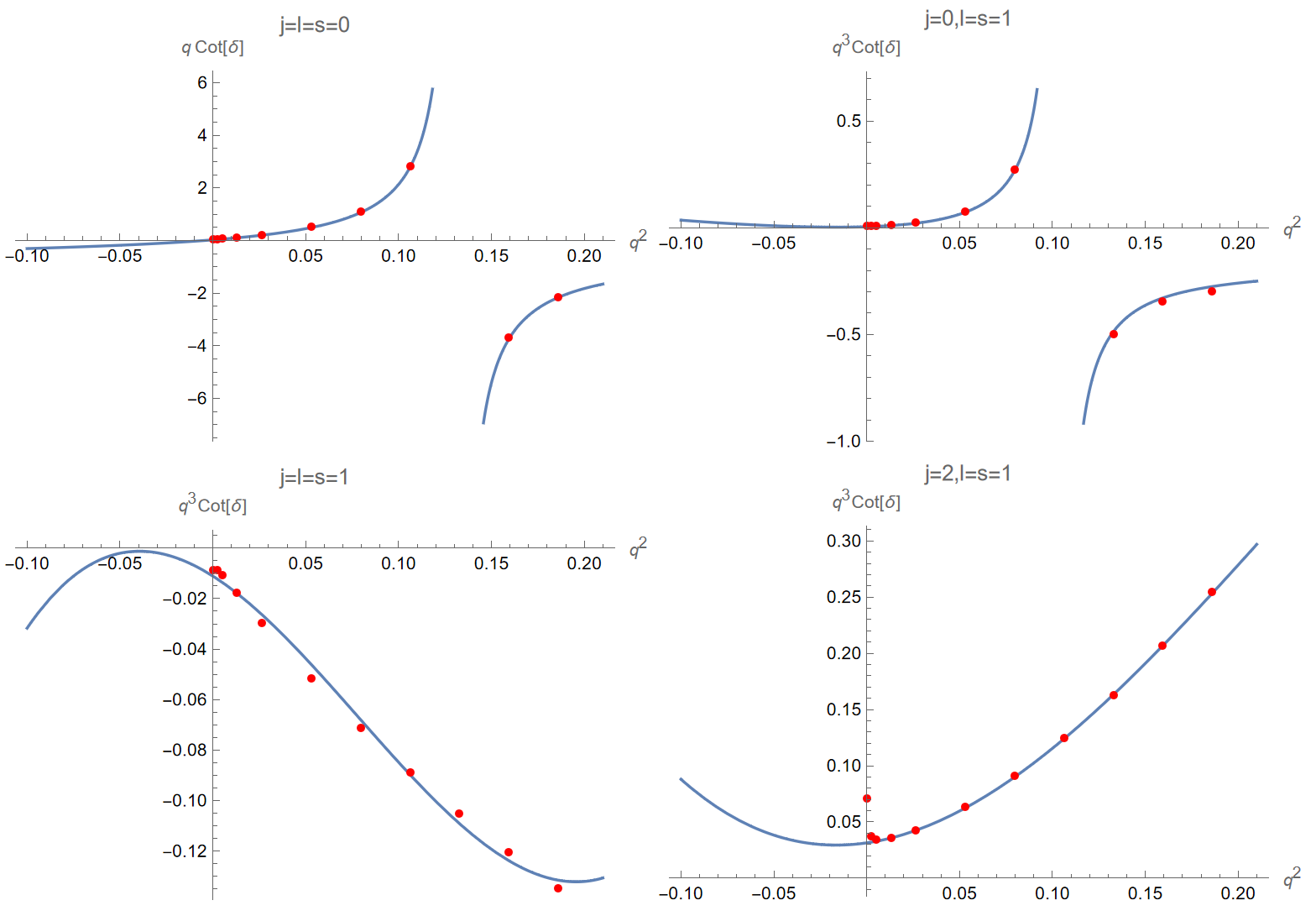}
    \caption{Form of the scattering phase shifts for each of the $nn$ channels that we consider. Data points are from Ref.~\cite{Stoks:1994wp}; the blue lines are our chosen forms. All quantities are expressed in units of the neutron mass. For further discussion, see text. 
    }
    \label{fig:sl}
\end{figure}

In order to have results that are as close to physical as possible, we have used forms based on measurements of $I=1$ nucleon-nucleon scattering. Specifically, we use the results for proton-proton scattering phase shifts given
in Table IV of Ref.~\cite{Stoks:1994wp}, 
taking an approximate average of the analyses considered in that work.
We assume that isospin breaking effects (including Coulomb effects) are small, 
so that $pp$ phase shifts give a good estimate of those for $nn$.
The results for $q^{2\ell+1} \cot \delta(q)$ for the four channels of interest are shown as the (red) points in \Cref{fig:sl}.
We do not attempt to estimate errors as we will not be doing a fit.

Instead, we choose forms that are polynomials in $q_k^{*2}$, together with possible poles, that match the experimental data reasonably well, and that do not lead to unphysical behavior in the region slightly below threshold.
Our chosen forms are shown by the (blue) curves in the figures. They are given explicitly by
\begin{align}
q \cot \delta^{0,0,0}(q) &= -0.63576 - \frac{0.08711}{q^2 - 0.13201} + 0.46267 \, q^2\,,
\\
q^3 \cot \delta^{0,1,1}(q) &= - 0.08878 - \frac{0.00973}{q^2 - 0.10417} -0.62487\, q^2 + 1.40252\, q^4\,,
\\
q^3 \cot\delta^{1,1,1}(q) &= -0.01107 - 0.46700\, q^2-4.71890\, q^4 + 20.14850\, q^6\,,
\\
q^3 \cot\delta^{2,1,1}(q) &= 0.03137 + 0.23995\, q^2 + 7.03364\, q^4 - 10.26176\, q^6\,,
\end{align}
where $q = q_k^*/M_N$, and, following \Cref{eq:K2cotd}, the notation for the phase shift is $\delta^{(j,\ell,s)}$.
To get a sense of the appropriate horizontal scale in the figures, 
we note that the left-hand cut due to $t$-channel pion occurs at 
$q_{\rm min}^2= - M_\pi^2/(4 M_N^2) \approx -0.0056$, which is only just below the origin in the plots.
This is where our cutoff function reaches zero, and cuts off the subthreshold behavior.
At the other end, the inelastic threshold for $NN \pi$ production
occurs at $q^2 = M_\pi/M_N+ M_\pi^2/(4 M_N^2) \approx 0.156$, 
close to the upper end of the range shown in the plots.

In the subthreshold region, there are singularities in $\cM_2$ 
whenever 
\begin{equation}
q^{2\ell+1} \cot\delta = \pm (-q^2)^{(2\ell+1)/2}\,.
\end{equation}
These can correspond to bound states or virtual bound states depending on which branch of the right-hand side is crossed.
For the  $j=\ell=s=0$ channel, we find the expected  virtual bound state, 
which occurs at $q^2 \approx -0.0005$ with our chosen parameters.
For the other channels, we have chosen parameters such that any such singularities lie 
reasonably far below $q_{\rm min}^2$.
In particular, for $j=0$, $\ell=s=1$, there is a bound-state crossing with unphysical residue at 
$q^2 \approx -0.019 \approx 3 q_{\rm min}^2$,
while for $j=\ell=s=1$ there is a virtual bound-state crossing at $q^2 \approx -0.022 \approx 4 q_{\rm min}^2$.
There are no nearby singularities for the $j=2$, $\ell=s=1$ case.

Before describing the resulting three neutron spectra, it is interesting to know which of the irreps appearing in the free levels listed in \Cref{tab:free1} are shifted by the two-particle interactions in each of the four channels.
To address this, we turn on the channels one by one, and determine which irreps are contained in the 
resulting matrix $\cK_{2,L}$. The answer depends on the energy $E^*$, so we consider the energies of
the lowest few free states given in the table.
The results are collected in \Cref{tab:K2shift}.
Note that once an irrep is present at a certain $E^*$, it will always be present for higher energies.

\begin{table}[htb]
\centering
\begin{tabular}{c|ccc}
$n_P^2$ & $E^*/M_N$ & $\{j,\ell,s\}$ & irreps \\
\hline
0 & 3.1 & $\{0,0,0\},\ \{0,1,1\}$ & $ G_{1g} \oplus  H_g \oplus G_{1u}  \oplus H_u$
\\
 & 3.1 & $\{1,1,1\},\ \{2,1,1\}$ & $ G_{1g} \oplus  G_{2g}\oplus H_g \oplus G_{1u} \oplus G_{2u} \oplus H_u$
\\
 & 3.19 &  All &
 $G_{1g}\oplus  G_{2g} \oplus H_g \oplus  G_{1u} \oplus  G_{2u} \oplus  H_u$
 \\ \hline
1 & 3.032 &$\{0,0,0\},\ \{0,1,1\}$  & $ G_{1}  $
 \\
 & 3.032 &$\{1,1,1\},\ \{2,1,1\}$  & $ G_{1}\oplus G_2  $
 \\
 & 3.13 & All  & $ G_{1}\oplus  G_2  $
\\ \hline
2 & 3.06 & All  &$  G  $
\\ \hline
3 & 3.1 & All  &$ F_1\oplus F_2 \oplus G  $
 \\ \hline
4 & 3.03 &$\{0,0,0\},\ \{0,1,1\}$  & $ G_1  $
\\
 & 3.03 &$\{1,1,1\},\ \{2,1,1\}$  & $ G_1 \oplus G_2  $
\\
 & 3.12 & All & $ G_1 \oplus G_2  $
\\ \hline
5 & 3.07 & All & $ F_1 \otimes  F_2 $
\\ \hline
\end{tabular}
\caption{Irreps appearing in $\cK_{2,L}$ when one of the four channels is turned on,
for energies roughly corresponding to those of three neutron states (as given in \Cref{tab:free1}).
Channels are labeled by $\{j,\ell, s\}$, with ``All'' indicating that each of channels leads to the irreps shown.
\label{tab:K2shift}}
\end{table}

What we learn by comparing the results from the table is those in \Cref{tab:free1} is that, with one exception,
$\cK_{2,L}$ in each channel {\em alone} contains the irreps necessary to shift all the free energies.
The exception is the free level in $G_{2u}$ irrep in the rest frame at $E^*=3.1 M_N$, which is not shifted by
the two $j=0$ channels.
Thus, aside from this case, there are no irreps whose energy shifts can serve as direct measure of the strength
of an individual channel.
Instead, a global fit is needed.

\subsection{Results with $\Kdf=0$}
\label{sec:num:Kdf0}

\begin{figure}[htb!]
    \centering
    \includegraphics[width=0.9\linewidth]{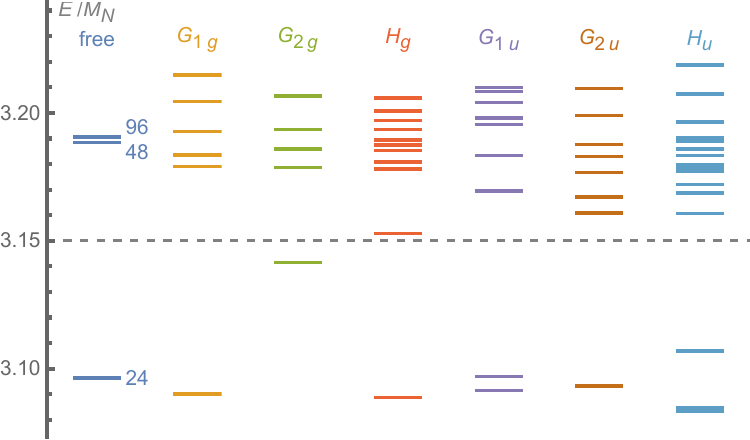}
    \caption{Three-neutron spectrum in the rest frame using $\Kdf=0$, broken down by irrep.
    Energies are in units of $M_N$;  numerical values of the energies are given in \Cref{tab:nP0noKdf}.
    Noninteracting energy levels are shown in the left-most column (denoted ``free''), along with their degeneracies 
    (which include the degeneracies within the irrep).
    Results are for $M_N L=20$ and $M_\pi/M_N=0.15$.
    The inelastic threshold is shown by the dashed horizontal line.
    }
    \label{fig:nP0noKdf}
\end{figure}

\begin{figure}[h!]
    \centering
        \begin{subfigure}{0.45\textwidth}
        \centering
        \includegraphics[width=\textwidth]{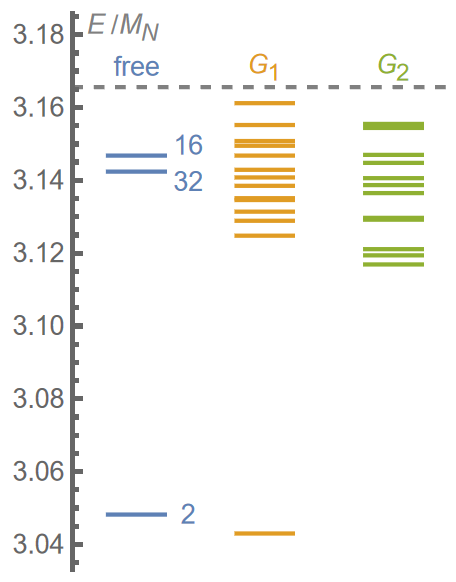}
        \caption{$n_P^2=1$ frame.
        Numerical values of the energies are given in \Cref{tab:nP1noKdf}.}
        \label{fig:nP1noKdf}
        \end{subfigure}
    \hfill
        \begin{subfigure}{0.45\textwidth}
        \centering
        \includegraphics[width=\textwidth]{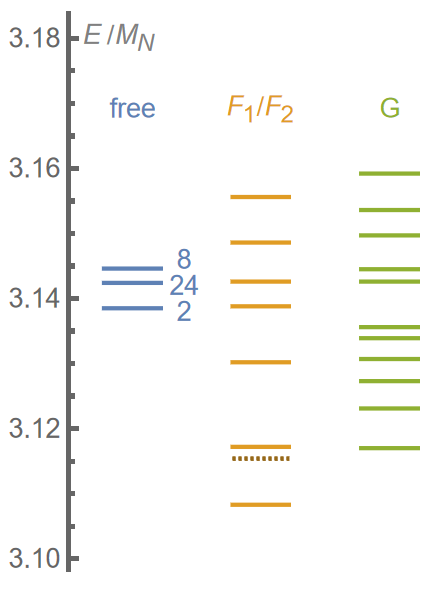}
        \caption{$n_P^2=3$ frame.
    Numerical values of the energies are given in \Cref{tab:nP3noKdf}.}
        \label{fig:nP3noKdf}
        \end{subfigure}
    
    \caption{As for \Cref{fig:nP0noKdf}, but for (a) $n_P^2=1$ and (b) $n_P^2=3$.
    In the right panel, the dotted (orange) line in the $F_1/F_2$ irrep indicates a level with unphysical residue, as discussed in the text.
    The inelastic threshold in the right panel lies above the plot range at $E=3.1967 M_N$.}
    \label{fig:nP13nokdf}
\end{figure}

We now solve the three-particle quantization condition, \Cref{eq:QC3}, while setting $\Kdf=0$.
This requires that eigenvalues of $F_3$ diverge, and so, given the definition \Cref{eq:F3def}, 
the quantization condition simplifies to
\begin{equation}
\det\left[\cK_{2,L}^{-1}+F+G\right] = 0\,.
\end{equation}
Using the forms for the two particle interactions described in the previous section, we have determined solutions to this equation for the frames with $n_P^2=0-5$,
Numerical values are collected in \Cref{app:moreframes}, and the results
for the frames with $n_P^2=0$, $1$ and $3$ are shown in
\Cref{fig:nP0noKdf,fig:nP1noKdf,fig:nP3noKdf}, respectively.

We see that, as expected, two-particle interactions completely break the degeneracies of the free levels,
although the ``bands'' of noninteracting levels that are degenerate in the nonrelativistic limit remain visible.
The spread of levels in each band is much greater than the splitting between the free levels making up the band.
In the rest frame, $n_P^2=0$, the second band lies mostly above the inelastic threshold at $E=E^*=3.15$,
so that the quantization condition is, strictly speaking, no longer valid.
However, practical applications of the three-particle formalism typically find that inelastic effects are small for some distance above the threshold (see, e.g., Ref.~\cite{\threepPRD}), 
so we expect that the displayed results will provide a reasonable guide to the spectrum.
In the other frames, the situation is better, with only a few levels having energies above the inelastic threshold.
Futhermore, as $M_\pi L$ increases, the energies of excited levels in the rest frame decrease, while the inelastic threshold stays fixed.

We also note that the none of the states lie close to the lower limit of the energy range of applicability of the quantization condition, given in \Cref{eq:Erange}. For our parameters, this limit occurs at $E=2.994 M_N$ in the rest frame, and $E=3.011 M_N$ and $3.043 M_N$ in the $n_P^2=1$ and $3$ frames, respectively. The fact that the three-neutron states must be completely antisymmetric pushes the states away from the lower limit, despite the fact that the limit lies close to the threshold energy.

In almost all cases the number of levels in each irrep is the same as that for the free levels shown in \Cref{tab:free1}. This is as expected in a nonresonant system with no bound states.
Levels are both lowered and raised by the two-particle interactions, which is also expected given that there is a mix of attractive and repulsive channels.
There is clearly a significant amount of information contained in the splittings between the levels.
However, it must also be noted that the splittings between adjacent levels are typically $5-25\;$MeV,
and thus will be challenging to determine in simulations.
The decomposition into irreps will be an essential tool in helping separate the levels.

\begin{figure}[htb!]
    \centering
      \includegraphics[width=0.8\linewidth]{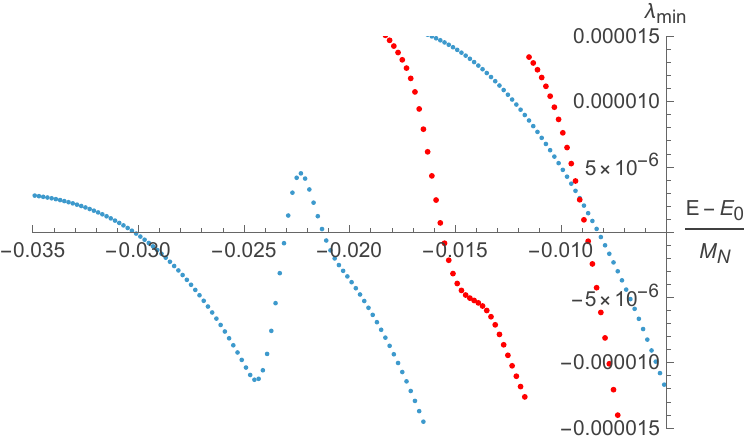}
    \caption{Eigenvalue of $(\cK_{2,L}^{-1}+F+G)M_N$ for $n_P^2=3$ in the $F_1$ irrep with the smallest absolute value plotted against $(E - E_0)/M_N$,   
    where $E_0$ is the lowest noninteracting energy in this irrep.
    Blue points are for $M_N L=20$ ($M_\pi L=3$), for which $E_0/M_N=3.1385$; 
    red points are for $M_N L = 80/3$ ($M_\pi L=4$), for which $E_0/M_N=3.0801$.
    A physical crossing corresponds to the eigenvalue passing through zero from above. 
    }
    \label{fig:unphysical}
\end{figure}

The exception to the equality of free and interacting levels occurs in $n_P^2=3$ frame.
Here, in the $F_1$ and $F_2$ irreps, there are two more solutions in the presence of interactions than in the free case, with one of these having an unphysical residue.
As discussed in Refs.~\cite{\BHSnum,\dwave}, the requirement that diagonal elements of a finite-volume correlator matrix have poles with a positive residue translates into the requirement that the eigenvalues of $\cK_{2,L}^{-1}+F+G$ cross zero in a specific direction (in our case, from positive to negative with increasing energy).
This holds also in the presence of spin.
Crossings in the opposite direction correspond to ``ghost'' states with unphysical residue, and are indicative of a breakdown of the formalism.
A plot of the smallest eigenvalue (in magnitude) versus $E$ in the relevant energy range is shown by the blue points in \Cref{fig:unphysical}. 
Starting at the left, we see a physical crossing followed by a closely-spaced physical-unphysical pair, and then by a further physical crossing.

Such unphysical crossings have been seen previously in Refs.~\cite{\BHSnum,\dwave}.
They typically occur in conjunction with an ``extra'' solution with physical residue,
similar to the central pair in \Cref{fig:unphysical}.
Here ``extra" means that it is in addition to the expected number of solutions based on the counting of noninteracting levels.
They have been found to occur when interactions are strong, as is the case here in several of the two-particle channels.
In addition, they have been found to disappear as $L$ increases, e.g.
by the central ``bump'' in the blue points in \Cref{fig:unphysical} dropping below the $x$-axis.
The most likely interpretation of these solutions is that they are due to the exponentially-suppressed effects
that are not incorporated into the quantization condition being enhanced by the combination of
the small value of $M_\pi L$ and a larger coefficient due to strong interactions.
Alternative hypotheses are that the choices of $\cK_2$ and $\Kdf$ are unphysical, or that the truncation in $\ell$ leads to unphysical solutions.

To test the interpretation that the unphysical solutions are due to enhanced exponentially suppressed effects,
we have made several further calculations of the $n_P^2=3$ spectrum.
First, we have increased $L$ by a factor of $4/3$, so that $M_\pi L=4$.
The resulting behavior of the minimal eigenvalue is shown by the red points in \Cref{fig:unphysical},
and the energies are listed in the right-hand part of \Cref{tab:nP3noKdf}.
We indeed find that the unphysical-physical pair disappears, with the bump being replaced by monotonic behavior.
The energy shifts are also, in general, smaller, as expected for effects that scale roughly as $1/L^3$.

We have also studied the impact of reducing the strength of the interaction in the $\ell=s=1$, $j=0$ channel, which is the most attractive, and leads, as seen above, to a virtual bound state within our energy range.
We find that if we scale up $q^3 \cot\delta^{0,1,1}$ by a factor of greater than about 6, then there are no unphysical solutions or extra physical crossings.

Finally, we have recalculated the spectrum using a cutoff function, \Cref{eq:cutoff}, with $\epsilon_H=1$ rather than
our canonical choice of $\epsilon_H=0$.
This choice compresses the change in $H$ from unity to zero to the lower half of the subthreshold region. 
We find that, at the level of $10^{-5} M_N$, almost all energy levels are unchanged, 
with the exceptions being the second and third levels in the $F_1$ irrep, 
which change from $\{3.1154^*, 3.1172\}$ to $\{3.1143^*,  3.1164\}$, 
and the first, fifth and six levels in the $G$ irrep, which change from
$\{3.1170,3.1339, 3.1356\}$ to $\{3.1157, 3.1337, 3.1361 \}$.
To interpret these results, we recall that changes in the cutoff function can, in principle, 
be compensated by changes in $\Kdf$. 
However, the fact that almost all levels are unchanged implies that any change in $\Kdf$ must be very small, 
and is, in particular, unlikely to be the cause of the significant shift in the five levels described above.
This leaves enhanced exponentially-suppressed effects (which we know depend on the cutoff function)
as the most likely explanation for the shifts.

All three results are consistent with the hypothesis that the unphysical solutions are due to exponentially-suppressed finite-volume effects. Given that we are working with $M_\pi L=3$, this is not a surprise. Indeed, the surprise is that such effects only show up in one of the frames we are considering. 

As a final note on the results for the spectrum, we observe an
exact degeneracy of levels in the $F_1$ and $F_2$ irreps for $n_P^2=3$ and $5$.
We have not found an analytic explanation of this degeneracy.

\subsection{Impact of nonzero $\Kdf$}
\label{sec:num:Kdf}

We now turn on $\Kdf$, keeping $\cK_{2,L}$ unchanged,
and determine the impact of the $\cK_A$ and $\cK_B$ terms of \Cref{eq:KdfEFT}
in turn. Although one might have expected the natural size of the coefficients of these terms
in pionless EFT to be of $\cO(1)$ (with $\Lambda_{\rm EFT}=M_\pi$),
one should keep in mind that much larger values are possible for three-particle interactions,
as the quantization condition and integral equations remain well defined even when $\Kdf$ diverges~\cite{\BHSnum}.
We thus consider a large range of values of the coefficients.

The specific form of the quantization condition that we implement is
\begin{equation}
\det\left[F_3^{-1} + \Kdf\right] = 0\,.
\label{eq:QC3a}
\end{equation}
In this form, both terms on the left-hand side are hermitian, so all eigenvalues are real, and one can scan for solutions by tracking the eigenvalue with the smallest magnitude as the energy is varied.
We also note that taking the inverse of $F_3$ is not problematic,
as it is invertible except at the energies of noninteracting states, and, as we have seen,
all energy levels are shifted from their noninteracting positions by $\cK_{2,L}$.

The original derivation of the RFT quantization condition required that, for each channel, the quantity
$\cK_2^{(j,s,\ell)}$ defined in \Cref{eq:K2cotd} should not have a pole within our kinematic range 
(i.e. the range of values of $\sigma_k$ that enter in the quantization condition).
Thus its inverse, given by the right-hand side of \Cref{eq:K2cotd}, should not vanish.
We have checked that, for our choice of phase shifts, this does not occur in any channel.
This means that we do not have to use the modified PV prescription introduced in ref.~\cite{\largera}.

A general feature of solving \Cref{eq:QC3a} is that there are double, or higher-order, zeros at the energies of the noninteracting levels. This was first noted in ref.~\cite{\dwave}, where it was explained how such unphysical zeros 
can be introduced by the truncation of $\Kdf$, and will, generically, be removed if higher order terms in $\Kdf$ were included.
The conclusion drawn is that one can simply ignore solutions at noninteracting energies, and we do so here.

\begin{figure}[htb!]
    \centering
    \includegraphics[width=0.85\linewidth]{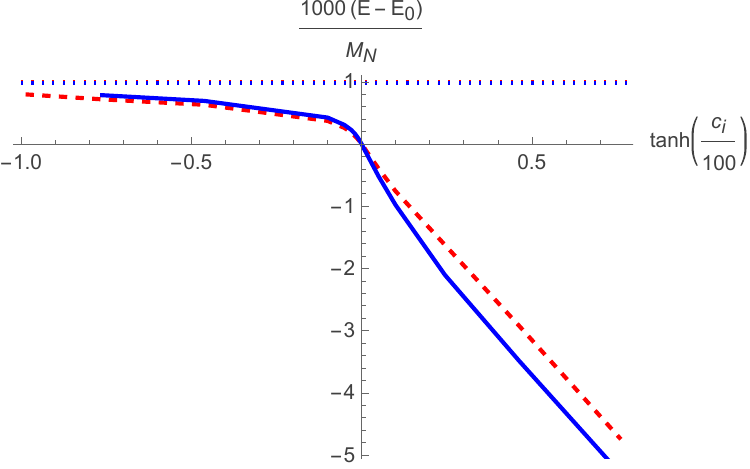}
    \caption{Shift in the energy of the lowest $G_1(1)$ level 
    when either the $\cK_A$ (solid blue curve) or $\cK_B$ term (dashed red curve) from $\Kdf$ is included. The horizontal axis is $\tanh(c_i/100)$, for $i=A,B$; the vertical axis is $10^3 (E-E_0)/M_N$, where $E_0=3.0430 M_N$ is the energy of the state when $\Kdf=0$.
    Parameters are $M_N L=20$ and $M_\pi/M_N=0.15$.
    The curves have been cut off at the maximal values of $|c_i|$ for which no unphysical solutions are present.
The blue and red dotted horizontal lines (barely distinguishable by eye) are explained in the text.
    }
    \label{fig:nP1G1KABband1}
\end{figure}

We have studied the impact of $\Kdf$ on several bands of levels for a number of irreps,
and have found the same qualitative behavior in all cases.
Thus we present results only for two representative examples.
The first is the simple case of the lowest band for $n_P^2=1$, 
which, as can be seen in \Cref{fig:nP1noKdf}, consists of a single level in the $G_1$ irrep.
We show in \Cref{fig:nP1G1KABband1}
how this level moves as the coefficients of $\cK_A$ and $\cK_B$ are individually varied.
We stress that the shifts, defined as $\delta E = E-E(\Kdf=0)$, 
have been multiplied by $1000$, so that they are roughly in units of MeV.
We consider coefficients with magnitudes up to $|c_i| = 1000$,
and plot the results against $\tanh(c_i/100)$ in order to fit them into a single panel.
This also  illustrates the logarithmic dependence on $|c_i|$ for large values of the magnitudes. 
We observe that both $\cK_A$ and $\cK_B$ lead to attractive interactions, with the energies lowered for $c_{A,B}>0$.
The scale of the overall variation is small, $\sim 6\;$MeV in total.
The dashed lines will be explained below.

\begin{figure}
    \centering
    \includegraphics[width=\textwidth]{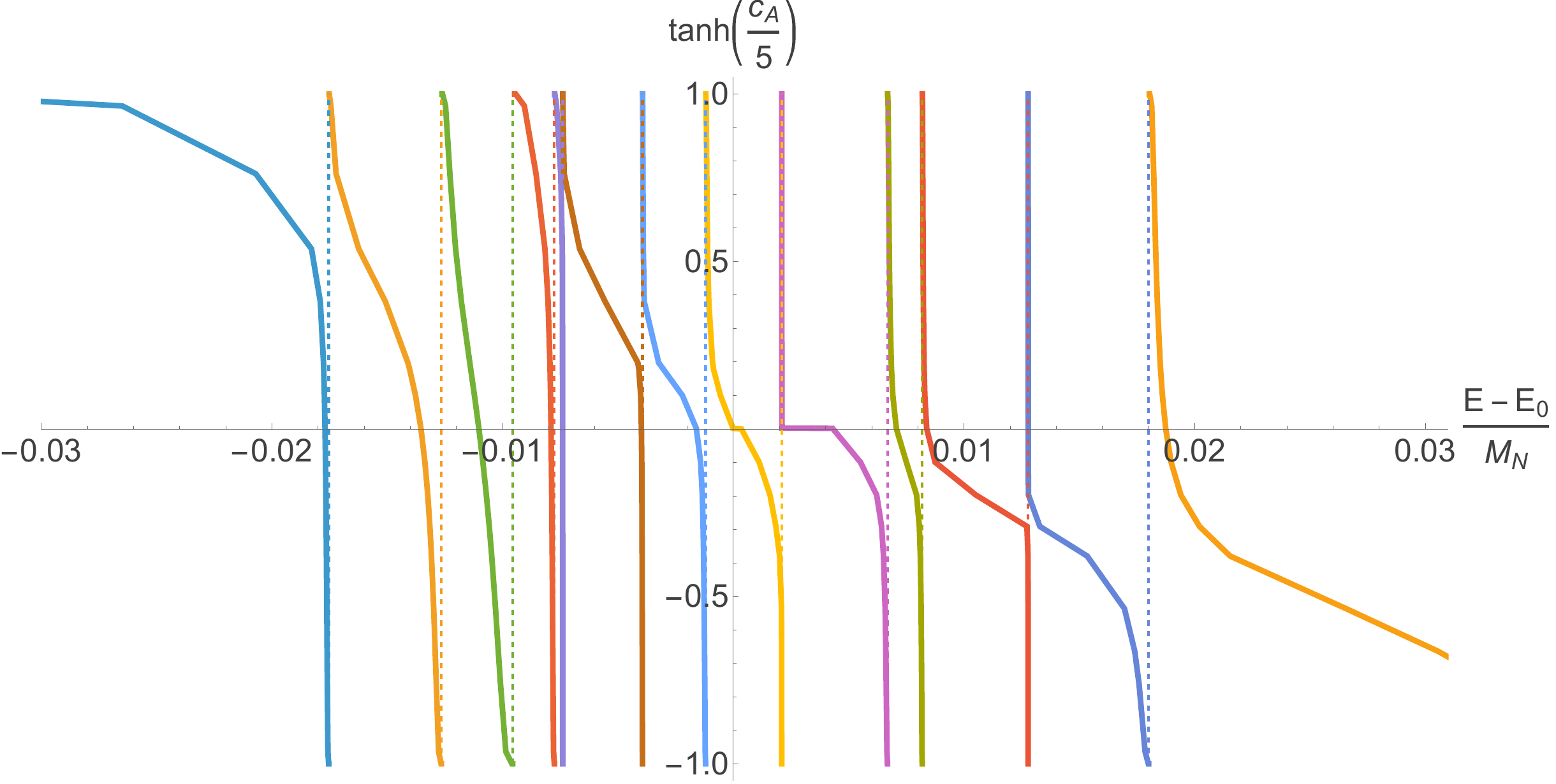}
    \caption{Dependence of the energies of the 13 $G_1(1)$ levels in the second band for $n_P^2=1$ upon $c_A$, the coefficient of $\cK_A$ (with $c_B=0$).
    The horizontal axis is $(E-E_0)/M_N$, where $E_0=3.1424 M_N$
    is the energy of the lowest noninteracting level in this band.
    The vertical axis is $\tanh(c_A/5)$. 
    Parameters are $M_N L=20$ and $M_\pi/M_N=0.15$.
    Vertical dashed lines are explained in the text.
    \label{fig:nP1G1KBband2}}
\end{figure}

The corresponding plot for the second $n_P^2=1$ band in the $G_1(1)$ irrep is shown in \Cref{fig:nP1G1KBband2}, except that here we only show the dependence on $c_A$. The dependence on $c_B$ is similar.
Note that, for the sake of clarity, the axes are interchanged relative to \Cref{fig:nP1G1KABband1}. The vertical axis is now $\tanh(c_A/5)$, where the denominator is chosen to make the variation with $c_A$ visible across all levels. We have used twenty values of $c_A$ in the range $-100$ to $100$;
jaggedness in the curves is due to linear interpolation between points.
The dashed (now vertical) lines will be explained shortly.

We see that, again, interaction is attractive. What is more notable, however,
is that the spectrum as $c_A\to \infty$ is the same as that at $c_A\to -\infty$,
with the levels have shifting up by one step, 
except for the levels at the ends of the band.
This pattern is shown by the vertical dashed lines, which are at the energies of the levels when
$c_A=-100$. This phenomenon implies that the largest shifts due to variation of $c_A$
are of order of the level splittings when $\Kdf=0$, which we have seen are as large as
$\sim 25\;$MeV. Said differently, the impact of nonzero $\Kdf$ can be of the same order of magnitude as the shifts due to two-particle interactions.

The agreement between energy levels for $c_i\to \pm\infty$ can be understood as follows.
Because both $\cK_A$ and $\cK_B$ lie in a subspace of the full matrix space
(since, as noted above, they have only two and six nonzero eigenvectors, respectively, while the matrix space has dimension 20 or more),
when their coefficients become large enough in magnitude, 
solutions to \Cref{eq:QC3a} that lie in the subspace orthogonal to $\Kdf$ will be independent of the sign of the coefficients.
Thus we conclude that the bulk of the solutions do lie in this subspace.

This raises the question of what happens to the lowest and highest levels in \Cref{fig:nP1G1KBband2}.
For example, the lowest (left-most) level has nothing to match onto within the band as $c_A\to \infty$.
A similar question applies for both ends of the curves in \Cref{fig:nP1G1KABband1}.
In the cases we have studied, the answer is that additional solutions appear at large values of the coefficients $c_{A,B}$, and these provide the matching solutions.
For example, in the lowest $n_P^2=1$ band, when either $c_A$ or $c_B$ exceed about $250$,
and new pair of solutions appears at higher energy, with the lower of the pair being
a physical crossing, while the upper one is unphysical. The position of the additional physical crossings for $c_{A,B}=1000$ is shown by the dotted horizontal lines in \Cref{fig:nP1G1KABband1}. As can be seen, these lines lie close to maximum values of
the solution curves as $c_{A,B} \to -\infty$.
Similarly, considering \Cref{fig:nP1G1KBband2},
we find for $c_A=-100$ (but not for $c_A=-50$) that there is an unphysical-physical pair of solutions to the left of the region shown in the plot, at $E-E_0= -0.08 M_N$ and $-0.072 M_N$, respectively.
The physical crossing plausibly matches onto the asymptote of the lowest level as $c_A\to\infty$. Indeed the energy of this level is rapidly decreasing as $c_A$ increases, reaching $-0.049$ at $c_A=100$.

This phenomenon of unphysical solutions appearing at large magnitudes of $\Kdf$
appears to be a fairly generic feature of our results.
For example, in the lower band shown in \Cref{fig:nP1G1KABband1},
they are present also for $c_{A,B} \gtrsim 250$ and $c_B\lesssim 1000$.
Our interpretation of at least some of these solutions is that they are
associated with unphysical choices for $\Kdf$.
We do not expect them all to disappear as $M_\pi L$ is increased because
some are needed to provide the ``missing matches'' between $c_i=\pm \infty$ solutions,  
as described in the examples above.
By contrast, for moderately large values of $|c_{A,B}|$, such unphysical solutions are absent and the resulting spectra appear trustworthy.

There is, however, another class of unphysical solutions that we have found.
These lie close to the noninteracting energies, and disappear as $M_\pi L$ is increased.
We thus assume that they are associated with exponentially-suppressed corrections to the formalism. 
They lead to the horizontal regions of the eighth (yellow) and ninth (purple) curves from the left in \Cref{fig:nP1G1KBband2}.
We describe these solutions in \Cref{app:unphys}.

\section{Conclusions and Outlook}
\label{sec:conc}

This work presents the detailed implementation of the three-neutron quantization condition, based on the formalism developed in ref.~\cite{Draper:2023xvu}. Specifically, if one were provided with the spectrum of a three-neutron system in a finite volume (via LQCD, for instance) the results described here could then be used to determine (constraints on) the two- and three-neutron K matrices. These could in turn be input into the integral equations described in ref.~\cite{\threeN}, the solution to which would yield the three neutron scattering amplitude.

The work presented here falls into three parts, described respectively in \Cref{sec:implement,sec:symm,sec:num}.
The first part, discussed in \Cref{sec:implement}, concerns the implementation of the four matrices that appear in the quantization condition.
The major new feature compared to previous implementations is the incorporation of the spin degrees of freedom, and, in particular, their transformations between different frames.
The most complicated matrix to implement is $K_{\mathrm{df},3}$, which here is parameterized by the two leading order operators an expansion around threshold. The coefficients of these two operators would be the free three-particle parameters in a fit to the finite-volume spectrum. 

The second part of the work, presented in \Cref{sec:symm}, decomposes solutions of the quantization condition into irreps of the appropriate finite little groups, 
which here are fermionic. 
This is a standard step in implementations of quantization conditions,
but is particularly important in systems of heavier hadrons,
where the number of levels that lie below the inelastic threshold is large.
By spreading levels between different irreps, the level-density is significantly reduced.

This brings us back to the question raised in the introduction: What is the precision required in the determination of energy levels so as to provide detailed information on two- and three-neutron interactions?
This is addressed in the final part of this work, described in \Cref{sec:num}.
Here, for choices of particle masses and box sizes that are likely close to those that will be used in the first LQCD studies,
we determine the spectrum by solving the quantization condition for realistic choices of the two-neutron K matrices, and for a wide range of the parameters entering $\Kdf$.
We find that the two-neutron interactions lead to level spacings in the $5-25\;$MeV range,
and that the three particle interactions can lead to splittings almost as large,
albeit for large values of the corresponding couplings.
If one is able to reach a precision of $\lesssim 5\;$MeV, which is obviously a major challenge, then, using the large number of levels available in different frames and irreps, it should be possible to disentangle interactions in the different two-particle channels, and also constrain the components of $\Kdf$.
In this regard, it will be important to combine spectra from two- and three-neutron systems in order to distinguish the effects of two- and three-particle interactions.
We have also found that there is no particular frame/irrep combination that picks out any two- or three-particle component---a global fit will be required.

Our numerical investigations also found several examples of unphysical solutions to the quantization condition. 
Such solutions have been observed and studied previously in simpler three-particle systems~\cite{\BHSnum,\dwave,\largera}. 
We have likely amplified their presence by working in small boxes with $M_\pi L=3$.
Indeed, in most cases they disappear quickly as $M_\pi L$ is increased, and are likely present due to the fact that the quantization conditions are derived ignoring terms that are exponentially suppressed in $M_\pi L$.
However, other unphysical solutions, which appear when $|\Kdf|$ is large, 
may be due to the use of unphysical choices of the interactions.
In any case, the lesson is that any future application of the three-neutron quantization condition should ensure that these solutions are absent for the parameters being used,
or find a clear rationale for dropping them from the predicted spectrum.

To complete the finite-volume formalism for three neutrons, the most pressing next step is to implement the integral equations that connect the K matrices to the scattering amplitude.
With that in hand, 
the tools will be available to use LQCD to predict the three-neutron scattering amplitude,
which can then be compared to the form predicted by chiral EFT,
allowing the determination of the EFT coefficients that describe three-neutron interactions. 
An alternative approach is to use chiral EFT to predict $\Kdf$, and thus relate the EFT coefficients to those that appear in the threshold expansion of $\Kdf$,
namely $c_A$ and $c_B$. Such a calculation is underway.
In that regard, it may prove necessary to extend the threshold expansion to higher terms.

\section*{Acknowledgements}

We thank Max Hansen and Fernando Romero-López 
for useful comments
and discussions. This work is supported in part by the U.S. Department of Energy grant No. DE-SC0011637. This work contributes to the goals of the USDOE ExoHad Topical Collaboration,
contract DE-SC0023598.

\appendix

\section{Determining the matrix $I (\bm k)$}
\label{app:Ibar}

In this appendix, we determine the matrices $I^Q(\bm k)$ and $I(\bm k)$, given in \Cref{eq:IQdef,eq:Idef},  respectively.  We first focus on $I$, since the conversion to $I^Q$ is straightforward.
We repeat its definition for clarity,
\begin{align}
I(\bm k)_{\ell' m' m'_a m'_b; \ell m m^*_a m^*_b} &=
\int_{\Omega_{\hat a^*}} Y^*_{\ell' m'}( \hat a^*)
\cD^{(1/2)}(R_{k,a}^{-1})_{m'_a m^*_a}
\cD^{(1/2)}(R_{k,b}^{-1})_{m'_b m^*_b} 
Y_{\ell m}(\hat a^*)  \,,
\label{eq:Idefb}
\end{align}
where the rotations $R_{k,a}$ and $R_{k,b}$ are given by \Cref{eq:Rpdef}.
Here we rewrite these rotations in terms of $\bm a^*= - \bm b^*$ rather an $\bm a$.
Introducing the shorthands $\bm P_k\equiv \bm P-\bm k$
and $\gamma_k\equiv \gamma_{P-k}$,
the axis $\hat n_a$ and angle $\theta_a$ of the rotation $R_{k,a}$ are
\begin{align}
\hat n_a &= - \hat v\,,\ \ \bm v =\bm P_k \times \bm a^*\,,
\ \
c_a \equiv \cos\theta_a = 
\frac{(1+\gamma_k+\gamma_{a^*}+\gamma_a)^2}
{(1+\gamma_k)(1+\gamma_{a^*})(1+\gamma_a)} - 1\,,
\end{align}
while those for $R_{k,b}$ are 
\begin{align}
\hat n_b &= - \hat n_a\,,\ \
c_b \equiv \cos\theta_b = 
\frac{(1+\gamma_k+\gamma_{a^*}+\gamma_b)^2}
{(1+\gamma_k)(1+\gamma_{a^*})(1+\gamma_b)} - 1\,.
\end{align}
We also need the results
\begin{align}
\gamma_a &= \gamma_k \gamma_{a^*} + \frac{\bm P_k \cdot \bm a^*}{M_N^2}\,, \ \
\gamma_b = \gamma_k \gamma_{a^*} - \frac{\bm P_k \cdot \bm a^*}{M_N^2}\,.
\end{align}
The required Wigner D matrices are then given by
\begin{align}
\cD^{(1/2)}(R_{k,a}^{-1}) &= \bar c_a + i \bar s_a \hat n_a \cdot \boldsymbol \sigma\,,
\quad
\cD^{(1/2)}(R_{k,b}^{-1}) =\bar c_b - i \bar s_b \hat n_a \cdot \boldsymbol \sigma\,,
\end{align}
where 
\begin{equation}
\bar c_a = \cos(\theta_a/2)\,,\qquad \bar s_a = \sin(\theta_a/2)\,,\ \  \text{etc.}
\end{equation}

The expression for the product of Wigner matrices simplifies if we evaluate the integral
with the orientation of $\bm a^*$ defined relative to $\bm P_k$. 
However, the spherical harmonics are determined relative to the fixed lab-frame $z$ axis. 
We can rectify this mismatch using
\begin{equation}
Y_{\ell m}(\hat a^*) = Y_{\ell m'}(R_k \hat a^*) \cD^{(\ell)}_{m' m}(R_k)\,,
\end{equation}
and  choosing the rotation $R_k$ such that $R_k \hat{P_k} = \hat z$.\footnote{%
If $\bm P_k$ has polar and azimuthal angles $\theta,\phi$, then, in terms of Euler angles,
\begin{equation}
R_k = R(\alpha, \beta=-\theta, \gamma=-\phi)\,,
\end{equation}
with $\alpha$ arbitrary. The $\alpha$ dependence cancels in $I$, so we set $\alpha=0$.
}
Then $Y_{\ell m'}(R_k \hat a^*)$ are the
spherical harmonics in which the $z$ axis is aligned with $\bm P_k$.
A similar argument holds for the $\cD^{(1/2)}$ matrices, which must be conjugated by $\cD^{(1/2)}(R_k)$.
In this way we find (dropping the argument $\bm k$ for brevity)
\begin{multline}
I_{\ell' m' m'_a m'_b; \ell m m^*_a m^*_b} =   \cD^{(\ell')}_{m' m'''}(R_k^{-1})
 \cD^{(1/2)}_{m'_a m''_a}(R_k^{-1}) \cD^{(1/2)}_{m'_b m''_b}(R_k^{-1})
I'_{\ell' m''' m''_a m''_b; \ell m'' m'^*_a m'^*_b} 
\\
\times
\cD^{(\ell)}_{m'' m}(R_k) \cD^{(1/2)}_{m'^*_a m^*_a}(R_k) \cD^{(1/2)}_{m'^*_b m^*_b}(R_k)
\,,
\label{eq:Ires}
\end{multline}
where 
\begin{multline}
I'_{\ell' m' m'_a m'_b; \ell m m^*_a m^*_b} =  
\\
  \int_{\Omega_{\hat a^*}} Y^*_{\ell' m'}(R_k \hat a^*)
(\bar c_a \!+\! i \bar s_a (R_k \hat n_a)\cdot\boldsymbol\sigma)_{m'_a m^*_a}
(\bar c_b \!-\! i \bar s_b (R_k \hat n_a)\cdot\boldsymbol\sigma)_{m'_b m^*_b}
Y_{\ell m}(R_k\hat a^*)\,.
\label{eq:Iprime}
\end{multline}

We now change the integration variable to $R_k\hat a^*$,
which amounts to using polar and azimuthal angles, $\theta,\ \phi$, relative to $\bm P_k$, 
which lies along the $z$ axis, so that $R_k \hat n_a  = (s_\phi, - c_\phi, 0)$.
Then we find
\begin{equation}
R_k \hat n_a \cdot \boldsymbol \sigma = \frac1{2i} \left( e^{i\phi} \sigma_- - e^{-i\phi} \sigma_+\right)\,,\quad
\sigma_\pm = \sigma_x \pm i\sigma_y\,,
\end{equation}
and (using $c_\theta \equiv \cos\theta$ and $s_\theta=\sin\theta$)
\begin{align}
\bar c_a &= \frac{(A + B + d_k  c_\theta)}{\sqrt{2(A+B)}} \frac1{\sqrt{A+d_k c_\theta}}\,,
\quad
\bar c_b = \frac{(A + B - d_k  c_\theta)}{\sqrt{2(A+B)}} \frac1{\sqrt{A-d_k c_\theta}}\,,
\\
\bar s_a &= \frac{d_k  s_\theta}{\sqrt{2(A+B)}}  \frac1{\sqrt{A+d_k c_\theta}}\,,
\quad
\bar s_b = \frac{d_k  s_\theta}{\sqrt{2(A+B)}}  \frac1{\sqrt{A-d_k c_\theta}}\,,
\end{align}
where
\begin{equation}
A = 1 + \gamma_k \gamma_{a^*}\,, \quad B= \gamma_k+\gamma_{a^*}\,, \quad
d_k = \frac{P_k a^*}{\sqrt{\sigma_k} M} = \sqrt{(\gamma_k^2-1)(\gamma_{a^*}^2-1)}\,.
\end{equation}
Note that, while $\bar c_b(d_k) = \bar c_a(-d_k)$, we have $\bar s_b(d_k)= - \bar s_a(-d_k)$, with the sign flip
needed to ensure that $\bar s_b \ge 0$, which follows since $s_b\ge 0$ by definition.

We now restrict to $\ell_{\rm max}=1$, and revert, for now, to using standard complex spherical harmonics.
The resulting integrals over $\phi$ are elementary, leaving nontrivial integrals over $\theta$,
results for which are collected in \Cref{tab:Iintegrals}. In terms of these, the matrix $I'$ in \Cref{eq:Iprime} is given by
\begin{multline}
I'_{\ell' m' m'_a m'_b; \ell m m^*_a m^*_b} =
\\
\begin{pmatrix} \left[I_1 \Sigma_{1,1} + \tfrac14 I_7 \Sigma_{-,+}^{(+)}\right] &
-\sqrt{\tfrac38} I_3 \Sigma_{1,+}^{(-)} & 
0 &
-\sqrt{\tfrac38} I_3 \Sigma_{1,-}^{(-)}
\\
\sqrt{\tfrac38} I_3 \Sigma_{1,-}^{(-)} & 
 \left[ 
 \substack{
 \tfrac32 (I_1\!-\!I_2) \Sigma_{1,1} \\ + \tfrac38 (I_7\!-\!I_8) \Sigma_{-,+}^{(+)} }
  \right]
 &
\tfrac3{\sqrt8} I_4 \Sigma_{1,-}^{(+)} &
 \tfrac38 (I_7\!-\!I_8) \Sigma_{-,-}
\\ 
0 & 
-\tfrac3{\sqrt8} I_4 \Sigma_{1,+}^{(+)} &
\left[ \substack{3 I_2 \Sigma_{1,1} \\ + \tfrac34 I_8 \Sigma_{-,+}^{(+)} }\right] &
-\tfrac3{\sqrt8} I_4 \Sigma_{1,-}^{(+)}
\\
 \sqrt{\tfrac38} I_3 \Sigma_{1,+}^{(-)} & 
 \tfrac38 (I_7-I_8) \Sigma_{+,+} &
 \tfrac3{\sqrt8} {I_4} \Sigma_{1,+}^{(+)} &
\left[ \substack{\tfrac32 (I_1\!-\!I_2) \Sigma_{1,1}\\ + \tfrac38 (I_7\!-\!I_8) \Sigma_{-,+}^{(+)}}\right]
\end{pmatrix}\,.
\label{eq:Ipres}
\end{multline}
Here the outer $4\times4$ matrix describes the $\ell, m$ structure, using the index order
\begin{equation}
(\ell, m) = \left\{ (0,0),\ (1,1),\ (1,0),\ (1,-1)\right\}\,,
\label{eq:ellmorder}
\end{equation}
while the spin index dependence is contained with the $\Sigma$s,
which are defined as
\begin{multline}
\Sigma_{1,1} = \bm 1\otimes \bm 1\,,\ \
\Sigma_{+,+} = \sigma_+ \otimes \sigma_+\,,\ \
\Sigma_{-,-} = \sigma_- \otimes \sigma_-\,,\ \
\Sigma_{-,+}^{(+)} = \sigma_-\otimes \sigma_+ + \sigma_+\otimes \sigma_-\,,
\\
\Sigma_{1,\pm}^{(+)} = \bm 1 \otimes \sigma_\pm + \sigma_\pm\otimes \bm 1 \,,\ \
\Sigma_{1,\pm}^{(-)} = \bm 1 \otimes \sigma_\pm - \sigma_\pm\otimes \bm 1\,,
\end{multline}
with the explicit form of the indices being exemplified by
\begin{equation}
\sigma_- \otimes \sigma_+ \longrightarrow (\sigma_-)_{m'_a m^*_a} (\sigma_+)_{m'_b m^*_b}\,.
\end{equation}

\begin{table}[h]
\centering
\renewcommand{\arraystretch}{2.7}
\begin{tabular}{|l|}
\hline
$I_i =  \int_{-1}^{+1}\frac{dc_\theta}{2} \left\{ 
\bar c_a \bar c_b,\ \bar c_a \bar c_b c_\theta^2,\
\bar c_a \bar s_b s_\theta,\  \bar c_a \bar s_b s_\theta c_\theta,\ 
\bar s_a \bar c_b s_\theta,\  \bar s_a \bar c_b s_\theta c_\theta,\ 
\bar s_a \bar s_b,\ \bar s_a \bar s_b c_\theta^2
\right\}$
\\
\hline \hline
$
I_1 = \displaystyle 
 \frac{  \sqrt{\widetilde A^2 - 1} + (\widetilde A^2+4 \widetilde A \widetilde B+2 \widetilde B^2) 
 \sin ^{-1}\!(\widetilde A^{-1})}{4 (\widetilde A+\widetilde B)}
$ \\
\hline
$
I_2 = \displaystyle 
\frac{ \sqrt{\widetilde A^2-1} \left(-\widetilde A^2-8 \widetilde A \widetilde B-4 \widetilde B^2+2 \right)
+\widetilde A^2  \left(\widetilde A^2+8 \widetilde A \widetilde B+4 \widetilde B^2\right) \sin ^{-1}\!\widetilde A^{-1}}
  {16 (\widetilde A+\widetilde B)}
$
\\ \hline
$
I_3 = I_5 = \displaystyle 
\frac14 \left[\sqrt{\widetilde A^2 - 1} + (2 - \widetilde A^2 ) \sin^{-1}\!\widetilde A^{-1}\right]
$
\\
\hline
$
I_4 = - I_6 = \displaystyle 
\frac{(3 \widetilde A^2 - 2 ) \sqrt{\widetilde A^2 - 1} + (-3 \widetilde A^4 + 4 \widetilde A^2 ) \sin^{-1}\! \widetilde A^{-1}}
{16 (\widetilde A + \widetilde B) }
$ 
\\ \hline
$
I_7 = \displaystyle 
\frac{\sqrt{\widetilde A^2 - 1}+ (2 - \widetilde A^2) \sin^{-1}\! \widetilde A^{-1}}{4 (\widetilde A + \widetilde B)}
$
\\ \hline 
$
I_8 = \displaystyle 
\frac{ (3 \widetilde A^2-2) \sqrt{\widetilde A^2-1}+(4 \widetilde A^2 - 3 \widetilde A^4) \sin ^{-1}\!\widetilde A^{-1}}
{16 (\widetilde A+\widetilde B)}
$
\\
\hline \hline
\shortstack{
\vspace{4pt}\\
$ 
\widetilde A = \displaystyle 
\frac{A}{d_k} = \frac{1 + \gamma_k \gamma_{a^*}}{\sqrt{(\gamma_k^2-1)(\gamma_{a^*}^2-1)} }\,,
\quad
\widetilde B = \frac{B}{d_k} = \frac{\gamma_k + \gamma_{a^*}}{ \sqrt{(\gamma_k^2-1)(\gamma_{a^*}^2-1)}}
$
\\
\vspace{-4pt}
}
\\
\hline
\end{tabular}
\renewcommand{\arraystretch}{1.0}
\caption{Analytic expressions for the integrals appearing in the evaluation of $I'$, \Cref{eq:Iprime}.
\label{tab:Iintegrals}}
\end{table}

The behavior of these functions as $a^*\to 0$ (so that $d_k \propto a^*$, $\gamma_{a^*}\to 1$, $B\to A$) is
\begin{multline}
I_1=1+\cO(a^{*2}),\   I_2=\tfrac13 + \cO(a^{*2}),\ I_3 = \tfrac{1}{3A} d_k + \cO(a^{*3}), 
\\
I_4 =\tfrac{1}{30 A^2}  d_k^2 + \cO(a^{*4}),\
I_7 =\tfrac1{6A^2} d_k^2 + \cO(a^{*4}),\ 
I_8 = \tfrac1{30A^2} d_k^2 + \cO(a^{*4})\,.
\label{eq:threshI}
\end{multline}
These are consistent with the general result that $I'\propto (a^*)^{|\ell' - \ell|}$,
which one can show from the definition \Cref{eq:Iprime}.

We now convert the result for $I$ to that for $I^Q$. As seen from \Cref{eq:IQdef}, this is
obtained by conjugating $I(\bm k)$ with appropriate powers of $q_k^*=a^*$.
 Since these factors commute with the Wigner matrices in \Cref{eq:Ires},
the $Q$-form of $I'$ is given by
\begin{equation}
I'^Q(\bm k)_{\ell' m' m'_{sa} m'_{sb}; \ell m m^*_{sa} m^*_{sb}} =
(a^*)^{\ell'}
I'(\bm k)_{\ell' m' m'_{sa} m'_{sb}; \ell m m^*_{sa} m^*_{sb}} (a^*)^{-\ell}\,,
\label{eq:IpQdefa}
\end{equation}
and is related to $I^Q$ by \Cref{eq:IQresa}.
The conjugation only impacts terms offdiagonal in $\ell$, and one obtains
\begin{multline}
I'^!_{\ell' m' m'_a m'_b; \ell m m^*_a m^*_b} =
\\
\begin{pmatrix} \left[I_1 \Sigma_{1,1} + \tfrac14 I_7 \Sigma_{-,+}^{(+)}\right] &
-\sqrt{\tfrac38} \frac{I_3}{a^*} \Sigma_{1,+}^{(-)} & 
0 & 
-\sqrt{\tfrac38} \frac{I_3}{a^*} \Sigma_{1,-}^{(-)}
\\
 \sqrt{\tfrac38} a^* I_3 \Sigma_{1,-}^{(-)} & 
 \left[ 
 \substack{
 \tfrac32 (I_1\!-\!I_2) \Sigma_{1,1} \\ + \tfrac38 (I_7\!-\!I_8) \Sigma_{-,+}^{(+)} }
  \right]
 &
\tfrac3{\sqrt8} I_4 \Sigma_{1,-}^{(+)} & 
\tfrac38 (I_7\!-\!I_8) \Sigma_{-,-}
\\ 
0 & 
-\tfrac3{\sqrt8} I_4 \Sigma_{1,+}^{(+)} &
\left[ \substack{3 I_2 \Sigma_{1,1} \\ + \tfrac34 I_8 \Sigma_{-,+}^{(+)} }\right] &
-\tfrac3{\sqrt8} I_4 \Sigma_{1,-}^{(+)}
\\
\sqrt{\tfrac38} a^*I_3 \Sigma_{1,+}^{(-)} & 
\tfrac38 (I_7-I_8) \Sigma_{+,+} &
\tfrac3{\sqrt8} {I_4} \Sigma_{1,+}^{(+)} &
\left[ \substack{\tfrac32 (I_1\!-\!I_2) \Sigma_{1,1}\\ + \tfrac38 (I_7\!-\!I_8) \Sigma_{-,+}^{(+)}}\right]
\end{pmatrix}\,.
\label{eq:IpQres}
\end{multline}
Since $I_3$ is an odd function of $a^*$ [see \Cref{eq:threshI}], 
the quantities $I_3/a^*$ and $a^* I_3$ are both even, nonsingular functions of $a^{*2}$.
The same holds for all other entries in the matrix.
In particular, this implies that the functions appearing in all entries remain real when $a^{*2} < 0$, 
i.e. when the pair goes below threshold.

Since we use real spherical harmonics when implementing the quantization condition, we need to convert
the above-described expression for $I(\bm k)$ into the real basis. This is achieved by the replacement
\begin{equation}
\cD^{(\ell)}(R_k) \longrightarrow \cD^{(\ell)}(R_k) T^{c\to r}\,,
\end{equation}
where, using the ordering given in \Cref{eq:ellmorder},
\begin{equation}
T^{c\to r} = \frac1{\sqrt2} \begin{pmatrix}
\sqrt2 & 0 & 0 & 0
\\ 
0 & -1 & 0 &i
\\
0 & 0 & \sqrt2 & 0 
\\
0 & 1 & 0 & i
\end{pmatrix}
\,.
\label{eq:ctor}
\end{equation}

\section{Form of $\cK_B$ in the lab frame}
\label{app:KB}

In this appendix we sketch the determination of the form of the
matrix $[\cK_{\rm df,3}^{Q, \rm lab}]$ arising from the $\cK_B$ term, \Cref{eq:KB}.
We first determine the result without $Q$ factors, and comment at the end about the (simple) changes
needed to include these factors.

Carrying out the antisymmetrization explicitly, and converting to spin indices, we have
\begin{multline}
\frac12 \left[\cK_B\right]_{m_p m_{a'} m_{b'}; m_k m_a m_b } =
\frac12 \overline \cA [\bm p\cdot \bm k \ \chi_{p}^\dagger \chi_k\ \chi_{a'}^\dagger \chi_a\ \chi_{b'}^\dagger \chi_b 
]_{m_p m_{a'} m_{b'}; m_k m_a m_b }  =
\\
\delta_{m_p m_k} \delta_{m_{a'} m_a} \delta_{m_{b'} m_b} 
(\bm p\cdot \bm k + \bm a' \cdot \bm a + \bm b' \cdot \bm b)
-
\delta_{m_p m_k} \delta_{m_{a'} m_b} \delta_{m_{b'} m_a} 
(\bm p \cdot \bm k + \bm a' \cdot \bm b + \bm b' \cdot \bm a) +
\\
\delta_{m_p m_a} \delta_{m_{a'} m_b} \delta_{m_{b'} m_k} 
(\bm p\cdot \bm a + \bm a' \cdot \bm b + \bm b' \cdot \bm k)
-
\delta_{m_p m_a} \delta_{m_{a'} m_k} \delta_{m_{b'} m_b} 
(\bm p\cdot \bm a + \bm a' \cdot \bm k + \bm b' \cdot \bm b) +
\\
\delta_{m_p m_b} \delta_{m_{a'} m_k} \delta_{m_{b'} m_a} 
(\bm p\cdot \bm b + \bm a' \cdot \bm k + \bm b' \cdot \bm a)
-
\delta_{m_p m_b} \delta_{m_{a'} m_a} \delta_{m_{b'} m_k} 
(\bm p\cdot \bm b + \bm a' \cdot \bm a + \bm b' \cdot \bm k)\,.
\end{multline}
To project onto $\ell,m$ indices, we need to rexpress lab-frame pair momenta in terms of their boosted versions
$\bm a^* = - \bm b^*$ and $\bm a'^* = - \bm b'^*$.
This is done using 
\begin{align}
\bm a &= \bm a^* + (\gamma_{P-k}-1) (\hat \beta_{P-k}\cdot \bm a^*) \hat \beta_{P-k} + \omega_{a^*} \gamma_{P-k}\boldsymbol \beta_{P-k}\,,
\label{eq:boosta}
\\
\bm b &= - \bm a^* - (\gamma_{P-k}-1) (\hat \beta_{P-k}\cdot \bm a^*) \hat \beta_{P-k} + \omega_{a^*} \gamma_{P-k} \boldsymbol \beta_{P-k}\,,
\label{eq:boostb}
\\
\bm a' &=  \bm a'^* + (\gamma_{P-p}-1) (\hat \beta_{P-p}\cdot \bm a'^*) \hat \beta_{P-p} + \omega_{a'^*}\gamma_{P-p} \boldsymbol \beta_{P-p}\,,
\label{eq:boostap}
\\
\bm b' &=  -\bm a'^* - (\gamma_{P-p}-1) (\hat \beta_{P-p}\cdot \bm a'^*) \hat \beta_{P-p} + \omega_{a'^*} \gamma_{P-p}\boldsymbol \beta_{P-p}\,.
\label{eq:boostbp}
\end{align}
where the notation is defined in \Cref{eq:betas}.

Thus the inner products we need are
\begin{align}
\bm a'\cdot \bm a &=  a'^*_i T_{ij}  a^*_j + \bm a'^* \cdot \bm V + \bm V' \cdot \bm a^* +  C_0\,,
\\
\bm b'\cdot \bm b &=  a'^*_i T_{ij} a^*_j - \bm a'^* \cdot \bm V - \bm V' \cdot \bm a^* +  C_0\,,
\\
\bm a'\cdot \bm b &=  - a'^*_i T_{ij} a^*_j + \bm a'^* \cdot \bm V - \bm V' \cdot \bm a^* +  C_0\,,
\\
\bm b'\cdot \bm a &=  -a'^*_i T_{ij} a^*_j - \bm a'^* \cdot \bm V + \bm V' \cdot \bm a^* +  C_0\,,
\end{align}
where
\begin{align}
\begin{split}
T_{ij} &= \delta_{ij} + \hat\beta_{p;i} (\gamma_{P-p}-1) \hat \beta_{p;j} + \hat\beta_{k;i} (\gamma_{P-k}-1) \hat \beta_{k;j}
\\ &\ \ 
+\hat\beta_{p;i} (\gamma_{P-p}-1)(\hat\beta_{P-p}\cdot \hat\beta_{P-k})(\gamma_{P-k}-1) \hat \beta_{k;j}\,,
\end{split}
\\
\bm V &= \omega_{a^*} \gamma_{P-k}
\left[ \boldsymbol \beta_{P-k} + (\gamma_{P-p}-1) \hat \beta_{P-p} \, \hat\beta_{P-p} \cdot \boldsymbol \beta_{P-k}\right]\,,
\\
\bm V' &= \omega_{a'^*} \gamma_{P-p}
\left[ \boldsymbol \beta_{P-p} + (\gamma_{P-k}-1) \hat \beta_{P-k} \, \hat\beta_{P-k} \cdot \boldsymbol \beta_{P-p}\right]\,,
\\
C_0 &=  \omega_{a'^*} \omega_{a^*} \gamma_{P-p} \gamma_{P-k} \boldsymbol \beta_{P-p}\cdot \boldsymbol \beta_{P-k}\,,
\end{align}
as well as
\begin{align}
\bm p \cdot \bm a &= \bm V'_p \cdot \bm a^* + C_p\,,
\qquad
\bm p \cdot \bm b = -\bm V'_p \cdot \bm a^* + C_p\,,
\\
\bm a' \cdot \bm k &= \bm a'^* \cdot \bm V_k + C_k\,,
\qquad
\bm b' \cdot \bm k = -\bm a'^* \cdot \bm V_k + C_k\,,
\end{align}
where
\begin{align}
\bm V'_p &= \bm p + \hat\beta_{P-k} (\gamma_{P-k}-1) \bm p \cdot \hat\beta_{P-k}\,,
\qquad
\bm V_k = \bm k  + \hat\beta_{P-p} (\gamma_{P-p}-1) \bm k \cdot \hat\beta_{P-p}\,,
\\
C_p &= \omega_{a^*} \gamma_{P-k} \bm p \cdot \boldsymbol \beta_{P-k}\,,
\qquad
C_k = \omega_{a'^*} \gamma_{P-p} \bm k \cdot \boldsymbol \beta_{P-p}\,.
\end{align}

Combining the above results, we can project onto the $\ell,m$ indices as in  \Cref{eq:Kdflab},
using the projection operators of \Cref{eq:projectellm}.
Since the manipulations above result in terms that are at most linear in $\bm a'^*$ and $\bm a^*$, 
only $\ell',\ell=0,1$ contribution will result from the projections.
The nontrivial projections we need are
\begin{align}
[ \cP_{1 m}^{\hat a^*}]^\dagger \circ a^*_j &= \sqrt{\frac13} a^* [V^\dagger]_{jm} \,,
\quad
[\cP_{1 m'}^{\hat a'^*}] \circ a'^*_j =  \sqrt{\frac13} a'^* V_{m' j}\,,
\label{eq:project1m}
\end{align}
where, for real spherical harmonics,
\begin{align}
V &= \begin{pmatrix}
1 & 0 & 0 \\
0 & 0 & 1 \\
0 & 1 & 0
\end{pmatrix} = V^\dagger\,.
\end{align}
with the $m$ values are ordered $\{1, 0, -1\}$.

The final step is to convert to the $Q$ form, which we recall is $[Q^{-1} [\cK_{\rm df,3}^{\sf lab}] Q^{-1}] $.
This is simply achieved by dropping the factors of $a^*$ and $a'^*$ on the right-hand sides of
\Cref{eq:project1m} when implementing the projections.

The above-described steps lead to long algebraic expressions that have been implemented in two independent  {\sf Mathematica } codes and cross checked.

\section{Form of $\cK'_A$ in the lab frame}
\label{app:KA}

In this appendix we extend the results of the previous appendix to the $\cK'_A$ term, \Cref{eq:KAp}.

We can piggy back on the work for $\cK_B$ by recognizing that
\begin{equation}
i\boldsymbol \sigma\cdot \bm p \times \bm k = p_i r_{jl} k_l\,,
\quad
r_{j l} = i \epsilon_{jln} \sigma_n\,,
\end{equation}
where $r$ is an hermitian tensor that replaces the $\delta_{jl}$ appearing in $\cK_B$.
Thus we introduce the notation for a new product of vectors,
\begin{equation}
[\bm p \bm k] \equiv p_j r_{jl} k_l\,,
\end{equation}
which is linear in both entries, and has an implicit $2\times 2$ matrix structure.
We stress that it is not an inner product, but all we need is linearity in the following.
We refer to it as the $r$-product.

Using this, we can write out the $\cK'_A$ term explicity as
\begin{multline}
\frac12  \left[\cK'_A\right]_{m_p m_{a'} m_{b'}; m_k m_a m_b } =
[\bm p \bm k ]_{m_p m_k}  [\delta_{m_{a'} m_a} \delta_{m_{b'} m_b}-\delta_{m_{a'} m_b}\delta_{m_{b'} m_a}]
\\
+ 
[\bm p \bm a]_{m_p m_a}  [\delta_{m_{a'} m_b} \delta_{m_{b'} m_k}-\delta_{m_{a'} m_k}\delta_{m_{b'} m_b}]
 +
 [\bm p \bm b ]_{m_p m_b}  [\delta_{m_{a'} m_k} \delta_{m_{b'} m_a}-\delta_{m_{a'} m_a}\delta_{m_{b'} m_k}]
\\
+ 
[\bm a' \bm k]_{m_{a'} m_k}  [\delta_{m_{b'} m_a} \delta_{m_p m_b}-\delta_{m_{b'} m_b}\delta_{m_p m_a}]
 +
 [\bm a' \bm a ]_{m_{a'} m_a}  [\delta_{m_{b'} m_b} \delta_{m_p m_k}-\delta_{m_{b'} m_k}\delta_{m_p m_b}]
\\
+
[\bm a' \bm b]_{m_{a'} m_b}  [\delta_{m_{b'} m_k} \delta_{m_p m_a}-\delta_{m_{b'} m_a}\delta_{m_p m_k}]
 +
 [\bm b' \bm k ]_{m_{b'} m_k}  [\delta_{m_p m_a} \delta_{m_{a'} m_b}-\delta_{m_p m_b}\delta_{m_{a'} m_a}]
\\
+
[\bm b' \bm a]_{m_{b'} m_a}  [\delta_{m_p m_b} \delta_{m_{a'} m_k}-\delta_{m_p m_k}\delta_{m_{a'} m_b}]
+
 [\bm b' \bm b ]_{m_{b'} m_b}  [\delta_{m_p m_k} \delta_{m_{a'} m_a}-\delta_{m_p m_a}\delta_{m_{a'} m_k}]\,.
\end{multline}

Using the boosts given in \Cref{eq:boosta,eq:boostb,eq:boostap,eq:boostbp} we can write the $r$ products
quadratic in pair momenta as
\begin{align}
[\bm a' \bm a] = a'^*_i [T_r]_{ij} a^*_j + \bm a'^* \cdot \bm V_r + \bm V'_r \cdot \bm a^* + C_{0,r}\,,
\\
[\bm a' \bm b] = -a'^*_i [T_r]_{ij} a^*_j + \bm a'^* \cdot \bm V_r - \bm V'_r \cdot \bm a^* + C_{0,r}\,,
\\
[\bm b' \bm a] = -a'^*_i [T_r]_{ij} a^*_j - \bm a'^* \cdot \bm V_r + \bm V'_r \cdot \bm a^* + C_{0,r}\,,
\\
[\bm b' \bm b] = a'^*_i [T_r]_{ij} a^*_j - \bm a'^* \cdot \bm V_r - \bm V'_r \cdot \bm a^* + C_{0,r}\,,
\end{align}
where the generalized tensors of $2\times 2$ matrices are
\begin{align}
\begin{split}
[T_r]_{ij} &= r_{ij} + \hat\beta_{p;i} (\gamma_{P-p}-1) [\hat \beta_{p}\cdot r]_j 
+ [r\cdot \hat\beta_{k}]_i (\gamma_{P-k}-1) \hat \beta_{k;j}
\\
&\ \ \ +\hat\beta_{p;i} (\gamma_{P-p}-1)[\hat\beta_{P-p} \hat\beta_{P-k}] (\gamma_{P-k}-1) \hat \beta_{k;j}\,,
\end{split}
\\
\bm V_r &=  \omega_{a^*} \gamma_{P-k}
 \left\{ r \cdot \boldsymbol \beta_{P-k} + (\gamma_{P-p}-1) [\hat\beta_{P-p} \boldsymbol \beta_{P-k}] \hat\beta_{P-p}\right\}
\\
\bm V_r' &=  \omega_{a'^*} \gamma_{P-p} 
\left\{ \boldsymbol \beta_{P-p} \cdot r + (\gamma_{P-k}-1) [\boldsymbol \beta_{P-p} \hat \beta_{P-k}] \hat\beta _k \right\}
\\
C_0 &=  \omega_{a'^*} \omega_{a^*} \gamma_{P-p} \gamma_{P-k} [\boldsymbol \beta_{P-p} \boldsymbol \beta_{P-k}]\,,
\end{align}
in which we have adopted the notation
\begin{equation}
[r \cdot \bm V]_j = r_{jk} \bm V_k\,,\qquad
[\bm V \cdot r]_k = \bm V_k r_{kj}\,.
\end{equation}
The $r$ products linear in pair momenta are
\begin{align}
[\bm p \bm a] &= \bm V'_{p,r} \cdot \bm a^* + C_{p,r}\,,
\qquad
[\bm p \bm b] = -\bm V'_{p,r} \cdot \bm a^* + C_{p,r}\,,
\\
[\bm a'  \bm k] &= \bm a'^* \cdot \bm V_{k,r} + C_{k,r}\,,
\qquad
[\bm b' \bm k] = -\bm a'^* \cdot \bm V_{k,r} + C_{k,r}\,,
\end{align}
where
\begin{align}
\bm V'_{p,r} &= \bm p \cdot r + \hat\beta_{P-k} (\gamma_{P-k}-1) [\bm p  \hat\beta_{P-k}]\,,
\quad
\bm V_{k,r} = r \cdot \bm k  + \hat\beta_{P-p} (\gamma_{P-p}-1) [\hat\beta_{P-p} \bm k]\,,
\\
C_{p,r} &= \omega_{a^*} \gamma_{P-k} [\bm p  \boldsymbol \beta_{P-k}]\,,
\qquad
C_{k,r} = \omega_{a'^*} \gamma_{P-p} [\boldsymbol \beta_{P-p} \bm k]\,.
\end{align}
The remainder of the implementation follows that for $\cK_B$.

\section{Irreps contributing to projection matrices}
\label{app:projdims}

In \Cref{tab:d1,tab:d2,tab:d3,tab:d4,tab:d5,tab:d6,tab:d7}
we list the irreps that appear, along with the dimension of the corresponding
projection matrices, in the different orbit types for each of the little groups for 
$\{\ell,s\}= \{0,0\}$ and  $\{1,1\}$, which are the choices we use in our numerical exploration.
If one knows how many orbits are active (which depends upon $E$, $\bm P$ and the spectator momentum),
then one can use these tables to determine how many eigenvalues are present in a given irrep.
This is not, however, the same as the number of solutions to the quantization condition, since many
eigenvalues will not lead to a solution~\cite{\dwave}. A simple example of this is that the $(000)$ orbit cannot
lead to solutions due to the antisymmetry of the state, but contains eigenvalues nevertheless.

\begin{table}[tbh]
\centering
\begin{tabular}{c|ccccccc}
 &\multicolumn{7}{c}{orbit types}\\
\text{irrep} &
$(000)_1$ & $(00a)_6$ & $(aa0)_{12}$ & $(aaa)_8$ & $(ab0)_{24}$ & $(aab)_{24}$ & $(abc)_{48}$ \\
 \hline
$ G_{1g}[2]$ & (2,0) & (2,10) & (2,18) & (2,12) & (4,36) & (4,36) & (8,72) \\
$ G_{2g}[2]$ & (0,0) & (0,8) & (2,18) & (2,12) & (4,36) & (4,36) & (8,72) \\
$ H_g[4]$ & (0,0) & (4,36) & (8,72) & (4,48) & (16,144) & (16,144) & (32,288) \\
$ G_{1u}[2]$ & (0,4) & (2,10) & (2,18) & (2,12) & (4,36) & (4,36) & (8,72) \\
$ G_{2u}[2]$ & (0,2) & (0,8) & (2,18) & (2,12) & (4,36) & (4,36) & (8,72) \\
$ H_u[4]$ & (0,12) & (4,36) & (8,72) & (4,48) & (16,144) & (16,144) & (32,288) \\
\hline
total & (2,18) & (12,108) & (24,216) & (16,144) & (48,432) & (48,432) & (96,864) \\
\end{tabular}
\caption{Dimension of irrep projection sub-blocks for each orbit type
and angular momentum for the frame with $\bm P=(0,0,0)$.
The triplets of results correspond to $\{\ell,s\}=\{0,0\}$, $\{1,1\}$, and $\{2,0\}$, respectively.
Only fermionic irreps of $O_h^D$ appear;
their dimensions are listed in square parentheses.
Note that the dimensions of the projectors includes the degeneracies of the representations.
Thus, for example, all entries in the $H_g[4]$ row must be multiples of $4$.
The bottom row gives the sum of the rows above,
 which equals $2 d_{\rm orbit}\times (1,3\times3)$,
 where $d_{\rm orbit}$ is the number of elements in the orbit,
 which is given as a subscript for each orbit type.
 In the labelling of orbit types, roman letters are all nonzero and different from one another.
\label{tab:d1}}
\end{table}

\begin{table}[tbh]
\centering
\begin{tabular}{c|cccc}
& \multicolumn{4}{c}{orbit types}\\
\text{irrep} & $(00z)_1$ & $(a0z)_4$ & $(aaz)_4$ & $(abz)_8$ \\
\hline
$G_1[1]$ & (2,10) & (4,36) & (4,36) & (8,72) \\
$G_2[1]$ & (0,8) & (4,36) & (4,36) & (8,72) \\
\hline
total & (2,18) & (8,72)& (8, 72)& (16,144)
\end{tabular}
\caption{As for Table~\ref{tab:d1} but for frames with $\bm P=(0,0,a)$.
\label{tab:d2}}
\end{table}

\begin{table}[tbh]
\centering
\begin{tabular}{c|cccc}
 &\multicolumn{4}{c}{orbit types}\\
\text{irrep} & $(xx0)_1$ & $(xxa)_2$ & $(xy0)_2$ & $(xya)_4$ \\
\hline
$G[2]$ & (2,18) & (4,36) & (4,36) & (8,72) \\
\end{tabular}
\caption{As for Table~\ref{tab:d1} but for frames with $\bm P=(a,a,0)$.
\label{tab:d3}}
\end{table}

\begin{table}[tbh]
\centering
\begin{tabular}{c|ccc}
& \multicolumn{3}{c}{orbit types}\\
\text{irrep} & $(xxx)_1$ & $(xxy)_3$ & $(xyz)_6$ \\
\hline
$F_1[1]$ & (0,3) & (1,9) & (2,18) \\
$F_2[1]$ & (0,3) & (1,9) & (2,18) \\
$G[2]$ & (2,12) & (4,36)& (8,72) \\
\hline
total & (2,18) & (6,54)& (12,108)
\end{tabular}
\caption{As for Table~\ref{tab:d1} but for frames with $\bm P=(a,a,a)$.
\label{tab:d4}}
\end{table}

\begin{table}[tbh]
\centering
\begin{tabular}{c|cc}
& \multicolumn{2}{c}{orbit types}\\
\text{irrep} & $(xy0)_1$ & $(xya)_2$ \\
\hline
$F_1[1]$ & (1,9) & (2,18) \\
$F_2[1]$ & (1,9) & (2,18) \\
\hline
total & (2,18) & (4,36)
\end{tabular}
\caption{As for Table~\ref{tab:d1} but for frames with $\bm P=(a,b,0)$.
\label{tab:d5}}
\end{table}

\begin{table}[tbh]
\centering
\begin{tabular}{c|cc}
 &\multicolumn{2}{c}{orbit types}\\
\text{irrep} & $(xxz)_1$ & $(xyz)_2$ \\
\hline
$F_1[1]$ & (1,9) & (2,18) \\
$F_2[1]$ & (1,9) & (2,18) \\
\hline
total & (2,18) & (4,36)
\end{tabular}
\caption{As for Table~\ref{tab:d1} but for $\bm P=(a,a,b)$.
\label{tab:d6}}
\end{table}

\begin{table}[tbh]
\centering
\begin{tabular}{c|c}
 &\multicolumn{1}{c}{orbit types}\\
\text{irrep} & $(abd)_1$ \\
\hline
$F[1]$ & (2,18) \\
\end{tabular}
\caption{As for Table~\ref{tab:d1} but for $\bm P=(a,b,c)$.
\label{tab:d7}}
\end{table}

\clearpage
\newpage
\section{Numerical results for $\mathcal K_{\rm df,3}=0$}
\label{app:moreframes}

In this appendix, we collect numerical results for the spectrum
with $\Kdf=0$ in the frames with $n_P^2=0-5$, so as to allow cross-checking of our implementation of the quantization condition.
These are given in \Cref{tab:nP0noKdf,tab:nP1noKdf,tab:nP2noKdf,tab:nP3noKdf,tab:nP4noKdf,tab:nP5noKdf,tab:nP1noKdfL}.
All results are for $M_\pi/M_N=0.15$, with the box size chosen so that $M_N=20$ or $80/3$,
as noted in the captions.
The results for $n_P^2=0$, $1$, and $3$ are displayed in \Cref{sec:num}---see \Cref{fig:nP0noKdf,fig:nP1noKdf,fig:nP3noKdf}, respectively.

\begin{table}[htb]
\centering
\begin{tabular}{l | cccccc}
free  & $G_{1g}$ & $G_{2g}$ & $H_g$ & $G_{1u}$ & $G_{2u}$ & $H_u$ \\
\hline
3.0964 & 3.0901 & $-$ & 3.0887 & 3.0914  & 3.0932 & 3.0834, 3.0848\\
            &             & &             & 3.0968 &             & 3.1069 \\
\hline
3.1885 & 3.1789 & 3.1415 & 3.1529, 3.1781 & 3.1694 & 3.1608 & 3.1607, 3.1687\\
3.1906 & 3.1835 & 3.1786 & 3.1808, 3.1852 & 3.1833 & 3.1671 & 3.1721, 3.1770\\
            & 3.1927 & 3.1859 & 3.1874, 3.1894 & 3.1954 & 3.1767 & 3.1785, 3.1797\\
            & 3.2045 & 3.1936 & 3.1936, 3.1969 & 3.1980 & 3.1830 & 3.1832, 3.1859\\
            & 3.2148 & 3.2069 & 3.2007, 3.2056 & 3.2041 & 3.1877 & 3.1889, 3.1903\\
            &             &             &                          & 3.2083 & 3.1989 & 3.1964, 3.2071\\
            &             &             &                          & 3.2099 & 3.2094 & 3.2075, 3.2187\\
\hline
\end{tabular}
\caption{Lowest two bands of energy levels in the $n_P^2=0$ frame, in units of $M_N$, after inclusion of two-particle interactions but with $\Kdf=0$. Results are for $M_N L=20$, $M_\pi/M_N=0.15$.
Also shown are the energies of the corresponding free levels, which are grouped into clusters whose energies are equal in the nonrelativistic limit (see \Cref{tab:free1}). 
\label{tab:nP0noKdf}}
\end{table}

\begin{table}[htb]
\centering
\begin{tabular}{l | cc}
free  & $G_{1}$ & $G_{2}$ \\
\hline
3.0482 & 3.0430 & $-$\\
\hline
3.1424 & 3.1248, 3.1289, 3.1314, 3.1346  & 3.1169, 3.1194, 3.1211, 3.1292\\
3.1446 & 3.1351, 3.1385, 3.1408, 3.1428 & 3.1297, 3.1365, 3.1387, 3.1406 \\
 & 3.1468, 3.1495, 3.1508, 3.1552 & 3.1448, 3.1470 ,3.1544, 3.1555\\
 & 3.1612 & \\
\hline
\end{tabular}
\caption{As in \Cref{tab:nP0noKdf}, but for the $n_P^2=1$ frame.
\label{tab:nP1noKdf}}
\end{table}

\begin{table}[htb]
\centering
\begin{tabular}{l | c}
free  & $G$  \\
\hline
3.0943 &  3.0708, 3.0830, 3.0881 \\
3.0964 &  3.0975, 3.1022 \\
\hline
\end{tabular}
\caption{As in \Cref{tab:nP0noKdf}, but for the $n_P^2=2$ frame. 
\label{tab:nP2noKdf}}
\end{table}

\begin{table}[htb]
\centering
\begin{tabular}{l | c c || l | c c }
\multicolumn{3}{c||}{$M_N L=20,\ M_\pi L=3$} &
\multicolumn{3}{c}{$M_N L = 80/3,\ M_\pi L = 4$} \\
\hline
free   & $F_1/F_2$ & $G$ & free & $F_1/F_2$ & $G$ \\
\hline
3.1385 & 3.1083, 3.1154* & 3.1170, 3.1231 & 3.0801 & 3.0645, 3.0713 & 3.0650, 3.0698\\
3.1424 & 3.1172, 3.1302  & 3.1273, 3.1307 & 3.0814 & 3.0779, 3.0800 & 3.0712, 3.0723\\
3.1446 & 3.1388, 3.1426  & 3.1339, 3.1356 & 3.0822 & 3.0816, 3.0861 & 3.0754, 3.0774\\
            & 3.1486, 3.1556  & 3.1426, 3.1445 & & & 3.0799, 3.0812\\
            &                           & 3.1497, 3.1536 & & & 3.0831, 3.0849\\
            &                           & 3.1592 & & &3.08833 \\
\hline
\end{tabular}
\caption{As in \Cref{tab:nP0noKdf}, but for the $n_P^2=3$ frame.
Levels in the $F_1$ and $F_2$ irreps are degenerate.
The asterisk denotes a level with an unphysical residue, which is discussed in the text. 
Here we show results both for $M_N L=20$ and $M_N L=80/3$.
\label{tab:nP3noKdf}}
\end{table}

\begin{table}[htb]
\centering
\begin{tabular}{l | c c }
free  & $G_1$ & $G_2$  \\
\hline
3.0964 &  3.0911 & $-$ \\
\hline
3.1810 & 3.1699, 3.1752, 3.1768, 3.1786 & 3.1649, 3.1661, 3.1685, 3.1761 \\
3.1885 & 3.1802, 3.1856, 3.1884, 3.1896 & 3.1792, 3.1842, 3.1864, 3.1876 \\
3.1906 & 3.1921, 3.1958, 3.1966, 3.2006 & 3.1900, 3.1919, 3.2005, 3.2029\\
             & 3.2079 & \\
\hline
\end{tabular}
\caption{As in \Cref{tab:nP0noKdf}, but for the $n_P^2=4$ frame.
\label{tab:nP4noKdf}}
\end{table}

\begin{table}[htb]
\centering
\begin{tabular}{l | c  }
free  & $F_1/F_2$  \\
\hline
3.1424 &  3.1174, 3.1299, 3.1361 \\
3.1446 &  3.1438, 3.1487  \\
\hline
\end{tabular}
\caption{As in \Cref{tab:nP0noKdf}, but for the $n_P^2=5$ frame. 
Levels in the $F_1$ and $F_2$ irreps are degenerate.
\label{tab:nP5noKdf}}
\end{table}

\begin{table}[htb]
\centering
\begin{tabular}{l | cc}
free  & $G_{1}$ & $G_{2}$ \\
\hline
3.0274& 3.0229 & $-$\\
\hline
3.0814 
& 3.0678, 3.0698, 3.0717, 3.0742
& 3.0644, 3.0693, 3.0710, 3.0735
\\
3.0822 
& 3.0744, 3.0757, 3.0778, 3.0802 
& 3.0749, 3.0765, 3.0794, 3.0797
\\
& 3.0816, 3.0830, 3.0835, 3.0865
& 3.0801, 3.0820, 3.0854, 3.0869
 \\
& 3.0890 
& \\
\hline
\end{tabular}
\caption{As in \Cref{tab:nP0noKdf}, but for the $n_P^2=0$ frame with $M_N L=80/3$.
\label{tab:nP1noKdfL}}
\end{table}

\clearpage
\section{Unphysical levels introduced by $\Kdf$}
\label{app:unphys}

In this appendix we describe the unphysical solutions that we have found in the vicinity of the noninteracting energies when $\Kdf$ in nonzero.
These are different from those described in \Cref{sec:num:Kdf} which are associated with large values of $\Kdf$. In particular, in all cases we have studied, the unphysical solutions described here disappear as $M_\pi L$ is increased.

The presence of unphysical solutions near to noninteracting energies for small enough $M_\pi L$ appears to be generic, although the detailed form (e.g. the number of unphysical crossings, and the relative order of physical and unphysical crossings) varies.
To be concrete, the examples we show are for the second band of $G_1$ levels for $n_P^2=1$ as we turn on the $\cK_A$ term. The physical levels for this band are shown in \Cref{fig:nP1G1KBband2}.

\begin{figure}[htb!]
    \centering
    \includegraphics[width=\textwidth]{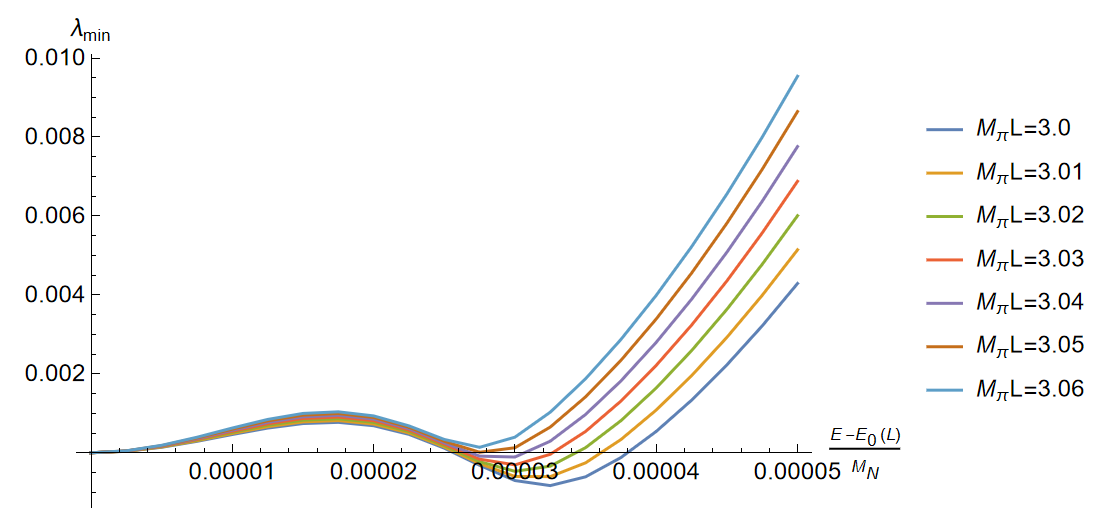}
    \caption{Dependence of the smallest magnitude eigenvalue of $(\Kdf+F_3^{-1})/M_N$ as a function of $[E-E_0(L)]/M_N$, for $c_A=-50$, $M_\pi/M_N=0.15$, and values of $M_\pi L$ shown in the legend. This is for the second band of levels in the $G_1$ irrep, with $c_B=0$. $E_0(L)$ is the lower of the noninteracting energies in this band, given in \Cref{eq:E0lower}.
    \label{fig:cAm50lower}}
\end{figure}

We show in \Cref{fig:cAm50lower} the behavior of the smallest magnitude eigenvalue of 
$\Kdf+F_3^{-1}$ just above the lowest of the two noninteracting energies, which is given by
\begin{equation}
    E_0(L) = M_N + \sqrt{M_N^2 + (2\pi/L)^2} + \sqrt{M_N^2 + 2 (2\pi/L)^2}\,.
    \label{eq:E0lower}
\end{equation}
In this case (unlike that discussed in \Cref{sec:num:Kdf0}) a physical solution occurs when the eigenvalue passes through zero from below.
We see that there is a double zero at the noninteracting energy---a phenomenon discussed in \Cref{sec:num:Kdf} and which we ignore for reasons given there.
For $M_\pi L \lesssim 3.04$ we then find an unphysical-physical pair of crossings higher in energy. As $M_\pi L$ increases to $3.05$ and above, the crossings coalesce into a double-zero and then disappear.
This is the behavior expected for solutions that are due to exponentially-suppressed errors,
as already seen in \Cref{fig:unphysical}.

Similar behavior is seen near the lower noninteracting
energy for all values of $c_A$ except for the
range $-5 \lesssim c_A \lesssim 0.008$, within which there is only the double zero at the noninteracting energy.
As $c_A$ approaches the lower limit of this range from below, the unphysical and physical solutions approach one another, combine into a double zero, and then disappear.
By contrast, as $c_A$ approaches the upper limit of the range from above, the unphysical solution approaches $E_0$, while the physical one approaches the value of a physical crossing at $\Kdf=0$ ($E_0+0.000353 M_N$).
At the same time, the next lowest physical crossing approaches $E_0$ from below, and it is
this crossing that combines with the unphysical one and disappears.
This phenomenon leads to the small horizontal ``jump'' seen in the (yellow) eighth curve from the left in
\Cref{fig:nP1G1KBband2}.
The lack of smoothness in this curve is indicative of the fact that, in the presence of unphysical solutions, the choice of which solutions are physical is ambiguous.

\begin{figure}[htb!]
    \centering
    \includegraphics[width=\textwidth]{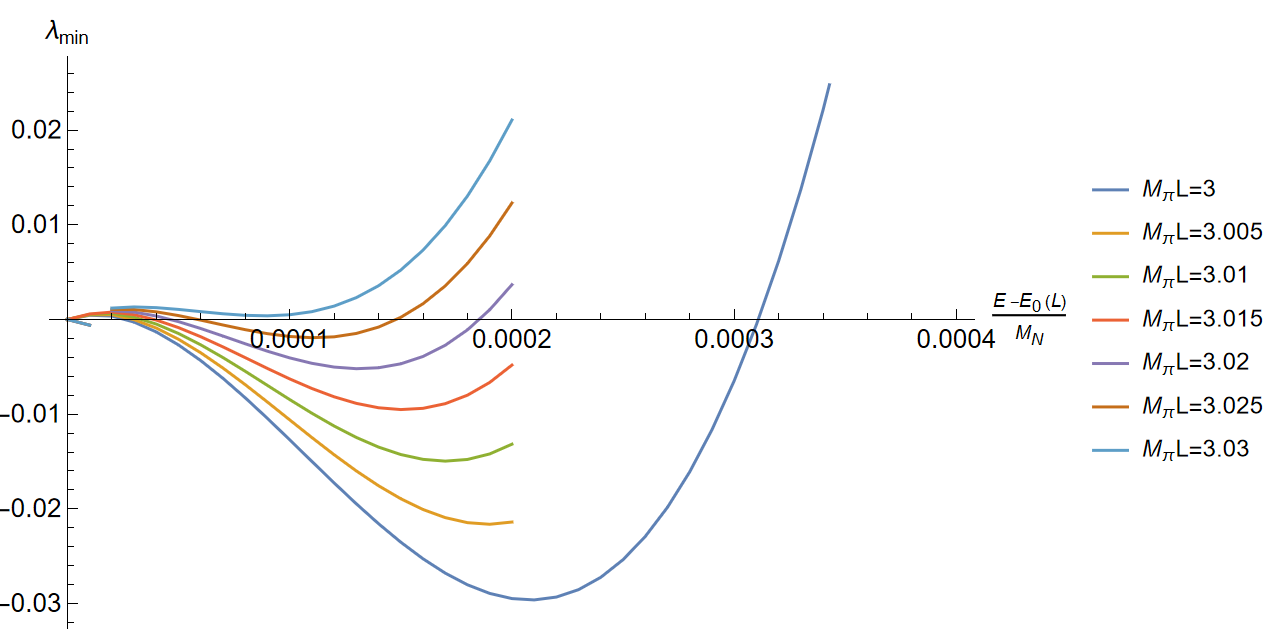}
    \caption{Dependence of the smallest magnitude eigenvalue of $(\Kdf+F_3^{-1})/M_N$ as a function of $[E-E'_0(L)]/M_N$, for $c_A=-50$, $M_\pi/M_N=0.15$, and values of $M_\pi L$ shown in the legend. This is for the second band of levels in the $G_1$ irrep, with $c_B=0$. $E'_0(L)$ is the upper of the noninteracting energies in this band, given in \Cref{eq:E0upper}.
    \label{fig:cAm50upper}}
\end{figure}

The behavior just above the upper noninteracting energy is qualitatively similar, as shown in \Cref{fig:cAm50upper}. Here the noninteracting energy is given by
\begin{equation}
    E'_0(L) = 3\sqrt{M_N^2 + (2\pi/L)^2}\,.
    \label{eq:E0upper}
\end{equation}
The unphysical-physical pair disappear in this case for $M_\pi L\gtrsim 3.03$.
The pair are present for all $c_A$ except for $-5 \lesssim c_A \lesssim 0$,
a range that is slightly smaller than that for the lower noninteracting level.
Again, the pair coalesce as $c_A$ approaches the lower end of this range, while the
unphysical level coalesces with the next lower physical level as $c_A$ approaches the upper end.
This leads to horizontal jump in the (purple) ninth curve from the left in 
\Cref{fig:nP1G1KBband2}.

In the examples we have shown, the unphysical solutions disappear for $M_\pi L$ only slightly larger than $3$. In other cases, one needs to go to higher values, in one example as high as $M_\pi L \approx 6$. Clearly, in any future application of the formalism to actual lattice QCD spectra, it will be necessary to perform a thorough search for unphysical solutions at the parameters being used.

\bibliographystyle{JHEP}
\bibliography{ref.bib}

@article{Jackura:2025wbw,
    author = "Jackura, Andrew W. and Chambers, Nicholas C. and Brice{\~n}o, Ra{\'u}l A.",
    title = "{Symmetrizing relativistic three-body partial wave amplitudes}",
    eprint = "2507.14098",
    archivePrefix = "arXiv",
    primaryClass = "hep-ph",
    month = "7",
    year = "2025"
}

@article{Briceno:2024ehy,
    author = "Brice{\~n}o, Ra{\'u}l A. and Costa, Caroline S. R. and Jackura, Andrew W.",
    title = "{Partial-wave projection of relativistic three-body amplitudes}",
    eprint = "2409.15577",
    archivePrefix = "arXiv",
    primaryClass = "hep-ph",
    reportNumber = "JLAB-THY-24-4199",
    doi = "10.1103/PhysRevD.111.036029",
    journal = "Phys. Rev. D",
    volume = "111",
    number = "3",
    pages = "036029",
    year = "2025"
}

@article{Dawid:2023kxu,
    author = "Dawid, Sebastian M. and Islam, Md Habib E. and Briceno, Raul A. and Jackura, Andrew W.",
    title = "{Evolution of Efimov states}",
    eprint = "2309.01732",
    archivePrefix = "arXiv",
    primaryClass = "nucl-th",
    doi = "10.1103/PhysRevA.109.043325",
    journal = "Phys. Rev. A",
    volume = "109",
    number = "4",
    pages = "043325",
    year = "2024"
}

@article{Briceno:2025yuq,
    author = "Brice{\~n}o, Ra{\'u}l A. and Hansen, Maxwell T. and Jackura, Andrew W. and Edwards, Robert G. and Thomas, Christopher E.",
    title = "{Isotensor $\pi\pi\pi$ scattering with a $\rho$ resonant subsystem from QCD}",
    eprint = "2510.24894",
    archivePrefix = "arXiv",
    primaryClass = "hep-lat",
    reportNumber = "JLAB-THY-25-4592",
    month = "10",
    year = "2025"
}

@article{Hoppe:2019uyw,
    author = "Hoppe, J. and Drischler, C. and Hebeler, K. and Schwenk, A. and Simonis, J.",
    title = "{Probing chiral interactions up to next-to-next-to-next-to-leading order in medium-mass nuclei}",
    eprint = "1904.12611",
    archivePrefix = "arXiv",
    primaryClass = "nucl-th",
    doi = "10.1103/PhysRevC.100.024318",
    journal = "Phys. Rev. C",
    volume = "100",
    number = "2",
    pages = "024318",
    year = "2019"
}

@article{Drischler:2020hwi,
    author = "Drischler, C. and Furnstahl, R. J. and Melendez, J. A. and Phillips, D. R.",
    title = "{How Well Do We Know the Neutron-Matter Equation of State at the Densities Inside Neutron Stars? A Bayesian Approach with Correlated Uncertainties}",
    eprint = "2004.07232",
    archivePrefix = "arXiv",
    primaryClass = "nucl-th",
    doi = "10.1103/PhysRevLett.125.202702",
    journal = "Phys. Rev. Lett.",
    volume = "125",
    number = "20",
    pages = "202702",
    year = "2020"
}

@article{Machleidt:2024bwl,
    author = "Machleidt, Ruprecht and Sammarruca, Francesca",
    title = "{Recent advances in chiral EFT based nuclear forces and their applications}",
    eprint = "2402.14032",
    archivePrefix = "arXiv",
    primaryClass = "nucl-th",
    doi = "10.1016/j.ppnp.2024.104117",
    journal = "Prog. Part. Nucl. Phys.",
    volume = "137",
    pages = "104117",
    year = "2024"
}

@inproceedings{Aoki:2025abn,
    author = "Aoki, Sinya",
    title = "{Interaction between two hadrons in lattice QCD}",
    booktitle = "{11th International Workshop on Chiral Dynamics}",
    eprint = "2502.20671",
    archivePrefix = "arXiv",
    primaryClass = "hep-lat",
    reportNumber = "YITP-25-33",
    month = "2",
    year = "2025"
}

@inproceedings{Green:2025rel,
    author = "Green, Jeremy R.",
    title = "{Status of two-baryon scattering in lattice QCD}",
    booktitle = "{11th International Workshop on Chiral Dynamics}",
    eprint = "2502.15546",
    archivePrefix = "arXiv",
    primaryClass = "hep-lat",
    reportNumber = "DESY-25-027",
    month = "2",
    year = "2025"
}

@article{BaSc:2025yhy,
    author = "Bulava, John and others",
    collaboration = "BaSc",
    title = "{Di-nucleons do not form bound states at heavy pion mass}",
    eprint = "2505.05547",
    archivePrefix = "arXiv",
    primaryClass = "hep-lat",
    reportNumber = "LLNL-JRNL-2005660",
    month = "5",
    year = "2025"
}

@article{Detmold:2024iwz,
    author = "Detmold, William and Illa, Marc and Jay, William I. and Parre{\~n}o, Assumpta and Perry, Robert J. and Shanahan, Phiala E. and Wagman, Michael L.",
    collaboration = "NPLQCD",
    title = "{Constraints on the finite volume two-nucleon spectrum at m{\ensuremath{\pi}}{\ensuremath{\approx}}806{\,}{\,}MeV}",
    eprint = "2404.12039",
    archivePrefix = "arXiv",
    primaryClass = "hep-lat",
    reportNumber = "FERMILAB-PUB-24-0126-T, MIT-CTP/5700",
    doi = "10.1103/PhysRevD.111.114501",
    journal = "Phys. Rev. D",
    volume = "111",
    number = "11",
    pages = "114501",
    year = "2025"
}

@article{Ishii:2012ssm,
    author = "Ishii, Noriyoshi and Aoki, Sinya and Doi, Takumi and Hatsuda, Tetsuo and Ikeda, Yoichi and Inoue, Takashi and Murano, Keiko and Nemura, Hidekatsu and Sasaki, Kenji",
    collaboration = "HAL QCD",
    title = "{Hadron{\textendash}hadron interactions from imaginary-time Nambu{\textendash}Bethe{\textendash}Salpeter wave function on the lattice}",
    eprint = "1203.3642",
    archivePrefix = "arXiv",
    primaryClass = "hep-lat",
    doi = "10.1016/j.physletb.2012.04.076",
    journal = "Phys. Lett. B",
    volume = "712",
    pages = "437--441",
    year = "2012"
}

@article{Ishii:2006ec,
    author = "Ishii, N. and Aoki, S. and Hatsuda, T.",
    title = "{The Nuclear Force from Lattice QCD}",
    eprint = "nucl-th/0611096",
    archivePrefix = "arXiv",
    reportNumber = "UTCCS-P-30, TKYNT-06-19, UTHEP-534",
    doi = "10.1103/PhysRevLett.99.022001",
    journal = "Phys. Rev. Lett.",
    volume = "99",
    pages = "022001",
    year = "2007"
}

@article{Yan:2025mdm,
    author = {Yan, Haobo and Mai, Maxim and Garofalo, Marco and Feng, Yuchuan and D{\"o}ring, Michael and Liu, Chuan and Liu, Liuming and Mei{\ss}ner, Ulf-G. and Urbach, Carsten},
    title = "{Emergence of the $\pi(1300)$ Resonance from Lattice QCD}",
    eprint = "2510.09476",
    archivePrefix = "arXiv",
    primaryClass = "hep-lat",
    month = "10",
    year = "2025"
}

@article{Hansen:2025oag,
    author = "Hansen, Maxwell T. and Romero-L{\'o}pez, Fernando and Sharpe, Stephen R.",
    title = "{Finite-volume formalism for $N\pi\pi$ at maximal isospin}",
    eprint = "2509.24778",
    archivePrefix = "arXiv",
    primaryClass = "hep-lat",
    month = "9",
    year = "2025"
}

@article{Dawid:2025zxc,
    author = {Dawid, Sebastian M. and Draper, Zachary T. and Hanlon, Andrew D. and H{\"o}rz, Ben and Morningstar, Colin and Romero-L{\'o}pez, Fernando and Sharpe, Stephen R. and Skinner, Sarah},
    title = "{QCD Predictions for Physical Multimeson Scattering Amplitudes}",
    eprint = "2502.14348",
    archivePrefix = "arXiv",
    primaryClass = "hep-lat",
    reportNumber = "MIT-CTP/5845",
    doi = "10.1103/6nql-yrhw",
    journal = "Phys. Rev. Lett.",
    volume = "135",
    number = "2",
    pages = "021903",
    year = "2025"
}

@article{Dawid:2025doq,
    author = {Dawid, Sebastian M. and Draper, Zachary T. and Hanlon, Andrew D. and H{\"o}rz, Ben and Morningstar, Colin and Romero-L{\'o}pez, Fernando and Sharpe, Stephen R. and Skinner, Sarah},
    title = "{Two- and three-meson scattering amplitudes with physical quark masses from lattice QCD}",
    eprint = "2502.17976",
    archivePrefix = "arXiv",
    primaryClass = "hep-lat",
    reportNumber = "MIT-CTP/5846",
    doi = "10.1103/bx16-lp3r",
    journal = "Phys. Rev. D",
    volume = "112",
    number = "1",
    pages = "014505",
    year = "2025"
}

@article{Dawid:2025wsn,
    author = "Dawid, Sebastian M. and Romero-L{\'o}pez, Fernando and Sharpe, Stephen R.",
    title = "{Comparison of integral equations used to study $ {T}_{cc}^{+} $ for a stable D$^{*}$}",
    eprint = "2505.05466",
    archivePrefix = "arXiv",
    primaryClass = "nucl-th",
    doi = "10.1007/JHEP09(2025)058",
    journal = "JHEP",
    volume = "09",
    pages = "058",
    year = "2025"
}

@article{Jackura:2023qtp,
    author = "Jackura, Andrew W. and Brice\~no, Ra\'ul A.",
    title = "{Partial-wave projection of the one-particle exchange in three-body scattering amplitudes}",
    eprint = "2312.00625",
    archivePrefix = "arXiv",
    primaryClass = "hep-ph",
    doi = "10.1103/PhysRevD.109.096030",
    journal = "Phys. Rev. D",
    volume = "109",
    number = "9",
    pages = "096030",
    year = "2024"
}

@article{Xiao:2024dyw,
    author = "Xiao, Qi-Chao and Pang, Jin-Yi and Wu, Jia-Jun",
    title = "{Lattice spectra of the DDK three-body system with Lorentz covariant kinematic}",
    eprint = "2408.16590",
    archivePrefix = "arXiv",
    primaryClass = "hep-lat",
    doi = "10.1103/PhysRevD.110.094517",
    journal = "Phys. Rev. D",
    volume = "110",
    number = "9",
    pages = "094517",
    year = "2024"
}

@article{Bubna:2023oxo,
    author = {Bubna, Rishabh and M\"uller, Fabian and Rusetsky, Akaki},
    title = "{Finite-volume energy shift of the three-nucleon ground state}",
    eprint = "2304.13635",
    archivePrefix = "arXiv",
    primaryClass = "hep-lat",
    doi = "10.1103/PhysRevD.108.014518",
    journal = "Phys. Rev. D",
    volume = "108",
    number = "1",
    pages = "014518",
    year = "2023"
}

@article{Yan:2024gwp,
    author = "Yan, Haobo and Mai, Maxim and Garofalo, Marco and Mei\ss{}ner, Ulf-G. and Liu, Chuan and Liu, Liuming and Urbach, Carsten",
    title = "{\ensuremath{\omega} Meson from Lattice QCD}",
    eprint = "2407.16659",
    archivePrefix = "arXiv",
    primaryClass = "hep-lat",
    doi = "10.1103/PhysRevLett.133.211906",
    journal = "Phys. Rev. Lett.",
    volume = "133",
    number = "21",
    pages = "211906",
    year = "2024"
}

@article{Draper:2023boj,
    author = {Draper, Zachary T. and Hanlon, Andrew D. and H\"orz, Ben and Morningstar, Colin and Romero-L\'opez, Fernando and Sharpe, Stephen R.},
    title = "{Interactions of \ensuremath{\pi}K, \ensuremath{\pi}\ensuremath{\pi}K and KK\ensuremath{\pi} systems at maximal isospin from lattice QCD}",
    eprint = "2302.13587",
    archivePrefix = "arXiv",
    primaryClass = "hep-lat",
    reportNumber = "MIT-CTP/5536",
    doi = "10.1007/JHEP05(2023)137",
    journal = "JHEP",
    volume = "05",
    pages = "137",
    year = "2023"
}

@article{Dawid:2024dgy,
    author = "Dawid, Sebastian M. and Romero-L\'opez, Fernando and Sharpe, Stephen R.",
    title = "{Finite- and infinite-volume study of DD\ensuremath{\pi} scattering}",
    eprint = "2409.17059",
    archivePrefix = "arXiv",
    primaryClass = "hep-lat",
    reportNumber = "MIT-CTP/5774",
    doi = "10.1007/JHEP01(2025)060",
    journal = "JHEP",
    volume = "01",
    pages = "060",
    year = "2025"
}

@article{Hansen:2024ffk,
    author = "Hansen, Maxwell T. and Romero-L\'opez, Fernando and Sharpe, Stephen R.",
    title = "{Incorporating DD\ensuremath{\pi} effects and left-hand cuts in lattice QCD studies of the T$_{cc}$(3875)$^{+}$}",
    eprint = "2401.06609",
    archivePrefix = "arXiv",
    primaryClass = "hep-lat",
    reportNumber = "MIT-CTP/5667",
    doi = "10.1007/JHEP06(2024)051",
    journal = "JHEP",
    volume = "06",
    pages = "051",
    year = "2024"
}

@article{Draper:2024qeh,
    author = "Draper, Zachary T. and Sharpe, Stephen R.",
    title = "{Three-particle formalism for multiple channels: the \ensuremath{\eta}\ensuremath{\pi}\ensuremath{\pi} + $ K\overline{K}\pi $ system in isosymmetric QCD}",
    eprint = "2403.20064",
    archivePrefix = "arXiv",
    primaryClass = "hep-ph",
    doi = "10.1007/JHEP07(2024)083",
    journal = "JHEP",
    volume = "07",
    pages = "083",
    year = "2024"
}

@inproceedings{Schaaf:2024qer,
    author = "Schaaf, Wilder and Sharpe, Stephen R.",
    title = "{Implementation of the three-neutron quantization condition}",
    booktitle = "{41st International Symposium on Lattice Field Theory}",
    eprint = "2410.14037",
    archivePrefix = "arXiv",
    primaryClass = "hep-lat",
    month = "10",
    year = "2024"
}

@article{Stoks:1994wp,
    author = "Stoks, V. G. J. and Klomp, R. A. M. and Terheggen, C. P. F. and de Swart, J. J.",
    title = "{Construction of high quality N N potential models}",
    eprint = "nucl-th/9406039",
    archivePrefix = "arXiv",
    reportNumber = "THEF-NYM-93-05",
    doi = "10.1103/PhysRevC.49.2950",
    journal = "Phys. Rev. C",
    volume = "49",
    pages = "2950--2962",
    year = "1994"
}

@article{Draper:2023xvu,
    author = "Draper, Zachary T. and Hansen, Maxwell T. and Romero-L\'opez, Fernando and Sharpe, Stephen R.",
    title = "{Three relativistic neutrons in a finite volume}",
    eprint = "2303.10219",
    archivePrefix = "arXiv",
    primaryClass = "hep-lat",
    reportNumber = "MIT-CTP/5539",
    doi = "10.1007/JHEP07(2023)226",
    journal = "JHEP",
    volume = "07",
    pages = "226",
    year = "2023"
}

@article{Dawid:2023jrj,
    author = "Dawid, Sebastian M. and Islam, Md Habib E. and Brice\~no, Ra\'ul A.",
    title = "{Analytic continuation of the relativistic three-particle scattering amplitudes}",
    eprint = "2303.04394",
    archivePrefix = "arXiv",
    primaryClass = "nucl-th",
    doi = "10.1103/PhysRevD.108.034016",
    journal = "Phys. Rev. D",
    volume = "108",
    number = "3",
    pages = "034016",
    year = "2023"
}

@article{Garofalo:2022pux,
    author = "Garofalo, Marco and Mai, Maxim and Romero-L\'opez, Fernando and Rusetsky, Akaki and Urbach, Carsten",
    title = "{Three-body resonances in the \ensuremath{\varphi}$^{4}$ theory}",
    eprint = "2211.05605",
    archivePrefix = "arXiv",
    primaryClass = "hep-lat",
    reportNumber = "MIT-CTP/5487",
    doi = "10.1007/JHEP02(2023)252",
    journal = "JHEP",
    volume = "02",
    pages = "252",
    year = "2023"
}

@article{Muller:2022oyw,
    author = {M\"uller, Fabian and Pang, Jin-Yi and Rusetsky, Akaki and Wu, Jia-Jun},
    title = {{Three-particle Lellouch-L\"uscher formalism in moving frames}},
    eprint = "2211.10126",
    archivePrefix = "arXiv",
    primaryClass = "hep-lat",
    doi = "10.1007/JHEP02(2023)214",
    journal = "JHEP",
    volume = "02",
    pages = "214",
    year = "2023"
}

@article{Jackura:2022gib,
    author = "Jackura, Andrew W.",
    title = "{Three-body scattering and quantization conditions from S-matrix unitarity}",
    eprint = "2208.10587",
    archivePrefix = "arXiv",
    primaryClass = "hep-lat",
    reportNumber = "JLAB-THY-22-3664",
    doi = "10.1103/PhysRevD.108.034505",
    journal = "Phys. Rev. D",
    volume = "108",
    number = "3",
    pages = "034505",
    year = "2023"
}

@article{Konig:2017krd,
    author = {K\"onig, Sebastian and Lee, Dean},
    title = "{Volume Dependence of N-Body Bound States}",
    eprint = "1701.00279",
    archivePrefix = "arXiv",
    primaryClass = "hep-lat",
    doi = "10.1016/j.physletb.2018.01.060",
    journal = "Phys. Lett. B",
    volume = "779",
    pages = "9--15",
    year = "2018"
}

@article{Mandula:1983wb,
 author = "Mandula, Jeffrey E. and Shpiz, Edward",
 title = "{Doubled Valued Representations of the Four-dimensional Cubic Lattice Rotation Group}",
 reportNumber = "PRINT-83-0999 (WASH-U.,ST.LOUIS)",
 doi = "10.1016/0550-3213(84)90366-3",
 journal = "Nucl. Phys. B",
 volume = "232",
 pages = "180--188",
 year = "1984"
}

@article{Blanton:2021eyf,
 author = "Blanton, Tyler D. and Romero-L\'opez, Fernando and Sharpe, Stephen R.",
 title = "{Implementing the three-particle quantization condition for \ensuremath{\pi}$^{+}$\ensuremath{\pi}$^{+}$K$^{+}$ and related systems}",
 eprint = "2111.12734",
 archivePrefix = "arXiv",
 primaryClass = "hep-lat",
 reportNumber = "MIT-CTP/5360",
 doi = "10.1007/JHEP02(2022)098",
 journal = "JHEP",
 volume = "02",
 pages = "098",
 year = "2022"
}

@article{Jackura:2019bmu,
 author = "Jackura, A. W. and Dawid, S. M. and Fern\'andez-Ram\'\i{}rez, C. and Mathieu, V. and Mikhasenko, M. and Pilloni, A. and Sharpe, S. R. and Szczepaniak, A. P.",
 title = "{Equivalence of three-particle scattering formalisms}",
 eprint = "1905.12007",
 archivePrefix = "arXiv",
 primaryClass = "hep-ph",
 reportNumber = "JLAB-THY-19-2947",
 doi = "10.1103/PhysRevD.100.034508",
 journal = "Phys. Rev. D",
 volume = "100",
 number = "3",
 pages = "034508",
 year = "2019"
}

@article{Blanton:2021llb,
 author = {Blanton, Tyler D. and Hanlon, Andrew D. and H\"orz, Ben and Morningstar, Colin and Romero-L\'opez, Fernando and Sharpe, Stephen R.},
 title = "{Interactions of two and three mesons including higher partial waves from lattice QCD}",
 eprint = "2106.05590",
 archivePrefix = "arXiv",
 primaryClass = "hep-lat",
 doi = "10.1007/JHEP10(2021)023",
 journal = "JHEP",
 volume = "10",
 pages = "023",
 year = "2021"
}

@article{Mai:2021nul,
 author = {Mai, Maxim and Alexandru, Andrei and Brett, Ruair\'\i{} and Culver, Chris and D\"oring, Michael and Lee, Frank X. and Sadasivan, Daniel},
 collaboration = "GWQCD",
 title = "{Three-Body Dynamics of the a1(1260) Resonance from Lattice QCD}",
 eprint = "2107.03973",
 archivePrefix = "arXiv",
 primaryClass = "hep-lat",
 doi = "10.1103/PhysRevLett.127.222001",
 journal = "Phys. Rev. Lett.",
 volume = "127",
 number = "22",
 pages = "222001",
 year = "2021"
}

@article{NPLQCD:2020ozd,
 author = "Beane, S. R. and others",
 collaboration = "NPLQCD, QCDSF",
 title = "{Charged multihadron systems in lattice QCD+QED}",
 eprint = "2003.12130",
 archivePrefix = "arXiv",
 primaryClass = "hep-lat",
 reportNumber = "DESY-20-028, DESY 20-28, FERMILAB-PUB-20-123-T, ICCUB-20-007, MIT-CTP/5183,
 NT@UW-20-03",
 doi = "10.1103/PhysRevD.103.054504",
 journal = "Phys. Rev. D",
 volume = "103",
 number = "5",
 pages = "054504",
 year = "2021"
}

@article{Muller:2021uur,
    author = {M{\"u}ller, Fabian and Pang, Jin-Yi and Rusetsky, Akaki and Wu, Jia-Jun},
    title = "{Relativistic-invariant formulation of the NREFT three-particle quantization condition}",
    eprint = "2110.09351",
    archivePrefix = "arXiv",
    primaryClass = "hep-lat",
    doi = "10.1007/JHEP02(2022)158",
    journal = "JHEP",
    volume = "02",
    pages = "158",
    year = "2022"
}

@article{Jackura:2020bsk,
 author = "Jackura, Andrew W. and Brice\~no, Ra\'ul A. and Dawid, Sebastian M. and Islam, Md Habib E. and McCarty, Connor",
 title = "{Solving relativistic three-body integral equations in the presence of bound states}",
 eprint = "2010.09820",
 archivePrefix = "arXiv",
 primaryClass = "hep-lat",
 reportNumber = "JLAB-THY-20-3272",
 doi = "10.1103/PhysRevD.104.014507",
 journal = "Phys. Rev. D",
 volume = "104",
 number = "1",
 pages = "014507",
 year = "2021"
}

@article{Brett:2021wyd,
 author = {Brett, Ruair\'\i{} and Culver, Chris and Mai, Maxim and Alexandru, Andrei and D\"oring, Michael and Lee, Frank X.},
 title = "{Three-body interactions from the finite-volume QCD spectrum}",
 eprint = "2101.06144",
 archivePrefix = "arXiv",
 primaryClass = "hep-lat",
 doi = "10.1103/PhysRevD.104.014501",
 journal = "Phys. Rev. D",
 volume = "104",
 number = "1",
 pages = "014501",
 year = "2021"
}

@article{Hansen:2021ofl,
 author = {Hansen, Maxwell T. and Romero-L\'opez, Fernando and Sharpe, Stephen R.},
 title = "{Decay amplitudes to three hadrons from finite-volume matrix elements}",
 eprint = "2101.10246",
 archivePrefix = "arXiv",
 primaryClass = "hep-lat",
 doi = "10.1007/JHEP04(2021)113",
 journal = "JHEP",
 volume = "04",
 pages = "113",
 year = "2021"
}

@article{Alexandru:2020xqf,
 author = {Alexandru, Andrei and Brett, Ruair\'\i{} and Culver, Chris and D\"{o}ring, Michael and Guo, Dehua and Lee, Frank X. and Mai, Maxim},
 title = "{Finite-volume energy spectrum of the $K^-K^-K^-$ system}",
 eprint = "2009.12358",
 archivePrefix = "arXiv",
 primaryClass = "hep-lat",
 doi = "10.1103/PhysRevD.102.114523",
 journal = "Phys. Rev. D",
 volume = "102",
 number = "11",
 pages = "114523",
 year = "2020"
}

@article{Hansen:2020otl,
 author = {Hansen, Maxwell T. and Brice\~no, Ra\'ul A. and Edwards, Robert G. and Thomas, Christopher E. and Wilson, David J.},
 collaboration = "Hadron Spectrum",
 title = "{Energy-Dependent $\pi^+ \pi^+ \pi^+$ Scattering Amplitude from QCD}",
 eprint = "2009.04931",
 archivePrefix = "arXiv",
 primaryClass = "hep-lat",
 reportNumber = "CERN-TH-2020-147, JLAB-THY-20-3242",
 doi = "10.1103/PhysRevLett.126.012001",
 journal = "Phys. Rev. Lett.",
 volume = "126",
 pages = "012001",
 year = "2021"
}

@article{Culver:2019vvu,
 author = {Culver, Chris and Mai, Maxim and Brett, Ruair\'\i{} and Alexandru, Andrei and D\"{o}ring, Michael},
 title = "{Three pion spectrum in the $I=3$ channel from lattice QCD}",
 eprint = "1911.09047",
 archivePrefix = "arXiv",
 primaryClass = "hep-lat",
 doi = "10.1103/PhysRevD.101.114507",
 journal = "Phys. Rev. D",
 volume = "101",
 number = "11",
 pages = "114507",
 year = "2020"
}

@article{Blanton:2021mih,
 author = "Blanton, Tyler D. and Sharpe, Stephen R.",
 title = "{Three-particle finite-volume formalism for $\pi^+\pi^+K^+$ and related systems}",
 eprint = "2105.12094",
 archivePrefix = "arXiv",
 primaryClass = "hep-lat",
 doi = "10.1103/PhysRevD.104.034509",
 journal = "Phys. Rev. D",
 volume = "104",
 number = "3",
 pages = "034509",
 year = "2021"
}

@article{Romero-Lopez:2020rdq,
 author = {Romero-L\'opez, Fernando and Rusetsky, Akaki and Schlage, Nikolas and Urbach, Carsten},
 title = "{Relativistic $N$-particle energy shift in finite volume}",
 eprint = "2010.11715",
 archivePrefix = "arXiv",
 primaryClass = "hep-lat",
 doi = "10.1007/JHEP02(2021)060",
 journal = "JHEP",
 volume = "02",
 pages = "060",
 year = "2021"
}

@article{Fischer:2020jzp,
 author = {Fischer, Matthias and Kostrzewa, Bartosz and Liu, Liuming and Romero-L\'opez, Fernando and Ueding, Martin and Urbach, Carsten},
 title = "{Scattering of two and three physical pions at maximal isospin from lattice QCD}",
 eprint = "2008.03035",
 archivePrefix = "arXiv",
 primaryClass = "hep-lat",
 doi = "10.1140/epjc/s10052-021-09206-5",
 journal = "Eur. Phys. J. C",
 volume = "81",
 number = "5",
 pages = "436",
 year = "2021"
}

@article{Muller:2020wjo,
 author = {M\"{u}ller, Fabian and Rusetsky, Akaki},
 title = {{On the three-particle analog of the Lellouch-L\"{u}scher formula}},
 eprint = "2012.13957",
 archivePrefix = "arXiv",
 primaryClass = "hep-lat",
 doi = "10.1007/JHEP03(2021)152",
 journal = "JHEP",
 volume = "21",
 pages = "152",
 year = "2020"
}

@article{Muller:2020vtt,
 author = {M\"{u}ller, Fabian and Rusetsky, Akaki and Yu, Tiansu},
 title = "{Finite-volume energy shift of the three-pion ground state}",
 eprint = "2011.14178",
 archivePrefix = "arXiv",
 primaryClass = "hep-lat",
 doi = "10.1103/PhysRevD.103.054506",
 journal = "Phys. Rev. D",
 volume = "103",
 number = "5",
 pages = "054506",
 year = "2021"
}

@article{Blanton:2020gmf,
 author = "Blanton, Tyler D. and Sharpe, Stephen R.",
 title = "{Relativistic three-particle quantization condition for nondegenerate scalars}",
 eprint = "2011.05520",
 archivePrefix = "arXiv",
 primaryClass = "hep-lat",
 doi = "10.1103/PhysRevD.103.054503",
 journal = "Phys. Rev. D",
 volume = "103",
 number = "5",
 pages = "054503",
 year = "2021"
}

@article{Pang:2020pkl,
 author = "Pang, Jin-Yi and Wu, Jia-Jun and Geng, Li-Sheng",
 title = "{$DDK$ system in finite volume}",
 eprint = "2008.13014",
 archivePrefix = "arXiv",
 primaryClass = "hep-lat",
 doi = "10.1103/PhysRevD.102.114515",
 journal = "Phys. Rev. D",
 volume = "102",
 number = "11",
 pages = "114515",
 year = "2020"
}

@article{Blanton:2020gha,
 author = "Blanton, Tyler D. and Sharpe, Stephen R.",
 title = "{Alternative derivation of the relativistic three-particle quantization condition}",
 eprint = "2007.16188",
 archivePrefix = "arXiv",
 primaryClass = "hep-lat",
 doi = "10.1103/PhysRevD.102.054520",
 journal = "Phys. Rev. D",
 volume = "102",
 number = "5",
 pages = "054520",
 year = "2020"
}

@article{Blanton:2020jnm,
 author = "Blanton, Tyler D. and Sharpe, Stephen R.",
 title = "{Equivalence of relativistic three-particle quantization conditions}",
 eprint = "2007.16190",
 archivePrefix = "arXiv",
 primaryClass = "hep-lat",
 doi = "10.1103/PhysRevD.102.054515",
 journal = "Phys. Rev. D",
 volume = "102",
 number = "5",
 pages = "054515",
 year = "2020"
}

@article{Hansen:2020zhy,
 author = {Hansen, Maxwell T. and Romero-L\'opez, Fernando and Sharpe, Stephen R.},
 title = "{Generalizing the relativistic quantization condition to include all three-pion isospin channels}",
 eprint = "2003.10974",
 archivePrefix = "arXiv",
 primaryClass = "hep-lat",
 reportNumber = "CERN-TH-2020-045",
 doi = "10.1007/JHEP07(2020)047",
 journal = "JHEP",
 volume = "07",
 pages = "047",
 year = "2020"
}

@article{Mai:2019fba,
 author = {Mai, M. and D\"{o}ring, M. and Culver, C. and Alexandru, A.},
 archivePrefix = "arXiv",
 doi = "10.1103/PhysRevD.101.054510",
 eprint = "1909.05749",
 journal = "Phys.\ Rev.\ D",
 pages = "054510",
 primaryClass = "hep-lat",
 title = "{Three-body unitarity versus finite-volume $\pi^+\pi^+\pi^+$ spectrum from lattice QCD}",
 volume = "101",
 year = "2020"
}

@article{Blanton:2019vdk,
 author = "Blanton, Tyler D. and Romero-L\'opez, Fernando and Sharpe,
 Stephen R.",
 title = "{$I = 3$ three-pion scattering amplitude from lattice
 QCD}",
 journal = "Phys. Rev. Lett.",
 volume = "124",
 year = "2020",
 number = "3",
 pages = "032001",
 doi = "10.1103/PhysRevLett.124.032001",
 eprint = "1909.02973",
 archivePrefix = "arXiv",
 primaryClass = "hep-lat",
 SLACcitation = "%%CITATION = ARXIV:1909.02973;%%"
}

@article{Romero-Lopez:2019qrt,
 author = {Romero-L\'opez, Fernando and Sharpe, Stephen R. and
 Blanton, Tyler D. and Brice\~no, Ra\'ul A. and Hansen,
 Maxwell T.},
 title = "{Numerical exploration of three relativistic particles in
 a finite volume including two-particle resonances and
 bound states}",
 journal = "JHEP",
 volume = "10",
 year = "2019",
 pages = "007",
 doi = "10.1007/JHEP10(2019)007",
 eprint = "1908.02411",
 archivePrefix = "arXiv",
 primaryClass = "hep-lat",
 reportNumber = "JLAB-THY-19-3011, CERN-TH-2019-129",
 SLACcitation = "%%CITATION = ARXIV:1908.02411;%%"
}

@article{Pang:2019dfe,
 author = {Pang, Jin-Yi and Wu, Jia-Jun and Hammer, H. -W. and
 Mei{$\ss$}ner, Ulf-G. and Rusetsky, Akaki},
 title = "{Energy shift of the three-particle system in a finite
 volume}",
 journal = "Phys. Rev.",
 volume = "D99",
 year = "2019",
 number = "7",
 pages = "074513",
 doi = "10.1103/PhysRevD.99.074513",
 eprint = "1902.01111",
 archivePrefix = "arXiv",
 primaryClass = "hep-lat",
 SLACcitation = "%%CITATION = ARXIV:1902.01111;%%"
}

@article{Briceno:2019muc,
 author = {Brice\~no, Ra\'ul A. and Hansen, Maxwell T. and Sharpe,
 Stephen R. and Szczepaniak, Adam P.},
 title = "{Unitarity of the infinite-volume three-particle
 scattering amplitude arising from a finite-volume
 formalism}",
 journal = "Phys. Rev.",
 volume = "D100",
 year = "2019",
 number = "5",
 pages = "054508",
 doi = "10.1103/PhysRevD.100.054508",
 eprint = "1905.11188",
 archivePrefix = "arXiv",
 primaryClass = "hep-lat",
 reportNumber = "JLAB-THY-19-2945, CERN-TH-2019-078",
 SLACcitation = "%%CITATION = ARXIV:1905.11188;%%"
}

@article{Horz:2019rrn,
 author = {H\"{o}rz, Ben and Hanlon, Andrew},
 title = "{Two- and three-pion finite-volume spectra at maximal
 isospin from lattice QCD}",
 journal = "Phys. Rev. Lett.",
 volume = "123",
 year = "2019",
 number = "14",
 pages = "142002",
 doi = "10.1103/PhysRevLett.123.142002",
 eprint = "1905.04277",
 archivePrefix = "arXiv",
 primaryClass = "hep-lat",
 reportNumber = "MITP/19-033",
 SLACcitation = "%%CITATION = ARXIV:1905.04277;%%"
}

@article{Blanton:2019igq,
 author = "Blanton, Tyler D. and Romero-L\'opez, Fernando and Sharpe,
 Stephen R.",
 title = "{Implementing the three-particle quantization condition
 including higher partial waves}",
 journal = "JHEP",
 volume = "03",
 year = "2019",
 pages = "106",
 doi = "10.1007/JHEP03(2019)106",
 eprint = "1901.07095",
 archivePrefix = "arXiv",
 primaryClass = "hep-lat",
 SLACcitation = "%%CITATION = ARXIV:1901.07095;%%"
}

@article{Mai:2018djl,
 author = "Mai, Maxim and D{\"{o}}ring, Michael",
 title = "{Finite-Volume Spectrum of $\pi^+\pi^+$ and
 $\pi^+\pi^+\pi^+$ Systems}",
 journal = "Phys. Rev. Lett.",
 volume = "122",
 year = "2019",
 number = "6",
 pages = "062503",
 doi = "10.1103/PhysRevLett.122.062503",
 eprint = "1807.04746",
 archivePrefix = "arXiv",
 primaryClass = "hep-lat",
 reportNumber = "JLAB-THY-18-2767",
 SLACcitation = "%%CITATION = ARXIV:1807.04746;%%"
}

@article{Briceno:2018aml,
 author = {Brice\~no, Ra\'ul A. and Hansen, Maxwell T. and Sharpe,
 Stephen R.},
 title = "{Three-particle systems with resonant subprocesses in a
 finite volume}",
 journal = "Phys. Rev.",
 volume = "D99",
 year = "2019",
 number = "1",
 pages = "014516",
 doi = "10.1103/PhysRevD.99.014516",
 eprint = "1810.01429",
 archivePrefix = "arXiv",
 primaryClass = "hep-lat",
 reportNumber = "JLAB-THY-18-2819",
 SLACcitation = "%%CITATION = ARXIV:1810.01429;%%"
}

@article{Briceno:2018mlh,
 author = {Brice\~no, Ra\'ul A. and Hansen, Maxwell T. and Sharpe,
 Stephen R.},
 title = "{Numerical study of the relativistic three-body
 quantization condition in the isotropic approximation}",
 journal = "Phys. Rev.",
 volume = "D98",
 year = "2018",
 number = "1",
 pages = "014506",
 doi = "10.1103/PhysRevD.98.014506",
 eprint = "1803.04169",
 archivePrefix = "arXiv",
 primaryClass = "hep-lat",
 reportNumber = "JLAB-THY-18-2657, CERN-TH-2018-046",
 SLACcitation = "%%CITATION = ARXIV:1803.04169;%%"
}

@article{Mai:2017bge,
 author = "Mai, M. and {D\"{o}ring}, M.",
 title = "{Three-body Unitarity in the Finite Volume}",
 journal = "Eur. Phys. J.",
 volume = "A53",
 year = "2017",
 number = "12",
 pages = "240",
 doi = "10.1140/epja/i2017-12440-1",
 eprint = "1709.08222",
 archivePrefix = "arXiv",
 primaryClass = "hep-lat",
 reportNumber = "JLAB-THY-17-2554",
 SLACcitation = "%%CITATION = ARXIV:1709.08222;%%"
}

@article{Hammer:2017kms,
 author = "Hammer, H. -W. and Pang, J. -Y. and Rusetsky, A.",
 title = "{Three particle quantization condition in a finite
 volume: 2. General formalism and the analysis of data}",
 journal = "JHEP",
 volume = "10",
 year = "2017",
 pages = "115",
 doi = "10.1007/JHEP10(2017)115",
 eprint = "1707.02176",
 archivePrefix = "arXiv",
 primaryClass = "hep-lat",
 SLACcitation = "%%CITATION = ARXIV:1707.02176;%%"
}

@article{Hammer:2017uqm,
 author = "Hammer, Hans-Werner and Pang, Jin-Yi and Rusetsky, A.",
 title = "{Three-particle quantization condition in a finite
 volume: 1. The role of the three-particle force}",
 journal = "JHEP",
 volume = "09",
 year = "2017",
 pages = "109",
 doi = "10.1007/JHEP09(2017)109",
 eprint = "1706.07700",
 archivePrefix = "arXiv",
 primaryClass = "hep-lat",
 SLACcitation = "%%CITATION = ARXIV:1706.07700;%%"
}

@article{Briceno:2017tce,
 author = "Brice\~no, R. A. and Hansen, Maxwell T. and Sharpe,
 Stephen R.",
 title = "{Relating the finite-volume spectrum and the
 two-and-three-particle $S$ matrix for relativistic systems
 of identical scalar particles}",
 journal = "Phys. Rev.",
 volume = "D95",
 year = "2017",
 number = "7",
 pages = "074510",
 doi = "10.1103/PhysRevD.95.074510",
 eprint = "1701.07465",
 archivePrefix = "arXiv",
 primaryClass = "hep-lat",
 reportNumber = "JLAB-THY-17-2400",
 SLACcitation = "%%CITATION = ARXIV:1701.07465;%%"
}

@article{Hansen:2014eka,
 author = "Hansen, Maxwell T. and Sharpe, Stephen R.",
 title = "{Relativistic, model-independent, three-particle
 quantization condition}",
 journal = "Phys. Rev.",
 volume = "D90",
 year = "2014",
 number = "11",
 pages = "116003",
 doi = "10.1103/PhysRevD.90.116003",
 eprint = "1408.5933",
 archivePrefix = "arXiv",
 primaryClass = "hep-lat",
 SLACcitation = "%%CITATION = ARXIV:1408.5933;%%"
}

@article{Hansen:2015zga,
 author = "Hansen, Maxwell T. and Sharpe, Stephen R.",
 title = "{Expressing the three-particle finite-volume spectrum in
 terms of the three-to-three scattering amplitude}",
 journal = "Phys. Rev.",
 volume = "D92",
 year = "2015",
 number = "11",
 pages = "114509",
 doi = "10.1103/PhysRevD.92.114509",
 eprint = "1504.04248",
 archivePrefix = "arXiv",
 primaryClass = "hep-lat",
 SLACcitation = "%%CITATION = ARXIV:1504.04248;%%"
}

@article{Luscher:1986n2,
 author = {L\"{u}scher, M.},
 title = "{Volume Dependence of the Energy Spectrum in Massive
 Quantum Field Theories. 2. Scattering States}",
 journal = "Commun.Math.Phys.",
 volume = "105",
 pages = "153-188",
 doi = "10.1007/BF01211097",
 year = "1986",
 reportNumber = "DESY 86/034",
 SLACcitation = "%%CITATION = CMPHA,105,153;%%",
}

@article{Luscher:1991n1,
 author = {L\"{u}scher, Martin},
 title = "{Two particle states on a torus and their relation to the
 scattering matrix}",
 journal = "Nucl.Phys.",
 volume = "B354",
 pages = "531-578",
 doi = "10.1016/0550-3213(91)90366-6",
 year = "1991",
 reportNumber = "DESY-90-131",
 SLACcitation = "%%CITATION = NUPHA,B354,531;%%",
}

@article{Luscher:1991n2,
 author = {L\"{u}scher, Martin},
 title = "{Signatures of unstable particles in finite volume}",
 journal = "Nucl.Phys.",
 volume = "B364",
 pages = "237-254",
 doi = "10.1016/0550-3213(91)90584-K",
 year = "1991",
 reportNumber = "DESY-91-052",
 SLACcitation = "%%CITATION = NUPHA,B364,237;%%",
}

@article{Morningstar:2013bda,
 author = "Morningstar, C. and Bulava, J. and Fahy, B. and Foley, J. and Jhang, Y.C.
 and Juge, K.J. and Lenkner, D. and Wong, C.H.",
 title = "{Extended hadron and two-hadron operators of definite momentum for spectrum calculations in lattice QCD}",
 eprint = "1303.6816",
 archivePrefix = "arXiv",
 primaryClass = "hep-lat",
 doi = "10.1103/PhysRevD.88.014511",
 journal = "Phys. Rev. D",
 volume = "88",
 number = "1",
 pages = "014511",
 year = "2013"
}

\end{document}